\documentclass[12pt]{amsart}
\usepackage{amssymb}
\usepackage{latexsym}
\usepackage{epic, eepic, graphicx}
\usepackage{graphicx} \DeclareGraphicsExtensions{.ps}

\newcommand{\tB}{\widetilde B}

\newcommand{\tS}{\tilde S}

\newcommand{\tBh}{{\widetilde B_h}}

\newcommand{\1}{{\bold 1}}
\newcommand{\bet}{\boldsymbol{\eta}}

\newcommand{\ep}{\epsilon}
\newcommand{\bep}{\boldsymbol{\epsilon}}

\newcommand{\C}{{\mathbb C}}
\newcommand{\CC}{{\mathbb C}}
\newcommand{\HH}{{\mathbb H}}
\newcommand{\II}{{\mathbb I}}
\newcommand{\NN}{{\mathbb N}}

\newcommand{\RR}{{\mathbb R}}
\newcommand{\IR}{{\mathbb R}}

\newcommand{\TT}{{\mathbb T}}
\def\t2{{\mathbb T}^2}
\def\tn{{\mathbb T}^{2n}}
\newcommand{\ZZ}{{\mathbb Z}}

\newcommand{\CI}{{\mathcal C}^\infty }
\newcommand{\CIc}{{\mathcal C}^\infty_{\rm{c}} }
\newcommand{\CIb}{{\mathcal C}^\infty_{\rm{b}} }

\newcommand{\cD}{{\mathcal D}}
\newcommand{\F}{{\mathcal F}}

\newcommand{\Hh}{{\mathcal H}}

\newcommand{\cC}{{\mathcal C}}
\newcommand{\cJ}{{\mathcal J}}
\newcommand{\cL}{{\mathcal L}}

\newcommand{\cP}{{\mathcal P}}

\newcommand{\Oo}{{\mathcal O}} 
\newcommand{\cS}{{\mathcal S}}

\newcommand{\vol}{\operatorname{vol}}

\newcommand{\half}{\frac{1}{2}}

\newcommand{\supp}{\operatorname{supp}}

\newcommand{\tr}{\operatorname{tr}}
\newcommand{\Res}{\operatorname{Res}}
\newcommand{\Spec}{\operatorname{Spec}}

\newcommand{\rest}{|}

\renewcommand{\Re}{\mathop{\rm Re}\nolimits}
\renewcommand{\Im}{\mathop{\rm Im}\nolimits}

\newcommand{\Op}{\operatorname{Op}}
\newcommand{\WF}{\operatorname{WF}}
\newcommand{\ra}{\rangle}
\newcommand{\la}{\langle}

\newcommand{\bver}{\begin{verbatim}}
\newcommand{\defi}{\stackrel{\rm{def}}{=}}
\newcommand{\defeq}{\stackrel{\rm{def}}{=}}
\def\hto0{\xrightarrow{h\to 0}}
\newcommand{\cube}{H}

\theoremstyle{plain}
\newtheorem{thm}{Theorem}
\newtheorem{prop}{Proposition}[section]

\newtheorem{lem}[prop]{Lemma}

\theoremstyle{definition}

\newtheorem*{exmple}{Example}
\newtheorem{rem}{Remark}[section]
\newtheorem{defn}[prop]{Definition}
\numberwithin{equation}{section}

\newcommand{\secref}[1]{Section~\ref{#1}}

\newcommand{\propref}[1]{Proposition~\ref{#1}}
\newcommand{\bequ}{\begin{equation}}

\def\bbbone{{\mathchoice {1\mskip-4mu {\rm{l}}} {1\mskip-4mu {\rm{l}}}
{ 1\mskip-4.5mu {\rm{l}}} { 1\mskip-5mu {\rm{l}}}}}

\newcommand{\mn}[1]{\Vert#1\Vert}
\newcommand{\set}[1]{\left\{\,#1\,\right\}}

\def\squarebox#1{\hbox to #1{\hfill\vbox to #1{\vfill}}}

\usepackage{epic, eepic}

\title
[Distribution of resonances for open quantum maps]
{Distribution of resonances for open quantum maps}
\author[S. Nonnenmacher]
{St\'ephane Nonnenmacher}
\author[M. Zworski]
{Maciej Zworski}
\address{Service de Physique Th\'eorique, 
CEA/DSM/PhT, Unit\'e de recherche associ\'e CNRS,
CEA/Saclay,\\
91191 Gif-sur-Yvette, France}
\email{snonnenmacher@cea.fr}
\address{Mathematics Department, University of California \\
Evans Hall, Berkeley, CA 94720, USA}
\email{zworski@math.berkeley.edu}

\begin{document}

\maketitle

%%%%%%%%%%%%%%%%%%%%%%%%%%%%%%%%%%%%%%%%%%%%%%%%%%%%%%%%%%%%%%%%%%%%%%%%%%%%%%%
%%%%%%%%%%%%%%%%%%%%%%%%%%%%%%%%%%%%%%%%%%%%%%%%%%%%%%%%%%%%%%%%%%%%%%%%%%%%%%%

\section{Introduction}
\label{in}

\subsection{Statement of the results}
\label{sotr}
In this paper we analyze simple models of classical chaotic 
open systems and of their quantizations. They provide a numerical 
confirmation of the {\em fractal Weyl law} for the density 
of quantum resonances of such systems. The exponent in that 
law is related to the dimension of the classical repeller of the system.
In a simplified model, a rigorous argument gives the full
resonance spectrum, which satisfies
the fractal Weyl law. Our model is similar to models recently 
studied in atomic and  mesoscopic physics (see \S \ref{rmph} below).
Before stating the main result we 
remark that in this paper we use mathematicians' notation $h$ for
what the physicists call $\hbar$. That is partly to stress that 
our $ h$ is a small parameter in asymptotic analysis, not necessarily
interpreted as the Planck constant.

\begin{thm}
\label{t:1}
 There exist families of symplectic relations, 
$\widetilde  B \subset \t2 \times \t2 $, and of their (subunitary) 
quantization, $ \tBh \in 
\cL ( \CC^N ) $, $ N = ( 2 \pi h )^{-1} $, 
such that
\begin{gather*}
\# \big\{\lambda\in\Spec ( \tBh )\; :\; |\lambda|\geq r \big\}  = 
 c ( r )\,h^{-\nu} + o ( h^{-\nu} ) \,, \\ r > 0 \,, \ \ 
 h = h_k= ( 2\pi D^{k})^{-1}\to 
0, \ \ k \rightarrow \infty \,, \\
\nu = \dim \big(\Gamma_- ( \widetilde B ) \cap W_+ ( \widetilde B ) \big) \,, \ \ 
c ( r ) = ( 2 \pi)^{-\nu} \chi_{ [ 0 , r_0 ( \widetilde B) ] } ( r ) \,, 
\ \ 0 < r_0 ( \widetilde B ) < 1 \,, 
\end{gather*}
where the integer parameter $D>1$ 
depends on 
$\widetilde B $. The set  $ \Gamma_- ( \widetilde B ) \subset 
\t2$ is the forward trapped set of $ \widetilde B $ and 
$ W_+ ( \widetilde B ) $ is the unstable manifold of $ \widetilde B $ 
at any point of $ \Gamma_- ( \widetilde B )$. The eigenvalues are counted
with multiplicities.
\end{thm}

In the model discussed in detail we took $D=3$. 
The asymptotics are actually much more precise and include 
uniform angular distribution (see Prop.~\ref{p:last}). The resonances lie on a 
lattice, % (see Fig.~\ref{f:lat}) 
and some of this structure is also seen
in numerically computed more generic situations (some numerical results have
been presented in \cite{nzJPA,nQM9,NonRub06}).
Each symplectic relation $\tB$ (or ``multivalued symplectic map'') 
is defined together with the probabilities, for any point, to be mapped to each of its
images: $\tB$ thus represents a certain stochastic process.
The quantizations $ \tBh$ quantize the relations
together with their jump probabilities in the precise sense given in \S \ref{s:qwr}.

In the models used in Theorem~\ref{t:1} we can compute the 
conductance and the shot noise power (or the closely related Fano
factor) --- see \S\ref{s:ben} and references given there for 
physics background and \S\ref{cWm} for precise definitions.

\begin{thm}
\label{t:2}
Suppose that the models in Theorem~\ref{t:1} have the openings
consisting of two "leads" of equal width (see \S 6.1 for a 
detailed description), so that each lead carries the 
same number, $ M (h) \sim h^{-1} $, of scattering channels. Then, 
the quantum conductance \eqref{eq:cond} between the two leads
satisfies 
\begin{equation}
\label{eq:t2.1}
g ( h ) = \frac{1}{2}\, M ( h )\, \big( 1 + o ( 1) \big) \,, \quad 
h = h_k \rightarrow 0 \,.
\end{equation}
The Fano
factor \eqref{e:power} is given by 
\begin{equation}
\label{eq:t2.2}
F ( h ) = \frac{11}{80}\; \frac{ M(h)^{\nu} }{g ( h ) }\, \big( 1 + o ( 1 ) \big)\,, 
\quad h=h_k \rightarrow 0 \,,
\end{equation}
where the exponent $\nu$ is the same as in Theorem~\ref{t:1}.
\end{thm}

The theorem should be interpreted as follows. In \eqref{eq:t2.1} 
we see that for a model of scattering through a chaotic cavity, 
approximately one half of the scattering channels get transmitted
from one lead to the other, the other half being reflected back 
(this is natural and well known).
Asymptotics in \eqref{eq:t2.2} are more interesting. We see that
the fractal Weyl law, $ h^{-\nu } $, appears in the expression for the Fano
factor. In the interpretation of the Fano factor in terms of 
``shot noise'' (see \S\ref{s:ben}),  $ 11/80 $ gives 
the average ``shot noise'' per ``nonclassical transmission channel''.
This number is close to the random matrix theory prediction for 
this quantity, namely $1/8$ \cite{jalabert,beenakker}. In fact,
had \eqref{eq:t2.2} come from a physical experiment rather than
an asymptotic computation, it would be regarded as being in 
a very good agreement with random matrix theory\footnote{We are grateful
to Yan Fyodorov for this amusing comment.}.

Much of the paper is devoted to rigorous definitions of the objects
appearing in the statements of the two theorems.
In this section we give some general indications,
with detailed references to previous works appearing below.

We consider the two-torus $ \t2 = [0,1) \times [0,1) $ as our classical phase 
(with coordinates $\rho=(q,p)$).
Classical observables are functions on $ \t2 $ and 
classical dynamics is given in terms of relations, $ B 
\subset \t2 \times \t2 $, which are unions of truncated graphs 
of symplectic (area and orientation preserving) maps $ \t2 \rightarrow \t2 $.
An example is given by the baker's relation
\begin{equation}
\label{eq:exb}  (\rho';\rho)=( q',p'; q, p ) \in B 
\Longleftrightarrow \left\{ \begin{array}{lll} 
q' = 3 q\,, & p' = p/3\,, & 0 \leq q \leq 1/3 \\
q' = 3 q - 2\,, & p' = (p+2) /3\,, & 2/3 \leq q < 1 \,.  \end{array}
\right. \end{equation}
This is a ``rectangular horseshoe'' modeling a Poincar\'e map
of a chaotic open system: some points (here $ \set{\rho\;:\;1/3 < q < 2/3} $) 
are thrown out ``to infinity'' at each iteration.

For relations such as $ B $ we can define the {\em forward and backward trapped sets}
(see \eqref{eq:trap} for the definition in the case of flows):
\[ \begin{split} 
&  \rho \in \Gamma_- \; \Leftrightarrow \;  \exists
\{ \rho_j \}_{j=0}^\infty \,, \ \ \ \ \,
 \rho_0 = \rho \,, \ \ 
( \rho_j ; \rho_{j-1} ) \in B \,, \ j >0  \,,\\
&  \rho \in \Gamma_+ \; \Leftrightarrow \;  \exists
\{\rho_j \}_{j=-\infty}^0 \,, \ \ \rho_0 = \rho \,, \ \ 
( \rho_{j} ; \rho_{j-1} ) \in B \,, \ j \leq 0 \,.
\end{split} \]
In the example \eqref{eq:exb}, $ \Gamma_- = C \times [0,1) $, 
$ \Gamma_+ = [0,1) \times C $, where $ C $ is the usual $ \frac13 -$Cantor
set.

We also define the {\em trapped set} $ K = \Gamma_+ \cap \Gamma_- $
and, at points of $ K $, the stable and unstable manifolds, $ W_\pm $.
In the case of the above baker's relation,
\[ 
\nu = \dim \Gamma_- \cap W_+ = \half \dim K = \dim \Gamma_+ \cap W_- = 
\frac{\log 2}{ \log 3} \,, 
\]
but for general (possibly multivalued) relations these equalities do not hold.

A quantization (in the sense made rigorous in \S \ref{s:qbr}) of
$ B $ is given by 
\begin{equation}
\label{eq:bh} 
B_h = \F_N^* \left( \begin{array}{lll} \F_{N/3} & 0 & \ 0 \\
\ 0 & 0 &  \ 0 \\
\ 0 & 0 & \F_{N/3} \end{array} \right) \,, \ \ h = ( 2 \pi N )^{-1} \,, 
\ 3 | N \,, 
\end{equation}
where $ \F_M $ is the discrete Fourier transform on $ \CC^M $.

\begin{table}[htbp]
\begin{center}
\begin{tabular}{||l|l|l|l|l|l|l|l|l||}
\hline
\  &  \ &   \  &    \  &    \  &    \  &   \  &    \  &   \ \\
$ N = 3^k $ & $ r = 0.1$ & $ r = 0.2$  & $ r = 0.3 $ & $ r = 0.4 $  & 
$ r = 0.5 $ & $ r = 0.6$  & $ r = 0.7$  & $ r = 0.8 $  \\
\  &  \ &   \  &    \  &    \  &    \  &   \  &    \  &   \ \\
\hline
$ k=1 $  &  5 &   5  &    5  &    5  &    5  &   4  &    3  &   3 \\
\hline
$ k = 2 $ & 14 & 14 & 10 & 9 & 8 & 8 & 7 & 6 \\
\hline
$ k = 3 $ & 32 & 26 & 23 & 19 & 16 & 16  &  14  &   5 \\
\hline
$ k = 4 $ &  63 &  53 &  45 &   40 &   33 &   33 &   30 &    6 \\
\hline
$ k = 5 $ & 124 &  103 &   85 &   78 &   71 &   65 &   63 &   11 \\
\hline
$ k = 6 $ &  237 &  196 &  161 &  150 &  142 &  131 &  128 &   12\\
\hline
\end{tabular}
\bigskip
\caption{Number of eigenvalues of $ B_h $ in the regions
$ \set{| \lambda| > r} $, for $  2 \pi h = 1/N $, $ N $ given by powers of $ 3 $.}
\label{table2}
\end{center}
\end{table}

Table~\ref{table1} shows the analogies between the eigenvalues of
this subunitary quantum map and the resonances of a Schr\"odinger 
operator for a scattering situation (see \S\ref{mso}).

For $ B_h $ given by \eqref{eq:bh}
 we are unable to prove the fractal Weyl law presented in the
last line of Table~\ref{table1}, but numerical results strongly support
its validity \cite{nzJPA}. A striking illustration is given by tripling $N$, in which 
case the number of eigenvalues approximately doubles, in agreement with 
the fractal Weyl law --- see Table~\ref{table2}.

The family of subunitary quantum maps in the main theorem is 
obtained by simplifying $ B_h$, and is described explicitly in \eqref{eq:ttoy}.
It is a quantization of a more complicated multivalued relation 
for which $ \Gamma_+ = \t2 $, $ \Gamma_- = C \times [0,1) $, 
and $ \Gamma_- \cap W_+ \simeq C $ --- 
see Proposition~\ref{p:class}. Theorem~\ref{t:1} follows from the
more precise Proposition~\ref{p:toy}.

%\begin{figure}[ht]
%$$
%\rotatebox{-90}{\includegraphics[height=7.5cm]{toy.eps}}
%\quad
%\rotatebox{-90}{\includegraphics[height=7.5cm]{toy-log.eps}}
%$$
%\caption{The illustration of Theorem \ref{t:1}: on the left, eigenvalues 
%$  \lambda \in \Spec( \tBh )$
%for $ (2\pi h)^{-1} = N = 3^{10} $ (circles) and for $ N= 3^{15} $ (crosses). 
%The four circles correspond to radii  $1, | \lambda_+|, 
%| \lambda_+ \lambda_-|^{\frac12} |, |\lambda_-| $, where $ \lambda_\pm $
%are given in Proposition \ref{p:last}. The critical radius in the statement of the theorem
%is here $ r_0 ( \widetilde B )= |\lambda_+ \lambda_-|^{\frac12} $ (dotted line).
%On the right, we plot the phase (horizontal) and modulus (vertical, log scale)
%of these eigenvalues. The eigenvalues
%appear with high, varying,  multiplicities which peak on the circle 
%$ |\lambda| = r_0 (\widetilde B )$, but the angular distribution is asymptotically
%uniform.}
%\label{f:lat}
%\end{figure}

\subsection{Organization of the paper}
In \S \ref{mab} we present related results from recent mathematical,
numerical, and physics literature. In particular, in \S\ref{s:ben} 
we give the physical motivation for the objects appearing in 
Theorem~\ref{t:2} above.
\S\ref{rcr} is devoted to the review of classical dynamics used in 
our models, stressing the dynamics of open baker's relations.

In \S\ref{oqm} we first review the quantization of tori. We assume
the knowledge of semiclassical quantization in $ T^*\RR^n $ (pseudodifferential
operators) but otherwise the presentation is self contained. 
The definitions of Lagrangian states associated to smooth and singular
Lagrangian submanifolds is based on the ideas of Guillemin, H\"ormander,
Melrose, and Uhlmann in microlocal analysis but, partly due to 
technical differences, we give direct proofs of the properties we need
in this paper. These properties are used to analyze the quantizations 
of the baker's relation coming from the work of Balazs, Voros, Saraceno and Vallejos.

Numerical results for the (usual) quantization of the 
open baker's relation have been presented in \cite{nzJPA}, we briefly 
summarize them in \S\ref{nr}. In \S\ref{s:tm} we discuss the toy model $\tBh$, with
two different interpretations. That section contains the proof of
Theorem~\ref{t:1}. Finally in \S\ref{cWm} we give precise definitions
of objects appearing in Theorem~\ref{t:2} and in a lengthy computation 
we give its proof.

\subsection*{Acknowledgments}
We are grateful to Christof Thiele and
Terry Tao for pointing out the ``Walsh'' interpretation of our toy 
model, and to the anonymous referee for his comments.
The first author thanks Marcos Saraceno for his insights on that
model, and Andr\'e Voros for interesting questions. 
He is also grateful to UC Berkeley for the hospitality 
in April 2004. Generous support of both authors by the National Science
Foundation under the grant DMS-0200732 is also gratefully acknowledged.

%%%%%%%%%%%%%%%%%%%%%%%%%%%%%%%%%%%%%%%%%%%%%%%%%%%%%%%%%%%%%%%%%%%%%%%%%%%%%%%
%%%%%%%%%%%%%%%%%%%%%%%%%%%%%%%%%%%%%%%%%%%%%%%%%%%%%%%%%%%%%%%%%%%%%%%%%%%%%%%
\section{Motivation and background}
\label{mab}

In this section we discuss motivating topics in mathematics
and theoretical physics, and survey related results.

\subsection{Schr\"odinger operators}\label{mso}
The original motivation comes from the study of resonances in 
potential scattering. The simplest case is given by considering 
the following quantum Hamiltonian:
\begin{equation}
\label{eq:ham}
 H = - h^2 \Delta + V( q ) \,, \qquad  V \in \CIc ( \RR^n ; \RR) \,.
\end{equation}
By assuming that the potential vanishes near infinity and that
it is infinitely differentiable, 
we eliminate the need for technical assumptions --- see \cite{HeSj}
and \cite{SjZw04} for more general settings, in the analytic and
 $ \CI $ categories respectively. For instance, 
as noted in \cite[(c.32)-(c.33)]{SjDuke} 
the theory of \cite{HeSj} applies to arbitrary homogeneous polynomial 
potentials at nondegenerate energy levels.

\begin{table}[htbp]
\begin{center}
\begin{tabular}{||c|c||}
\hline
\ & \ \\
\hspace{0.6in} $ h \rightarrow 0 $ & \hspace{0.4in} 
$ N = ( 2 \pi h )^{-1} \rightarrow \infty  $ \\
\ & \ \\
\hline
\ & \ \\
\hspace{0.3in} $ \chi 
\exp \big( - i t (-h^2 \Delta + V ) /h \big) \chi \,,\ \ t\geq 0\,, $ & 
\hspace{0.3in} $ B_{h}^t$, $ t = 0, 1, \cdots $ \\
\ & \ \\
 $ \chi $ a cut-off on the interaction region & 
 $ B_h$ a subunitary matrix\\
\ & \ \\
\hline
\ & \ \\
$ \hspace{0.3in} 
e^{ - i t z / h } $, $z$ a resonance of $H=-h^2 \Delta + V$  & 
\hspace{0.1in} $  \lambda^t $, 
 $\lambda$ an eigenvalue of $ B_{h} \in \cL ( \CC^N ) $
\\
\ & \ \\
\hline
\ & \ \\
\hspace{0.3in} $ z \in [E-h , E+h ] - i [ 0 , \gamma h ]   $ &
\hspace{0.3in} $  1 \geq | \lambda | > r > 0  $ \\
\ & \ \\
\hline
\ & \ \\
$ \# \{ z \in [E-h , E+h ] - i [ 0 , \gamma h ] \}   \simeq \
C(\gamma)\, h^{-\mu_E }  $ & $ \# \{ \lambda, | \lambda | > r \}  \simeq  \ C(r)\, 
N^\nu  $ \\
\ & \ \\
\hline
\end{tabular}
\bigskip
\caption{Analogies between Schr\"odinger propagators and open quantum 
maps.}
\label{table1}
\end{center}
\end{table}
Before discussing open systems we recall the well known results 
for closed systems, obtained for instance by considering $ H $ above
on a bounded domain $ \Omega \subset \RR^n $ and imposing a
self-adjoint boundary condition at $ \partial \Omega $ (Dirichlet or
Neumann). Then the spectrum, $ \Spec ( H ) $, of $ H $ is discrete and,
at a non-degenerate energy level $ E $ its density is described by 
the celebrated {\em Weyl law}:
\begin{equation}
\label{eq:weyl} 
\# \set{\Spec( H ) \cap [ E- \delta, E + \delta ]}  =  
\frac{1}{ ( 2 \pi h )^n } \int \! \! \int_{ | p^2 + V ( q ) - E | < \delta } 
dq\, dp + {\mathcal O} ( h^{1-n} )\,, 
\end{equation}
see \cite{DiSj,Ivr} and references given there. We note that this implies
a precise upper bound 
\begin{equation}
\label{eq:weylh}
 \#\set{ \Spec( H ) \cap [ E- C h, E + Ch  ] } = {\mathcal O} ( h^{1-n} ) \,,
\end{equation}
which can be improved further by making assumptions on the classical flow of the Hamiltonian
$ p^2 + V ( q ) $ on $ \Omega $, see \cite{DiSj,Ivr}.

For open systems, with the simplest example given by the Hamiltonian 
in \eqref{eq:ham}, real eigenvalues are replaced by complex {\em resonances}.
The simplest definition (easily made rigorous in the case \eqref{eq:ham}) 
comes from considering the meromorphic continuation of the resolvent. 
Defining the Green's function $G( z , q', q )$ for $ \Im z > 0 $ through
\[ 
( z - H )^{-1} u ( q' ) = \int_{\RR^n } G( z , q', q )\, u ( q )\, d q \,, \quad
u \in \CIc ( \RR^n ) \,, 
\]
then $ G( z, q', q ) $ admits a meromorphic continuation in $ z $ across the real axis. 
Its poles for $ \Im z < 0 $ (which do not depend on $(q',q)$)
are the {\em quantum resonances} of $ H $.

Counting of resonances is affected by the dynamical structure of 
the scatterer much more dramatically than counting of eigenvalues of
closed systems. Since we are now counting points in the complex plane we need
to make geometric choices dictated by dynamical and physical 
considerations. Here we consider scatterers and energies
exhibiting a hyperbolic classical flow, and regions in the lower half-plane
which lie at a
distance proportional to $ h $ from the real axis. This choice is
motivated as follows. Quantum mechanics interprets a resonance
at $ z = E - i \gamma $ in terms of a {\em metastable state}, which decays proportionally to 
$ \exp ( - t \gamma / h ) $. Hence for $ \gamma \gg h $ the decay
is so rapid that the state is invisible. On the other hand, 
for many chaotic scatterers there are no resonances with $ \gamma \ll h $.
One class for which this is known rigorously consists in the Laplacian on co-compact
quotients $ {\mathbb H}^n/ \Gamma $, $ H = - h^2 \Delta_{ {\mathbb H}^n/ \Gamma} $, 
when the dimension of the limit set satisfies $ \delta ( \Gamma ) < (n-1)/2 $.
This follows from  the work of Patterson and Sullivan --- see the discussion 
below and \cite{Naud}.

After a complex deformation (see \cite{SjZw04} and references given there)
the long living quantum states should semiclassically concentrate
on the set of phase space points which do not escape to infinity,
that is on the {\em trapped set} $ K_E $ defined as follows:
let
$ \Xi_H $ be the Hamilton vector field of the Hamiltonian $H(q,p)=p^2/2+V(q)$:
$$ 
\Xi_H =  \sum_{j=1}^n  p_j \partial_{q_j} - \partial_{q_j} V( q ) 
\partial_{p_j } \,.
$$
Then 
\begin{gather}
\label{eq:trap}
\begin{gathered}
K_E \defeq \Gamma_+ (  E) \cap \Gamma_- ( E ) \,, \quad \text{with the forward/backward trapped sets}\\
\Gamma_{\pm} ( E ) \defeq  \{ 
\rho \in \Sigma_E \; : \; \exp t\, \Xi_H ( \rho ) \not \rightarrow \infty \,, \ 
\mp t \rightarrow \infty \} \,.
\end{gathered}
\end{gather}

Suppose that the flow generated by $\Xi_H$ is 
hyperbolic near $ K_{E'} $ for $ E'$ close to a non-degenerate energy $ E $. 
That means that the field $\Xi_H$ does not vanish on the energy surfaces 
$ \Sigma_{E'} = \{ p^2 + V ( q ) = E' \} 
\subset T^*\RR^n$ for $ E' \approx E $,
and that for $ \rho \in \Sigma_{ E' } $ near $ K_{E'} $, 
\begin{equation}
\label{eq:hyp}
\begin{split}
& T_\rho \Sigma_{E'}  = \RR\, \Xi_H (\rho ) 
\oplus E_+ ( \rho) \oplus E_- ( \rho ) \,,  \\
& \Sigma_{E'} \ni \rho \longmapsto E_{\pm} ( \rho ) \subset 
T_\rho \Sigma_{E'} \ 
\text{ is continuous,} \\
& d ( \exp t\, \Xi_H ) ( E_\pm ( \rho ) ) = E_\pm ( \exp t\, \Xi_H ( \rho )) \,,\\
& \| d (\exp t\, \Xi_H ) ( X ) \| \leq C e^{ \pm \lambda t } \| X \| \,, 
\ \text{ for all $ X \in E_\pm ( \rho ) $, $ \mp t \geq 0 $.}
\end{split}
\end{equation}
Weaker assumptions are possible --- see \cite[\S 5]{SjDuke} and 
\cite[\S 7]{SjZw04}.

Typically, the set $ K_E $ has a fractal structure and in the 
semiclassical estimates the Minkowski dimension naturally appears:
\[ 
\dim K_E  = 2n -1 - \sup \big\{ c \; : \; \limsup_{ \epsilon \rightarrow 0 }
\epsilon^{-c} \vol \{ \rho \in \Sigma_{E} \; : \; \mathrm{dist} ( K_E , \rho ) 
< \epsilon \}  < \infty \big\} \,. 
\]
We say that $ K_E $ is of {\em pure dimension}  if the supremum is
attained. For simplicity of the presentation 
we assume that this is the case.

Under these assumptions the estimate \eqref{eq:weylh} has an analogue
for chaotic open systems \cite{SjZw04}. For  $ C_0 > 0 $
there exists $ C_1 $ such that
\begin{equation}
\label{eq:fweylh}
\#\big\{ \Res ( H ) \cap \set{ z \; : \; \ |z - E | < C_0 h  } \big\}
\leq  C_1 h^{- \mu_E } \,, \qquad  \dim K_E = 2 \mu_ E + 1 \,.
\end{equation}
We notice that for a closed system the trapped set is the entire
energy surface, so that in that case
 $ \mu_E = n-1 $,  hence \eqref{eq:fweylh}
is consistent with \eqref{eq:weylh}. 
In this note we use open quantum maps to provide the first evidence that this
precise estimate is optimal. 

We should also mention that, as was already stressed in the work of
Sj\"ostrand \cite{SjDuke}, the estimates involving the dimension are only 
reasonable when the flow is strictly hyperbolic. In the case of
more complicated flows the estimates should be stated in terms of
properties of {\em escape} or {\em Lyapunov} functions associated 
to the flow -- see \cite{SjDuke,SjZw04}. For expository reasons
the estimates involving the dimension are however most persuasive.

%\begin{figure}[htbp]
%\begin{center}
%\includegraphics[angle=-90,width=4.0in]{pot.eps}
%\end{center}
%\caption
%{\label{Fblock} A three bump potential exhibiting hyperbolic dynamics
%in a range of energies.}
%\end{figure}

%%%%%%%%%%%%%%%%%%%%%%%%%%%%%%%%%%%%%%%%%%%%%%%%%%%%%%%%%%%%%%%%%%%%%%%%%%%%%%%
%%%%%%%%%%%%%%%%%%%%%%%%%%%%%%%%%%%%%%%%%%%%%%%%%%%%%%%%%%%%%%%%%%%%%%%%%%%%%%%

\subsection{Survey of related results}
The first indication that fractal dimensions enter into counting 
laws for quantum resonances of chaotic open systems appears in
a result of Sj\"ostrand \cite{SjDuke}:
\begin{gather}
\label{eq:sjduke}
\begin{gathered}
\#\big\{ \Res ( H ) \cap \set{ z \; : \; \ |z - E | < \delta \,, \ \  
\Im z > - \gamma } \big\}
\leq  C_1 \delta \left( \frac{h}{\gamma} \right)^{-n} \gamma^{- \frac12 \widetilde 
m } \,, \\
Ch \leq \gamma \leq 1/C \,, \ \ \max( h^{\frac12} , h/\gamma ) \leq \delta \leq 2/ C \,,
\end{gathered}
\end{gather}
where $ \widetilde m $ is any number greater than the dimension of the trapped set in
the shell $H^{-1}(E-1/C_2, E + 1/C_2)$.
In a homogeneous situation, such as 
for instance obstacle scattering, the dimension of $ K_E $, $ 2 \mu_E +1 $, 
is independent of $ E $, so that $ \widetilde m > 2 ( \mu_E + 1 ) $.

The improvement in \cite{SjZw04} quoted in \eqref{eq:fweylh}  
lies in providing a bound for the number of resonances in a smaller region
$ D ( E , Ch )=\set{z\in\CC\,:\,|z-E|<Ch} $.
Heuristic arguments suggesting that the estimate \eqref{eq:sjduke} 
should be optimal were given in \cite{LZ} and \cite{LSZ}.

Another class of Hamiltonians with chaotic classical flows and 
fractal trapped sets is given by Laplacians on convex co-compact
quotients, $ \HH / \Gamma $. Here $ \Gamma $ is a discrete
subgroup of isometries of the hyperbolic plane $ \HH $, such that 
\begin{itemize}
\item All elements $ \gamma \in \Gamma $ are hyperbolic, which means that
their action on $ \HH $ can be represented as
\begin{gather} 
\label{eq:ach}
\begin{gathered}
 \alpha \circ \gamma \circ \alpha^{-1} ( x, y ) = e^{ \ell ( \gamma ) }
( x , y ) \,,  \ \ 
 ( x , y ) \in {\mathbb H} \simeq \RR_+ \times \RR \,, \ \
\alpha \in {\rm{Aut}}( \HH ) \,.
\end{gathered}
\end{gather}
\item If $ \pi : \HH \rightarrow \HH/\Gamma $, and $ \Lambda ( \Gamma ) 
\subset \partial \HH$
is the limit set of $ \Gamma $, that is the set of limit points of 
$ \{ \gamma ( z ) : \gamma \in \Gamma \} $, $ z \in \HH $, then 
$ \pi ( \rm{convex \  hull} \; \Lambda ( \Gamma ) ) $ is compact.
\end{itemize}
%An example is shown in Fig.~\ref{Fsh}.
The trapped set is determined by $ \Lambda ( \Gamma ) $: trapped
trajectories are given by geodesics connecting two points of
$ \Lambda ( \Gamma ) $ at infinity, and
\[ 
\dim K_E = 2 \delta_\Gamma + 1 \,, \quad \delta_\Gamma = \dim \Lambda ( \Gamma ) \,.
\]
The limit set is always of pure dimension, which coincides with 
its Hausdorff dimension.

A nice feature of this model is the exact correspondence between
the resonances of 
$$ 
H = h^2 ( - \Delta_{ \HH / \Gamma } - 1/4 ) \,, 
$$ 
and the zeros of the Selberg zeta function, 
$ Z_\Gamma ( s ) $\footnote{We refer to 
 \cite{PaPe} for this and a general treatment. The term $ \frac14 $
in the definition of the Hamiltonian $ H$ comes from requiring that
the bottom of the spectrum of $ H $ is $ 0 $, so that Green's function
$ ( H - \lambda^2 )^{-1} $ is meromorphic in $ \lambda \in \CC $}:
\begin{gather}
\label{eq:corr}
\begin{gathered}
z \in \Res ( H ) \ \Longleftrightarrow \  Z_\Gamma ( s ) = 0 \,, \ \
z = h^2 (  s  ( 1 - s )  - 1/4) \,, \ \ \Re s \leq \delta_\Gamma 
\,, \end{gathered}
\end{gather}
where the multiplicities of zeros and resonances agree.
The Selberg zeta function is defined by the analytic continuation of 
\[  
Z_\Gamma ( s ) =  \prod_{ \{ \gamma \} } \prod_{k \geq 0 }
\left( 1 - e^{ - ( s + k ) \, \ell( \gamma ) } \right) \,, \quad \Re s > \delta_\Gamma 
\,, 
\]
where $ \{ \gamma \} $ denotes a conjugacy class of a primitive element
$ \gamma \in \Gamma $ (an element which is not a power of another element),
and we take a product over distinct
primitive conjugacy classes (each of which corresponds to a primitive
closed orbit). The length $ \ell ( \gamma ) $ of the corresponding closed orbit
 appears in \eqref{eq:ach}.
%%%%%%%%%%%%%%%%%%%%%
%\begin{figure}[ht]
%\includegraphics[width=4in]{mz1.eps}
%\caption
%{\label{Fsh} An example of $ \HH^2 / \Gamma $ where $ \Gamma $ is
%generated by compositions of reflections in three discs.}
%\end{figure}
%%%%%%%%%%%%%%%%%%%%%
The exact analogue of \eqref{eq:fweylh} is given by 
\begin{equation}\label{e:Weyl-schottky} 
\# \set{ s \; : \; Z_\Gamma ( s ) = 0 \,, \ \ \Re s > - C_0 \,, \ \
r < \Im s < r + C_1 }  \leq C_2 \,r^{\delta_\Gamma } \,,
\end{equation}
which is a consequence of an estimate established by 
Guillop\'e-Lin-Zworski \cite{GLZ} in a more general setting
of convex co-compact Schottky groups in any dimension,
\begin{equation}
\label{eq:estzet}
| Z_\Gamma ( s ) | \leq C_K\, e^{ C_K\, |s|^{\delta_\Gamma }} \,, \ \ 
\Re s \geq -K \,, \ \text{ for any $ K $.}
\end{equation}
This improved earlier estimates of \cite{ZwIn}, the proof of which 
was largely based on \cite{SjDuke}.

In the (non-quantum) context of rational 
maps on the complex plane, similar results were obtained concerning the zeros
of associated zeta functions \cite{Ch,StZw}. Take $ f $ a uniformly expanding 
rational map on $\C$ (for instance 
$ z \mapsto z^2 + c $, $ c < - 2 $),
and call $ f^n $ its $n$-fold composition. The zeta function associated with
this map is given by 
\begin{equation}
\label{eq:dzet}
 Z ( s ) = \exp\left( - \sum_{n=1}^\infty n^{-1} \sum_{f^n(z)=z}
\frac{| (f^n)'(z)|^{-s}|}{ 1 - |(f^n)'(z)|^{-1}} \right) \,. 
\end{equation}
Then the number or resonances in a strip is also given by
a law of the type \eqref{e:Weyl-schottky}, where
$ \delta_\Gamma $  is replaced by the dimension of the 
{\em Julia set}: 
$$ 
J = \overline{ \bigcup_{ n\geq 1} \{ z \; : \; f^n ( z ) = z \} }\,.
$$
Note that this set is also made of ``trapped orbits''.

%%%%%%%%%%%%%%%%%%%%%%%%%%%%%%%%%%%%%%%%%%%%%%%%%%%%%%%%%%%%%%%%%%%%%%%%%%%%%%%
%%%%%%%%%%%%%%%%%%%%%%%%%%%%%%%%%%%%%%%%%%%%%%%%%%%%%%%%%%%%%%%%%%%%%%%%%%%%%%%

\subsection{Survey of numerical results}

The first model investigated numerically was perhaps the hardest to 
give definitive results. Lin \cite{L,LZ} 
studied the semiclassical Schr\"odinger
Hamiltonian \eqref{eq:ham} with a potential made of 3 Gaussian ``bumps''. %given in Fig.~\ref{Fblock}.
The semiclassical resonances were computed 
using the method of complex scaling and were counted in boxes of type
$ [ E - \delta , E + \delta ] - i [ 0 , h ] $ with $ \delta $ fixed.
The purpose was to verify optimality of Sj\"ostrand's estimate
\eqref{eq:sjduke} with these parameters. The results were encouraging but
not conclusive. Since for small values of $ h $ the method of 
\cite{L} required the use of large matrices to discretize the Hamiltonians,
the range of $ h$'s was rather limited. 

%%%%%%%%%%%%%%%%%%%
%\begin{figure}[ht]
%\begin{center}
%$$\includegraphics[width=8cm]{resplot4.eps}\qquad\includegraphics[height=8cm]{hist4.eps}$$
%\end{center}
%\caption{A sample of results of \cite{L}: the plot on the left shows 
%resonances for the 3-bump potential% of Fig.~\ref{Fblock}% 
%($ h = 0.017 $). 
%On the right is the 
%log-log plot of the number of resonances vs. $ h$; 
%$\triangleleft$ denote numerical data, $\ast$ the density predicted by the 
%fractal Weyl law, and $ \circ $ the least square 
%interpolants. }
%\label{f:3}
%\end{figure}
%%%%%%%%%%%%%%%%%%%
A different point of view was taken by Lu-Sridhar-Zworski 
\cite{LSZ} where resonances
for the three discs scatterer in the plane were computed using the semiclassical 
zeta function of Eckhardt-Cvitanovi\'{c}, Gaspard, and others 
(see for instance
\cite{Cv-E,G,wirzba99} and references therein).
The zeta function is computed using the cycle expansion 
method loosely based on the Ruelle theory of dynamical zeta
functions.
Although it is not rigorously known if the resonances computed 
by this 
method approximate resonances of the Dirichlet Laplacian 
in the exterior of the discs, or even if the semiclassical 
zeta function has an analytic continuation, proceeding this way is 
widely accepted in the physics literature.
Resonances $z=h^2\,k^2$ were counted in regions 
\begin{equation}
\label{eq:clsz}  \set{ k \in \CC \; : \; 1 \leq \Re k \leq r \,, \ \ \Im k 
\geq - \gamma }\,, \ \ r \rightarrow \infty \,, 
\end{equation}
which under semiclassical rescaling correspond to counting in
$ [ 1/2 , 2] - i[ 0, \gamma h / 2 ] $, $ h \rightarrow 0 $. Let us
denote the number of resonances (zeros of the semiclassical zeta 
function) in \eqref{eq:clsz} by $ N ( r , \gamma )$. The fractal
Weyl law corresponds to the claim that for $\gamma $ large enough,
\begin{equation}
\label{eq:glsz}  N ( r , \gamma ) \sim C(\gamma)\, r^{\mu+1} \,,\quad r\to\infty\,, 
\end{equation}
where $ 2 \mu + 1 $ is the dimension of the trapped set in the 
three dimensional energy shell (for such homogeneous problems, 
all energy shells are equivalent). 
In \cite{LSZ} the prediction \eqref{eq:glsz} was tested by linear fitting of 
$ \log N ( r , \gamma ) $ as a function of $ \log r $:
$$ 
\log N ( r , \gamma ) =\big(\alpha ( \gamma )+1\big)
\log r  + \Oo ( 1 ) \,.
$$
We found that the coefficient $ \alpha ( \gamma ) $
was independent of $ \gamma $ for 
$ \gamma $ large enough, and that it agreed with $ \mu $. The counting was
done for three different equilateral disc configurations, parametrized
by $ \rho = R/a $ where $ a $ is the radius of each disc, and $ R $ the
distance between them. We also noticed that if 
$ {\gamma_\rho} $ is the classical rate of decay for the $\rho$ configuration,  
then
$$
\frac{ \alpha_\rho ( x \gamma_\rho /2  )}{ \mu_\rho}
$$
is essentially independent of $ \rho $ for $ 1 < x < 1.5 $.
This corresponds to a numerical observation
that for each $ \rho $
the distribution of resonance widths (imaginary parts)
peaks near $ \gamma = \gamma_\rho/2 $.

Encouraged by the results of \cite{LSZ}, 
the cycle method was used in \cite{GLZ} to 
count the zeros of the Selberg zeta function for a certain Schottky quotient,
%depicted in Fig.~\ref{Fsh}, 
but the results were not definitive.
For the dynamical zeta function \eqref{eq:dzet} with $ f ( z ) = 
z^2 + c $, $ c < - 2 $, the resonances were computed by Strain-Zworski
\cite{StZw}, using 
a different method based on the theory of the transfer operator
on Hilbert spaces of holomorphic functions introduced in \cite{GLZ}.
The zeros were counted in a region of the same type as in 
\eqref{eq:clsz},
\[ 
\{ s \; : \; \Re s > -K \,, \ 0 \leq \Im s \leq r \} \,, 
\]
where real parts and imaginary parts exchange their
meaning due to different conventions\footnote{Although
frustrating, the different conventions of semiclassical, obstacle, and
hyperbolic scattering show how the same phenomenon appears in 
historically different fields.}. By reaching very high 
values of $ r $ we saw a very good agreement of the log-log
fit with the fractal 
Weyl, with $ \mu $ given by the dimension of the Julia set.

In the model considered in this paper, we can verify the
optimality of the fractal Weyl law on much smaller scales
(see Table~\ref{table2} and the numerics presented in \cite{nzJPA,nQM9,NonRub06}).
That could not be seen in the other approaches.

%%%%%%%%%%%%%%%%%%%%%%%%%%%%%%%%%%%%%%%%%%
%%%%%%%%%%%%%%%%%%%%%%%%%%%%%%%%%%%%%%%%%%

\subsection{Related models in physics}
\label{rmph}

The behaviour of quantum open systems has been recently investigated in 
situations where 
the classical dynamics has chaotic features. The physical motivation 
can originate from nuclear or atomic physics
(study the lifetime statistics of metastable
states, possibly leading to ionization), 
mesoscopic physics (study the conductance, conductance fluctuations,
shot noise in quantum dots or quantum wires), and from 
waveguides  (optical wave propagation
in an optical fiber with some dissipation, microwave propagation 
in an open microwave
cavity). 

%%%%%%%%%%%%%%%%%%%%%%%%%%%%%%%%%%%%%%%%%%

\subsubsection{Kicked rotator with absorbing boundaries.}
\label{s:absorbing}
In \cite{borgonovi,casati} 
a kicked rotator with absorbtion was used to model the process of ionization.
The classical
kicked rotator is Chirikov's {\em standard map} on the cylinder, which is a paradigmatic
model for transitions from regular to chaotic motion \cite{chirikov}.
Quantizing the map on $L^2(\TT^1)$ results in a unitary operator $U$, a
first instance of {\em quantum map}.
To model the ionization process which takes place at some
threshold momentum $p_{\rm{ion}}$, 
the authors truncate the map $U$ to the subspace
$\Hh_{\rm{ion}}={\rm span}\set{|p_j\ra\,:\,|p_j|\leq p_{\rm{ion}}}$: a particle
reaching that threshold is ionized, or equivalently ``escapes to infinity''. 
Here the discrete values $ p_j = 2 \pi h j $ are the 
eigenvalues of the momentum operator on $L^2(\TT^1)$; the space 
$\Hh_{\rm{ion}}$ is thus of dimension $N\approx p_{\rm ion}/\pi h$.
This projection leads to an  
{\em open quantum map}, namely the subunitary propagator
$U_{\rm ion} = \Pi_{\rm ion} U$,
where $\Pi_{\rm ion}$ is the orthogonal projector on 
$\Hh_{\rm{ion}}$. 
For the parameters used by \cite{borgonovi}, the classical dynamics is diffusive,
meaning that a particle starting from $p=0$ will need many kicks
to reach the ionization threshold. 

The matrix $U_{\rm ion}$ was numerically diagonalized for various values of 
$h$ with $p_{\rm ion}$ fixed, and
the distribution of the $N$ level widths $\gamma_i=-2\ln|\lambda_i|$,
$ \lambda_i \in \Spec( U_{\rm ion }) $ was found approximately
independent of $h$, such that the number of resonances 
$$
n(N,\gamma)= \# \{\gamma_i\leq\gamma\}
$$ 
scales like $ C(\gamma)N $ in this case. 
In subsequent works \cite{frahm,zycz,fyodorov}, this distribution was shown to 
correspond to an ensemble of random subunitary random matrices, more precisely the ensemble
formed by the $[\alpha N]\times [\alpha N]$ upper-left corner ($0<\alpha<1$ fixed)
of a large $N\times N$ matrix drawn in the Circular Unitary Ensemble (that is, the set
$U(N)$ equipped with Haar measure).

%%%%%%%%%%%%%%%%%%%%%%%%%%%%%%%%%%%%%%%%%%
\subsubsection{Quasi-bound states in an open quantum map}\label{s:schomerus}
Recently, Schomerus and Tworzyd{\l}o 
\cite{schomerus} have performed a similar study for the quantized
kicked rotator on the torus (obtained from the map of the former section
by periodizing the momentum
variable). They also ``opened'' the map by assuming that particles 
reaching a certain position window $q\in L$ ``escape to infinity''. 
The quantum projector associated with
these ``escape window'' is denoted by $\Pi_L$, so that
the remaining subunitary quantum map reads 
$ U_{\rm op} = ( I - \Pi_L ) U $.
The main difference with the case studied in the previous section lies in the
strongly chaotic motion (as opposed to diffusive), due to a different choice of
parameters. The map has a positive
Lyapunov exponent $ \lambda$, and a typical trajectory
will escape after a few kicks: the average ``dwell time'',
called $\tau_D$, is of order unity.

The eigenmodes associated with 
eigenvalues bounded away from zero are called ``quasi-bound states'' , 
as opposed to the ``instantaneous decay modes''
associated with very small eigenvalues. 
The authors provide numerical and heuristic evidence that, in the semiclassical limit,
the number of quasi-bound states grows like $N_{\rm{eff}}=N^{1- 1/ ( \lambda \tau_D )} $. 
This shows that most 
eigenvalues of $U_{\rm op}$ are very close to zero, while only a small
fraction $N_{\rm{eff}}/N$ remains bounded away from zero. 
The authors also plot the distribution of the $\sim N_{\rm{eff}}$ quasi-bound eigenvalues:
again, it resembles the spectrum of a random subunitary matrix obtained by
keeping the upper-right corner block of size $N_{\rm{eff}}$ of
a $[\tau_D N_{\rm{eff}}]$-dimensional random unitary matrix.

The quantized baker's relation 
we will study in \S\ref{nr}--\ref{s:tm} will be of similar nature. For the
map \eqref{e:quantum-baker}, the
fractal dimension $\nu$ given in \eqref{e:hausdorff} can be shown to be close
to the formula $1- 1/ ( \lambda \tau_D )$, in the limit when the dwell time $\tau_D$
is large compared to unity (limit of small opening).

%%%%%%%%%%%%%%%%%%%%%%%%%%%%%%%%%%%%%%%%%%

\subsubsection{Conductance through an open chaotic cavity}
\label{s:ben}
The ``scattering approach 
to semiclassical quantization'' \cite{Bogo92,DorSmil92,Prosen96,OzoVall00}, consists in quantizing
the return map on a Poincar\'e surface of section of the Hamiltonian system
under study. Within this approach, the \emph{scattering matrix} of the open system can
be expressed as a ``multiple-scattering expansion'' in terms of the quantized return map. 

Using that framework, 
Beenakker {\it et al.} \cite{beenakker} study the quantum kicked rotator defined in the previous
section, in order
to understand the fluctuations of
conductance through a {\it quantum dot}. The evolution inside the
closed dot is represented by the
same unitary matrix $U$ as in last subsection, and its opening
$L$ is split into two intervals, $L_2$ and $L_1$, which
represent the two
``leads'' bringing in and taking out the charge carriers from the dot.
The orthogonal projector corresponding to these openings reads
$\Pi_L=\Pi_{L_1}\oplus \Pi_{L_2}$. The conductance
can then be analyzed from the
{\em scattering matrix} of the dot:
\begin{equation}\label{e:scattering}
\tS(\vartheta)
=\Pi_L\{e^{-i\vartheta}- U(1-\Pi_L)\}^{-1} U\Pi_L \,.
\end{equation}
Here $ \vartheta \in [0,2\pi) $ is called the {\em quasi-energy}. In terms of this parameter,
the ``physical half-plane'' corresponds to $ \Im \vartheta > 0 $: the matrix
$\tS(\vartheta)$ has no singularity in this region. On the opposite,
the resonances analyzed in the previous section,
which are the {\em poles} of $ \tS ( \vartheta) $, are situated in the 
region $ \Im \vartheta < 0 $.

While $\tS(\vartheta)$ is unitary, its subblock
$t \defeq \Pi_{L_2} \tS ( \vartheta) \Pi_{L_1}$
describes the {\em transmission} from the lead $ L_1 $ to the lead $ L_2 $. 
The dimensionless conductance (which depends on $\vartheta$) is given by the Landauer-B\"uttiker 
formula $g=\tr(tt^{*})$.
The eigenvalues of $ t t^* $ (called ``transmission eigenvalues'')
can be either close to $1$ (corresponding to a total transmission), or 
close to $0$ (corresponding to a total reflection), or inbetween. The last case corresponds to
genuinely quantum transmission eigenmodes, which are partly transmitted, partly reflected,
due to interference phenomena inside the dot. 
The ``quantum shot noise'' is due to these intermediate transmission eigenvalues.
A simple measure of that noise is given by the 
Fano factor \cite{buttiker}
$F={\tr(tt^*(1-tt^*))}/{\tr tt^*}$. 
Using similar arguments as in the former
section, the authors show that the number of intermediate transmission eigenvalues also scales
like $N_{\rm{eff}}$, and thereby estimate the Fano factor, by assuming that these eigenvalues
are distributed according to the prediction of random matrix theory.

In \S\ref{cWm} we will analytically compute both the conductance and the Fano factor 
in the case of the open quantum relation $\tBh$.

%%%%%%%%%%%%%%%%%%%%%%%%%%%%%%%%%%%%%%%%%%%%%%%%%%%%%%%%%%%%%%%%%%%%%%%%%%%%%%%
%%%%%%%%%%%%%%%%%%%%%%%%%%%%%%%%%%%%%%%%%%%%%%%%%%%%%%%%%%%%%%%%%%%%%%%%%%%%%%%

\section{Classical dynamics}
\label{rcr}

\subsection{Symplectic geometry on tori.\label{s:sympl-tori}}

We consider the simplest class of compact symplectic manifolds, the
tori,
$$  
\tn \defi \RR^{2n}/ \ZZ^{2n} \simeq 
(\II \times \II)^n \,, \quad \omega = \sum_{ \ell=1}^n dq_\ell \wedge d p_\ell \,, \ \
( q , p ) \in \tn \,.
$$
Here and in what follows, we identify 
the interval $\II=[0,1)$ with the circle $\TT^1=\RR/\ZZ$.
A Lagrangian (submanifold) $ \Lambda \subset \tn $ is a $n$-dimensional 
embedded submanifold of $ \tn $ such that 
$ \omega\rest_{\Lambda } = 0 $. We recall the following 
well known fact (see for instance \cite[Theorem 21.3.2]{Hor}):
\begin{prop}
\label{l:well}
Suppose that $ \Lambda \subset \tn $ is a Lagrangian submanifold, 
and that $ ( q_0, p_0 ) \in \Lambda $. Then, after a possible 
permutation of indices, there exists $ k $, 
$ 0 \leq k \leq n $, and a splitting
of coordinates:
$$
q = ( q',q'') \,, \quad p = ( p', p'') \,, \quad 
q' = ( q_1, \cdots q_{k} ) \,, \quad p'' = ( p_{k+1} , \cdots , p_n ) \,,
$$
such that the map
$$   
\Lambda \ni ( q, p ) \; \longmapsto \; ( q'', p') \in 
 \II^{n-k}\times\II^k
$$
is bijective from a neighbourhood $V$ of $ ( q_0, p_0 ) $ to a neighbourhood
$W$ of $( q''_0, p'_0 )$. Consequently there exists 
a function, $ S = S ( q'', p') $ defined on $W$, such that
$\Lambda\cap V$ is generated by the function $S$, that is, 
$$  
\Lambda\cap V = \set{ \Big(d_{p'} S ( q'', p'), q'' ; p', -d_{q''} S( q'',p') \Big) 
\,, \ \  ( q'', p') \in W}\,.
$$
\end{prop}
In this paper we will also consider {\em singular Lagrangian 
manifolds} obtained by taking finite unions of Lagrangians with 
piecewise smooth boundaries.

%%%%%%%%%%%%%%%%%%%%%%%%%%%%%%%%%%%%%%%%%%%%%%%%%%%%%%%%%%%%%%%%%%%%%%%%%%%%%%%

\subsection{Symplectic relations}\label{s:sympl.rela.}
\subsubsection{Symplectic maps}\label{s:sympl.maps}
A symplectic (or ``canonical'') diffeomorphism on the torus $\tn$ is a diffeomorphism
$\kappa \; : \; \tn \; \rightarrow \; \tn $ which leaves invariant the
symplectic form on $\tn$: $\kappa^* \omega = \omega$. An equivalent characterization of such
a map is through its {\em graph} $\Gamma$, which is 
the $2n$-dimensional embedded submanifold of $\tn\times\tn$, defined as
$$
\Gamma_\kappa=\set{(\rho';\rho)\;:\;\rho=(q,p)\in\tn,\ \rho'=\kappa(\rho)}\,.
$$
Using the identification $ \II^n = \RR^n / \ZZ^n $, we setup the reflection map
$ \II^n \ni p \mapsto - p \in \II^n$, and define the
{\em twisted graph} \cite[\S 25.2]{Hor}
\begin{equation}\label{e:twisted}
  \Gamma'_\kappa = \{ ( q', q ; p', - p ) \; : \; ( q', p'; q , p ) \in 
\Gamma_\kappa \}\subset \TT^{4n}\,.
\end{equation}
Then the diffeomorphism $\kappa$ is symplectic iff $\Gamma'_\kappa$ is a Lagrangian
submanifold of $\TT^{4n}$ (equipped with the symplectic form 
$\sum_{ j=1}^n dq'_j \wedge d p'_j+dq_j \wedge d p_j$). For this reason, we will sometimes
denote $\Gamma'_\kappa $ by $\Lambda_\kappa$.

The definition of the twisted graph is clearly dependent on the choice of the 
splitting of variables $ ( q, p ) $, which will be related to a 
choice of {\em polarization} in the quantization process.

More generally, one can consider invertible maps on $\tn$ which are smooth and symplectic
except on a negligible set of singularities (say, discontinuities on a hypersurface). 
The twisted graph of such a map is then
a singular Lagrangian submanifold of $\TT^{4n}$.

\noindent {\em Example.} The usual ``baker's map'' is the following 
piecewise-linear transformation 
$ \kappa $ on $ \t2 $:
\begin{equation}\label{e:2-baker}
\kappa ( q , p ) \defi \begin{cases} 
\ ( 2 q , p/2 ) \ &\mbox{if}\ 0 \leq q < 1/2 \\
( 2q -1 , p/2 + 1/2 ) &\mbox{if}\ 1/2 \leq q < 1\,. \end{cases}
\end{equation}
The twisted graph  of $ \kappa $:
$$  
\Lambda_\kappa \defi \set{ ( q',q; p',-p) \; : \;(q,p)\in \t2 ,\;
( q', p') = \kappa ( q, p ) } 
$$
is a singular Lagrangian submanifold of $\TT^4$. It can be decomposed into
$\Lambda_\kappa  = \Lambda_0 \cup \Lambda_1$, with the components
\begin{align*}
\Lambda_j &= \set{ ( 2 q - j , q ; \frac{p + j}{2} , - p ) \; : \; 
j/2 \leq q < j/2+1/2, \;  p\in\II }\\
&=\set{ ( 2 q - j , q ; p' , -2 p' + j ) \; : \; 
j/2 \leq q,p' < j/2+1/2}\,.
\end{align*}
Each $ \Lambda_j $ is locally Lagrangian in $ \TT^4 $ and, as
a manifold with corners, it 
is diffeomorphic to a 2-dimensional square. 

%%%%%%%%%%%%%%%%%%%%%%%%%%%%%%%%%%%%%%%%%%
\subsubsection{Multivalued symplectic maps}\label{s:multival}
A canonical (or symplectic) {\em relation} is an arbitrary subset 
$\Gamma \subset \TT^{2n} \times \TT^{2n} $, such that 
\[ 
\Gamma' = \set{ ( q',q; p',-p) \; : \; ( q', p'; q, p) \in \Gamma }
\]
is a Lagrangian 
submanifold of $\TT^{4n}$.
%Clearly, $\Gamma$ 
%is a canonical map if it is the graph of a map and $\Gamma' $ is Lagrangian.

We are interested in symplectic relations coming from 
multivalued symplectic maps. A multivalued map is the union of
finitely many components $\kappa_j$, where $\kappa_j$ is a canonical
diffeomorphism $\kappa_j$ between an open subset $\cS_j$ with piecewise smooth 
boundary of $\tn$ and its image 
$\cS_j'=\kappa_j(\cS_j)\in\tn$. A priori, the sets $\cS_j$ (respectively
 $\cS_j' $) 
can overlap, and
their union can be a proper subset of $\tn$. 

Each map $\kappa_j$ is associated to its graph
$$
\Gamma_j=\set{(\kappa_j(\rho);\rho)\;:\;\rho\in \cS_j}\,,
$$
and the symplectic relation can now be defined through its graph 
$$
\Gamma=\bigcup_{j}\Gamma_j\,,
$$
or equivalently its twisted graph 
$\Gamma'$ (defined from $\Gamma$ as in \eqref{e:twisted}). $\Gamma'$ is
a singular Lagrangian in $\TT^{4n}$. 

The inverse relation can be defined by
$$
\Gamma^{-1}\defeq \set{(\rho;\rho')\;:\;(\rho';\rho)\in\Gamma}
=\bigcup_j\set{(\kappa^{-1}_j(\rho);\rho)\;:\;\rho\in \cS'_j}
\,,
$$
and the composition of two relations by
$$
\widetilde{\Gamma}\circ\Gamma\defeq \set{ ( \rho''; \rho ) \in   \TT^{4n}
\; : \; \exists \, \rho' \in \tn \,, \ 
( \rho'; \rho) \in \Gamma \ \ \text{and}\ \ ( \rho''; \rho' ) \in \widetilde\Gamma }\,.
$$
Following \cite[Theorem 21.2.4]{Hor}, we note that $ \widetilde \Gamma 
\circ \Gamma $ will be a (locally)
smooth symplectic relation if 
$ \widetilde \Gamma
\times \Gamma \subset \TT^{4n} \times \TT^{4n} $ intersects 
$$
\{ ( \rho'' , \rho' , \rho' , \rho ) \; : \; 
\rho'', \rho', \rho \in \tn \} \subset  \TT^{4n} \times \TT^{4n} 
$$
cleanly, that is the intersections of tangent spaces are the 
tangent spaces of intersections. 

We can then iterate a relation $\Gamma$, defining a multivalued
dynamical system $\{\Gamma^n : n\in\ZZ\}$
on $\tn$. In \S\ref{s:wr} we will give a stochastic interpretation to 
this system.

%%%%%%%%%%%%%%%%%%%%%%%%%%%%%%%%%%%%%%%%%%

\subsection{Open baker's relation}\label{e:open-baker-clas}
The dynamics we will consider takes place on the 2-torus phase space, 
$$ 
\t2 = \set{ \rho=( q , p ) \; : \;  q,\,p\in\II} \,.
$$
On this phase space, we define two vertical
strips $\cS_j$ ($j=1,2$) from the data of four real numbers $D_1,\,D_2>1$ and
$\ell_1,\,\ell_2\geq 0$:
\begin{equation}
\cS_j=\set{(q,p)\,:\,q\in I_j,\ p\in\II},\qquad\text{with}
\quad I_j=\big(\frac{\ell_j}{D_j},\, \frac{\ell_j+1}{D_j}\big)\,\qquad j=1,2.
\end{equation}
The strips are assumed to be disjoint, which is the case if we impose the
conditions:
$$
\frac{\ell_1 + 1}{D_1}\leq\frac{\ell_2}{D_2}\quad\text{and}\quad
\frac{\ell_2 +1}{D_2}\leq 1\,.
$$
The corresponding baker's relation is made of two components $B_j$, $j=1,2$
associated with linear symplectic maps defined on the two strips:
\begin{equation}\label{e:classical-baker}
B_j=\set{ (\rho';\rho)\;:\;
( q', p' ) = \left( D_j q - \ell_j , \frac{p + \ell_j }{D_j} \right)
\,,\;\rho=(q,p)\in \cS_j}\,.
\end{equation}
The baker's relation is defined as the graph $B=B_1\cup B_2$.
One clearly notices that each component map is a hyperbolic diffeomorphism, 
with positive stretching exponent $\log D_1$ (resp. $\log D_2$). 
At all points where the map is defined, the unstable (stable) 
direction is the horizontal (vertical) one.

Since the two
strips are disjoint, each point $\rho\in\t2$ has at most one image. 
In the notations of Proposition~\ref{l:well} (taking $q''=q$, $q'=q'$), 
each Lagrangian component
$B_j'$ can be generated by the function 
\begin{equation}\label{e:generating}
S_j(q,p')=D_j \left(q-\frac{\ell_j}{D_j}\right)\left(
p'-\frac{\ell_j}{D_j}\right)\quad\text{defined on the square}\quad
\set{(q,p')\in I_j\times I_j}\,.
\end{equation}

Let 
$$  
\pi_L \,, \pi_R  \; : \;   \t2 \times \t2 
\; \longrightarrow \; \t2
$$ 
be the projections on the left and right factors respectively. 
From the definition \eqref{e:classical-baker},
the set $\pi_R(B)=\cS_1\cup\cS_2$ is made of points on $\rho\in\t2$ which have an image through the 
relation $B$.
Hence, a point 
$\rho\not\in \pi_R(B)$ is said to escape from the torus at time $1$. Similarly, a point 
$\rho\not\in \pi_L(B)=\pi_R(B^{-1})$
is said to escape from $\t2$ at time $-1$. This ``escape'' is the reason why we call
this relation an ``open'' relation: the system is not ``closed'' because it
sends particles ``to infinity'', both in the future and in the past.

We define
\begin{equation}\label{e:Gamma+}
\Gamma_\pm \defeq \bigcap_{n=1}^\infty
\pi_R  \left( B^{\mp n } \right) \,
\end{equation}
the set of points which do never escape from $\t2$ in the past, 
respectively in the future. 
One checks that these subsets have the form
$$
\Gamma_- = C \times {\mathbb I} \,, \ \  \Gamma_+ 
= {\mathbb I} \times C \,, 
$$
where $ C \subset {\mathbb I} $ is a ``cookie-cutter set'' in the 
sense of \cite{Fal}: if we consider the two contracting maps on $\II$
$$
f_j ( q ) = \frac{ q + \ell_j}{ D_j } \,,\quad q\in\II\,,\quad j=1,2\,,
$$
this closed set is defined as
$$ 
C = \overline{ \bigcup_{ n \in {\mathbb N}} \set{ q\in\II \; : \; f_{j_1} \circ \cdots 
\circ f_{j_n} ( q ) = q \,  \ \text{ 
for some sequence $ j_m \in \{1,2 \} $} } } \,.
$$
The Hausdorff 
dimension of $ C $ (which is equal to its Minkowski and box-counting dimensions)
is given by the unique $ 0 < \nu < 1 $ solving
\bequ\label{e:hausdorff}  
D_1^{-\nu} + D_2^{-\nu} = 1 \,. 
\end{equation}
The trapped set (or set of nonwandering points) 
is defined as the set of points which never escape from $\TT$:
$$ 
K = \Gamma_+ \cap \Gamma_- = C \times C \,, \quad
\dim K = 2\nu    \,. 
$$
The baker's relation is a hyperbolic invertible map on the set $K$, which is 
a ``fractal repeller''. This relation is a model of
Smale's horseshoe mechanism.

The simplest case consists in considering a symmetric baker's relation, with
$D_1=D_2=D$, $\ell=\ell_1=D-\ell_2-1$:
\begin{equation}
\begin{split}\label{e:symmetric}
\frac{\ell}{D}  < q < \frac{\ell + 1}{D} \quad & \Longrightarrow \quad
( q', p' ) = \left( D  q - \ell , \frac{p + \ell}{D} \right) \\
\frac{\ell}{D}< 1- q <  \frac{\ell + 1}{D} \quad&\Longrightarrow \quad
( q', p' )   = \left( D(q-1) + \ell +1  , 
\frac{p - \ell - 1}{D}+1 \right) \,. 
\end{split}
\end{equation}
Now $ C\subset {\mathbb I} $ is a symmetric $1/D-$Cantor set.
Notice that if we take $D=2$, $\ell_1=0$, $\ell_2=1$, we obtain the
usual (closed) baker's map described in the example of \secref{s:sympl-tori}, 
for which the trapped set ($=\t2$) has dimension $2$. The
``3-baker'' relation described in \eqref{eq:exb} corresponds to $D=3$, $\ell=0$.

For such a symmetric baker's relation, the analog of the 
fractal exponent of \eqref{eq:fweylh} is:
$$ 
\mu_E \ \longleftrightarrow \ \nu=\frac{\log 2 }{\log D } \,.
$$

%%%%%%%%%%%%%%%%%%%%%%%%%%%%%%%%%
\subsection{Weighted symplectic relations}
\label{s:wr}
%%%%%%%%%%%%%%%%%%%%%%%%%%%%%%%%%

To give a multivalued map $\Gamma$ a physical meaning, we  
assign {\em Markovian weights} $P_j(\rho)$ 
to the different ``jumps'', $\rho\mapsto \kappa_j(\rho)$.
The associated dynamical system is then stochastic, 
each point $\rho$ having finitely many images with well-prescribed 
transition probabilities $P_j(\rho)$.
The sum of all the probabilities from
$\rho$  must satisfy $0\leq P(\rho)\defi\sum_j P_j(\rho)\leq 1$, so that
$(1-P(\rho))$ is the probability that $ \rho$ ``escapes to infinity''. 

The weights associated with the inverse relation $\Gamma^{-1}$ are the same:
each point $\rho'$ jumps back to $\kappa_j^{-1}(\rho')$ with probability
$P'_j(\rho')=P_j(\kappa_j^{-1}(\rho'))$. Hence, the weights must also satisfy
$0\leq \sum_j P'_j(\rho')\leq 1$.

Such a weighted relation 
(in geometric optics one would speak of a ``ray-splitting'' map)
induces a discrete-time evolution of ``mass distributions'', which is 
in general \emph{dissipative}: the full mass
decrease at each step, the system 
expelling part of the mass  ``to infinity''.

\medskip

In more mathematical terms, 
we assume that the symplectic relation $\Gamma\subset
\tn\times \tn$ 
comes with a nonnegative measure (or weight) 
$ \mu $ on $\Gamma $, which for any 
$ \chi_\alpha \in \CI ( \tn, [0,1] ) $, $\alpha = L, R$, satisfies
\begin{gather}
\label{eq:meas}
\begin{gathered}
\pi_{\alpha *} ( \pi_L^* \chi_L \;   \pi_R^* \chi_R \; \mu  ) 
= g_\alpha^{\chi_L \chi_R} \;
 \frac{ \omega^n}{n!} \,, \quad  g_\alpha^{ \chi_L \chi_R }  \in \CI ( \tn ) \,, \quad  
0 \leq g_\alpha^{\chi_L \chi_R } \leq 1\,, 
\end{gathered}
\end{gather}
where $ \pi_L,\ \pi_R : \Gamma \rightarrow \tn $ 
are projections on left and right factors respectively, and $ \omega $
is the symplectic form on $ \tn $. 
The condition \eqref{eq:meas} implies that $ \pi_\alpha |_{\Gamma} $ 
is a local bijection, which
forces $ \Gamma $ to be a piecewise smooth union of graphs
of symplectic transformations, as defined in \S \ref{s:multival}.
When $ \Gamma$ is
singular, that is a union of smooth symplectic relations with 
boundaries, we demand that 
$$
g_\alpha^{\chi_L \chi_R} \in \CI ( \tn ) 
\quad \text{ if}\quad \supp ( \pi_L^* \chi_L \;   \pi_R^* \chi_R ) \cap 
\partial \Gamma = \emptyset\,,
$$
where $ \partial \Gamma  $ is the union of 
the boundaries of the smooth components.

The reason for introducing the measure $ \mu $ is to have
a quantity independent of the choice of coordinates
on $ \Gamma$. On $ \TT^{2n} $, an obvious intrinsic measure is
given by the symplectic form, hence $ g^{\chi_L \chi_R }_\alpha $
are well defined. Building an atlas of the manifold $ \Gamma $ 
we can use these functions to describe $\mu$ in local coordinates.

We denote a weighted relation by $( \Gamma , \mu )$ 
As explained above, one can invert such a relation, as well as compose them.

If $(\rho';\rho)\in\Gamma\setminus \partial \Gamma$, 
the probability of a transition from
$\rho$ to $\rho'=\kappa_{j_1}(\rho)$ is obtained by letting $\chi_R$ (resp. $\chi_L$) 
be supported in a sufficiently
small neighbourhood of $\rho$ (resp. of $\rho'$), with $\chi_R(\rho)=1$, $\chi_L(\rho')=1$.
This probability is then given by
\begin{equation}\label{e:probas}
P_{j_1}(\rho)=g_R^{\chi_L\,\chi_R}(\rho)= g_L^{\chi_L\,\chi_R}(\rho')=P'_{j_1}(\rho')\,.
\end{equation}

\medskip

\noindent {\em Examples.}
The simplest example is given by a graph of a symplectic 
transformation $ \kappa : \tn \rightarrow \tn $ in which 
case the density $ \mu $ is obtained by taking 
$ \mu = \pi_L^* (\omega^n/n!) = \pi_R^* (\omega^n/n!) $, where the
equality follows from $ \kappa^* \omega = \omega $. A slightly more 
complicated example is given by taking a union of two non-intersecting 
graphs $ \Gamma_j$ of $ \kappa_j $, $ j = 1,2 $, and putting 
$$
\mu =  (\pi_R |_{\Gamma_1})^* (g_1\,\omega^n/n!)
+  (\pi_R |_{\Gamma_2})^* (g_2\,\omega^n/n!) \,, 
$$
where $ g_j \in \CI ( \tn ; [0,1] ) $ satisfy $g_1 + g_2\leq 1$ and 
$g_1\circ\kappa_1^{-1}+g_2\circ\kappa_2^{-1}\leq 1$. 
In this case, $ g_j ( \rho )=P_j(\rho)$.

In the case of an open baker $B$ defined in \S \ref{e:open-baker-clas},
for instance the symmetric 3-baker \eqref{eq:exb},
a natural $\mu$ comes from pulling back the Liouville measure 
$\omega$ to each component $B_j$ given in \eqref{e:classical-baker}. One obtains
\begin{equation}
\label{eq:bmes}
\pi_{R *}\, \mu = \bbbone_{ I_1\cup I_2 } ( q )\; dq\, dp \,, \quad 
\pi_{L *}\, \mu = \bbbone_{ I_1\cup I_2 } ( p' )\; dq'\, dp' \,.
\end{equation}
These equations fully determine the measure $\mu$ on $B$.

%\renewcommand\thefootnote{\sharp}%

%\begin{figure}[ht]
%\includegraphics[width=13cm]{dense.eps}
%\caption{Plots of the left push-forwards $g_L^{1\,1}(\rho')=g_L^{1\,1}(p')$ for
%the densities $\mu$ on the symmetric 3-baker relation \eqref{eq:exb}
%(left)
%and $\tilde \mu$ on the multivalued baker relation \eqref{eq:tbc} (right).}
%\label{f:dense}
%\end{figure}

A more interesting example, which will be relevant in \S \ref{s:tm}, is given 
by the following multivalued generalization of the symmetric 3-baker:
\begin{gather}
\label{eq:tbc}
\begin{gathered}
 \tB = \bigcup_{\ell=0}^2 \big(B+(0,\ell/3;0,0)\big)
=\bigcup_{k=1}^2 \bigcup_{j=0}^2 \tB_{kj}\,,\quad \text{where}  \\
\tB_{kj} = \set{ \big(3q, \frac{p+j}{3}; q , p \big) \, : \,
q\in I_k,\,p\in\II }\,,\quad I_1=(0,1/3),\ I_2=(2/3,1)\,.
%\tB_{kj} \defeq  \set{ (3q - 2 k, (p+j)/{3}; q , p ) \; : \;
%q\in (2k/3 , (2k+1)/3)  \,,p\in\II }\,.
\end{gathered}
\end{gather}
Each point
$\rho\in\cS_1\cup\cS_2=(I_1\cup I_2)\times \II$
has $3$ images, and each point $\rho'\in\t2$ has two preimages.

The following measure on $\tB$ will arise in the quantum 
model studied in \S \ref{s:tm}.  We define it explicitely on each component
$\tB_{kj}$, using the right projection on $\cS_k$:
\begin{equation}
\label{eq:tbm}
\begin{split}
\pi_{R *}\,\tilde\mu\rest_{\tB_{1j}} & 
= \frac{\sin^2(\pi p)} {9 \sin^2(\pi(p + j)/3)}\,\bbbone_{I_1}(q)\; dq\,dp\,,
\\
\pi_{R *}\,\tilde\mu\rest_{\tB_{2j}} & 
= \frac{\sin^2(\pi p)} {9 \sin^2(\pi(p + j-2)/3)}\,\bbbone_{I_2}(q)\; dq\,dp\,,
\quad j=0,1,2 \,.
\end{split}
\end{equation}
The functions on the right hand sides are the probabilities $P_j(\rho)$. 
The sum of these components reads
\begin{gather*}
\pi_{R *}\, \tilde \mu = \left(\frac19\sum_{j=0}^2  \frac{\sin^2\pi p } 
{ \sin^2\pi(p/3 + j/3)} \right) \;  
\bbbone_{ I_1\cup I_2 } (q) \; dq\, dp 
= \bbbone_{ I_1\cup I_2 } (q) \; dq\, dp  \,.
\end{gather*}
Here we used the fact\footnote{The value of the sum at $x = 0$ is
equal to $D^2$, and the sum is invariant under translation 
$ x \mapsto x + k  \pi / D $. Fej\'er's formula for the Ces\`aro mean of 
the Fourier series shows that the sum is a trigonometric polynomial 
of degree $D -1$ in $x$, hence it is constant.} 
that $\sum_{ j=0}^{D-1} \sin^2 ( D x ) /\sin^2 ( x + j \pi / D)  = D^2$,
with $ D = 3 $ and $ x = \pi p/3 $.
This right pushforward is identical to that of \eqref{eq:bmes}: 
in both cases, any point $\rho\in (\cS_1\cup \cS_2)$ has an empty escape
probability, $1-P(\rho)=0$.

On the opposite, the left pushforward of $\tilde\mu$ is given by 
\begin{gather*}
\pi_{L *}\, \tilde \mu =    \frac{ \sin^2 3 \pi p' }{ 9 } \left( \frac1{ \sin^2 \pi p' } 
+  \frac{ 1 }{  \sin^2 \pi ( p' - 2/3 ) } \right)
\; dq' \,dp' \,.
\end{gather*}
Almost any point $\rho'\in\t2$ has a nonzero escape probability through 
$\tB^{-1}$. 
This left pushforward is obviously different from that of $\mu$. % (see Fig.~\ref{f:dense}).

%We see that the conditions in the definition of weighted relations
%are satisfied for $(\tB,\tilde \mu )$.

%%%%%%%%%%%%%%%%%%%%%%%%%%%%%%%%%%%%%%%%%%%%%%%%%%%%%%%%%%%%%%%%%
%%%%%%%%%%%%%%%%%%%%%%%%%%%%%%%%%%%%%%%%%%%%%%%%%%%%%%%%%%%%%%%%%

\section{Quantized maps and relations}
\label{oqm}

Before giving the definition of the quantized baker's relation, we need to define
the quantum Hilbert space corresponding to $\t2$, as well as the algebra of
quantum observables. 

%%%%%%%%%%%%%%%%%%%%%%%%%%%%%%%%%%%%%%%%%%%%%%%%%%%%%%%%%%%%%%%%%

\subsection{Quantized tori}\label{e:quantum-torus}
The quantization of tori $ \tn = \RR^{2n} / \ZZ^{2n} $ has
a long tradition in mathematical physics \cite{HB,DE,BouzDB}.
It can be considered as a special case of the Berezin-Toeplitz 
quantization of compact symplectic K\"ahler manifolds --- see
\cite{toeplitz} and references given there. Here we will give a 
self-contained presentation of the simplest case from the point
of view of pseudodifferential operators. 

We first recall from \cite{DiSj} the quantization of functions 
$ f \in  \CIb ( T^*\RR^n )$, 
$$  
\CIb  ( T^* \RR^n ) \defeq
\{ f \in \CI ( T^* \RR^n ) \; : \;  \forall \alpha,\beta\in\NN^n,\;
\sup_{ ( q, p ) \in T^* \RR^n } | \partial^\alpha_q 
\partial_p^\beta f ( q , p ) | <\infty \} \,.
$$
To any $ f \in \cS ( T^* \RR^n ) $ we associate its $h$-Weyl 
quantization,
that is the operator $f^w ( q, hD )$ acting as follows on $\psi\in\cS(\RR^n)$:
\begin{equation}
\label{eq:aw} [f^w ( q, hD )\, \psi](q)\defeq \frac{1}{( 2 \pi h )^n } 
\int \! \int f \Big( \frac{q+r}{2} , p \Big)\, e^{ \frac{i}{h}
\la q - r, p \ra}\, \psi (r)\, dr\, dp \,.
\end{equation}
This operator clearly has the mapping properties 
$$
f^w ( q , h D ) \; : \; \cS ( \RR^n ) \; \longrightarrow 
\; \cS ( \RR^n ) \,, \ \ 
  f^w ( q , h D ) \; : \; \cS' ( \RR^n ) \; \longrightarrow 
\; \cS' ( \RR^n ) \,.
$$
It can be shown \cite[Lemma 7.8]{DiSj} that $ f \mapsto f^w ( q , h D ) $ 
can be extended to any $ f\in \CIb ( T^* \RR^n ) $, and that the resulting 
operator has the same mapping properties. Furthermore, $f^w ( q , h D )$ is 
a bounded operator on $L^2(\RR^n)$.

We now introduce quantum spaces associated with the torus $\tn$. For that aim, we
fix our notations for the semiclassical Fourier transform on ${\mathcal S}' ( \RR^n )$: 
$$
\F_h \psi(p) \defeq \frac{1}{( 2 \pi h )^{n/2} }
\int \psi(q)\,e^{-\frac{i}{h} \langle q , p \rangle }\, dq \,,  
%\qquad \F_h^* = \F_h^{-1} \,, 
$$
and as usual in quantum mechanics, 
$\F_h \psi(p)$ is the ``momentum representation'' of the state $\psi$.
%The quantum spaces of the torus are indexed by a pair of
%Bloch angles $ ( \theta_p , \theta_q ) \in 
%\II^n \times \II^n $. Given any such pair, t
The torus quantum space is made of
distributions $\psi\in\cS'(\RR^n)$ which are both 
periodic in position
and momentum:
\begin{equation}
\label{eq:quasi}
\psi ( q + \ell ) = %e^{ 2 \pi i \langle \theta_p , \ell \rangle } 
\psi ( q ) \,, \qquad 
\F_h \psi (p + \ell ) = %e^{ 2 \pi i \langle \theta_q , \ell \rangle } 
\F_h \psi (p)\,.
\end{equation}
Let us denote by $ \Hh_h^n $ this space of distributions.
We have the following elementary
%%%%%%%%%%%%%%%
\begin{lem}
\label{l:elem}
$ \Hh_h^n \neq \{ 0 \} $ if and only if $ h = ( 2 \pi N )^{-1} $ for some positive
integer $N$, in which 
case $ \dim \Hh_h^n = N^n $ and $\Hh_h^n$ is generated by the following basis:
\begin{equation}
\label{eq:bas}  \Hh_h^n = {\rm{span}} \set{ \frac{1}{\sqrt{N^n}}
\sum_{ \ell \in \ZZ^n } 
\delta( q - \ell - j/N) \; : \; j \in (\ZZ / N \ZZ)^n } \,.
\end{equation}
\end{lem}
%%%%%%%%%%%%%%%
The distributions elements of this basis will be denoted by
\begin{equation}
\label{eq:Qjb} 
|Q_j\ra \,, \quad 
\text{$ Q_j = \frac{j}{N}\in\II^n$ is the position on which that state is microlocalized. }
\end{equation}
One can check that for such a value of $h$, the Fourier transform $\F_h$ maps
$\Hh_h^n$ to itself. In the above basis, it is represented by the discrete Fourier 
transform
\begin{equation}\label{e:DFT}
(\F_N)_{j\,j'}=\frac{e^{-2 i \pi \la j,j'\ra/N}}{N^{n/2}}\,,
\quad j,j'\in (\ZZ/ N \ZZ)^n\,.
\end{equation}
It is also easy to check the following
%%%%%%%%%%%%%%%
\begin{lem}
\label{l:elem'}
Suppose that $ f \in \CIb ( \RR^n \times \RR^n ) $ satisfies 
$ f ( q + \ell , p + m) = f ( q , p ) $ for any $ \ell, m \in \ZZ^n $.
Then the operator
$ f^w ( q , h D ) $ maps $ \Hh_h^n $ to itself.
\end{lem}
%%%%%%%%%%%%%%%
Identifying a function $ f \in \CI (\tn) $ with a 
periodic function on $ \RR^{2n} $, we will write $\Op_h ( f) $ for the restriction
of $ f^w ( q, h D)$ on $\Hh_h^n$, 
$$
\CI ( \tn ) \ni f \; \longmapsto \; \Op_h ( f ) \in 
\cL (\Hh_h^n)\,.
$$
We remark that $ \Op_h ( 1 ) = {\rm Id} $. 
The vector space $\Hh_h^n$ can be equipped with a natural
Hilbert structure.

%%%%%%%%%%%%%%%%
\begin{lem}
\label{l:hilb}
There exists a unique (up to a multiplicative constant) Hilbert
structure on $ \Hh_h^n $ for which all $ \Op_h(f) \; : \; \Hh_h^n \; 
\rightarrow \; \Hh_h^n $ with $ f\in\CI(\tn;\RR) $ are 
self-adjoint.

One can choose the constant such that the basis 
in \eqref{eq:bas} is orthonormal. 
This implies that the Fourier transform on $\Hh_h^n$ (represented by the
unitary matrix \eqref{e:DFT}) is unitary. 
\end{lem}
%%%%%%%%%%%%%%%%
\begin{proof}
Let $ \langle \bullet , \bullet \rangle_0 $ be the inner product 
for which the basis in \eqref{eq:bas} is orthonormal.
We write the operator $ f^w ( q , hD ) $ on $ \Hh_h^n $ 
explicitely in that basis using the Fourier expansion of its symbol:
\[
 f ( q, p ) = \sum_{ \ell, m \in \ZZ^n } \hat f ( \ell, m ) \,
e^{ 2 \pi i ( \la \ell , q \ra + \la m , p \ra ) } \,.
\]
For that let $L_{\ell, m } ( q, p ) = \la \ell , q \ra + \la m , p \ra $,
so that
\[ 
f^w ( q, h D ) = \sum_{ \ell, m \in \ZZ^n } \hat f ( \ell, m ) \,
\exp( 2 \pi i L^w _{ \ell, m } ( q, h D ) ) \,. 
\]
Applying this operator to the distributions \eqref{eq:Qjb}, we get
\[
 \exp\big( 2 \pi i L^w _{ \ell, m } ( q, h D ) \big)\, |Q_j\ra = 
\exp \Big( \frac{\pi i }N ( 2 \la j , \ell \ra - \la m , \ell \ra)\Big)\,  |Q_{j-m}\ra  \,,
\]
and consequently,
\begin{gather*}
 f^w ( q , hD)\, |Q_j\ra   = \sum_{ m \in \ZZ^n/( N \ZZ)^n } F_{mj} \, |Q_m\ra 
\,, \\
F_{mj} =  \sum_{ \ell , r \in \ZZ^n  } 
\hat f ( \ell, j - m - rN) (-1)^{\la r, \ell \ra}\, 
\exp \Big( \frac{\pi i}{ N}\, \la j + m , \ell \ra  \Big) \,. 
\end{gather*}
Since
\[ 
\begin{split} 
\bar F_{jm} & = 
 \sum_{ \ell , r \in \ZZ^n  } 
\hat {\bar f} (- \ell, j - m + rN) (-1)^{\la r, \ell \ra} \,
\exp \Big(- \frac{ \pi i }{ N}\,  \la j + m , \ell \ra  \Big) \\ 
& = 
\sum_{ \ell , r \in \ZZ^n  } 
\hat {\bar f} ( \ell, j - m - rN) (-1)^{\la r, \ell \ra} \,
\exp \Big( \frac{ \pi i }{ N}\,  \la j + m , \ell \ra \Big) \,, 
\end{split} 
\]
we see that for real $ f $, $ f = \bar f $, $ F_{jm} = 
\bar F_{mj} $. This means that $ f^w ( q , hD ) $ is 
self-adjoint for the inner product $ \langle \bullet , \bullet \rangle_0 $. 
We also see that the map 
$ f \mapsto (F_{jm} )_{ j , m \in ( \ZZ / N \ZZ )^n } $ is onto, 
from $ \CI ( \tn ; \RR ) $ to the space of Hermitian matrices. 

Any other metric on $ \Hh_h^n $ could be written as $ \langle u , 
v \rangle  = \langle B u , v \rangle_0 = \langle u , B v \rangle_0 $.
If $ \langle f^w u , v \rangle = \langle u , f^w v \rangle $ for all
$ f$'s, then $ B f^w = f^w B $ for all $ f$'s, and hence for all
Hermitian matrices. That shows that $ B = c\, {\rm Id} $, as claimed.
\end{proof}

This choice of normalization $ \langle \bullet , \bullet \rangle_0 $
can be obtained in a natural way, 
if we use the following periodization operator to construct $\Hh_h^n$ from
$\cS(\RR^n)$ \cite{BouzDB}:
%%%%%%%%%%%%%%%%
\begin{lem}
For any $h=(2\pi N)^{-1}$, 
the periodization operator $P_{\tn}:\cS(\RR^n)\to \Hh_h^n $ defined below is surjective:
\begin{equation}\label{eq:Ptn}
\forall \psi\in\cS(\RR^n),\qquad [P_{\tn}\, \psi](Q_j) \defeq 
\frac{1}{N^{n/2}}\sum_{\nu\in\ZZ^n} \psi(Q_j-\nu)\,,\qquad j\in(\ZZ/N\ZZ)^n\,.
\end{equation}
\end{lem}
%%%%%%%%%%%%%%%%
In the rest of this article we will always assume that $h=(2\pi N)^{-1}$ for some
$N\in\NN$, so the semiclassical limit  corresponds to $N\to\infty$. The scalar product
on $\Hh_h^n$ will be $ \la\bullet , \bullet \ra_0 $.
From now on we will omit the subscript $0$, and 
also often use Dirac's notation $\la \bullet |\bullet\ra$ for this product. For instance,
the $j$-th component of a state $\psi\in \Hh_h^n$ in the basis \eqref{eq:Qjb} will be
denoted by $\psi(Q_j)=\la Q_j|\psi\ra$.
The 
Hilbert norm associated with $\la \bullet , \bullet \ra$ 
will simply be written $\mn{\bullet}$.

%%%%%%%%%%%%%%%%%%%%%%%%%%%%%%%%%%%%%%%%%%%%%%%%%%%%%%%%%%%%%%%%%%%%%%%%%%%%%%%
%%%%%%%%%%%%%%%%%%%%%%%%%%%%%%%%%%%%%%%%%%%%%%%%%%%%%%%%%%%%%%%%%%%%%%%%%%%%%%%

\subsection{Lagrangian states}
\label{s:lagrangian}
We want to characterize the semiclassical localization in phase space 
of sequences of states of the form $ \psi = \{\psi_h \in \Hh^n_h\}_{h\to 0} $. 
In general we will assume that
each element of this sequence is normalized,  $ \| \psi_h \| = 1$, but
all definitions can be extended to sequences such that the norms 
satisfy $\mn{\psi_h}=\Oo( h^K )$ as $h\to 0$, for some fixed $K\in\RR$
(the sequence $\psi$ is then said to be {\em tempered}).
 
The localization of this sequence is first characterized 
through its  {\em microsupport}, or
{\em wave front set}, which is the following subset of $\tn$:
\begin{equation}
\label{eq:defwf} 
\WF_h ( \psi ) = \complement \set{ \rho \in \tn \; :\; 
\exists \; f \in \CI ( \tn ) \,, \ f( \rho ) \neq 0 \,, \ \
\mn{ \Op_h ( f ) \psi_h } 
= {\Oo} ( h^\infty )} \,, 
\end{equation}
where $ \complement $ stands for the set theoretical complement.
It is not hard to
show \cite[Proposition IV-8$'$]{robert} that this definition is equivalent
to the following: $\rho\not\in \WF_h ( \psi )$ if and only 
if there exists a neighbourhood
$W_\rho$ of $\rho$ such that, for any $f\in \CI ( \tn )$ supported in $W_\rho$,
$\mn{ \Op_h ( f ) \psi_h } = {\Oo} ( h^\infty )$. This yields the following
%%%%%%%%%%%%%%%%%%%%%%%%%%%
\begin{lem}\label{l:wfs}
For any function $f\in \CI ( \tn )$ with $f\equiv 0$ in an open neighbourhood of
$ \WF_h ( \psi )$, we have $\mn{ \Op_h ( f ) \psi_h } = {\Oo} ( h^\infty )$.
As a consequence, 
the microsupport of a sequence 
$\psi=\{\psi_h\}$, $\mn{\psi_h}\asymp h^K$, cannot be empty.
\end{lem}
%%%%%%%%%%%%%%%%%%%%%%%%%%%
\begin{proof}
The (compact) support of $f$ can be covered by finitely many $W_{\rho_i}$, and
using a partition of unity associated with these sets we can decompose it as
$f=\sum_i f_i$, with $\supp(f_i)\subset W_{\rho_i}$. We get the result
by linearity, and using the second definition of $\WF_h(\psi)$.
\end{proof}
We also make the following observation:
%%%%%%%%%%%%%%%%%%%%%%%%%%%
\begin{lem}
\label{l:wfo}
Let $ \psi = \{\psi_h \in \Hh^n_h\}_{h\to 0} $ be a tempered sequence. 
Considering $\psi_h$ as a $N^n$-component
vector in the basis \eqref{eq:bas}, we define $\bar\psi_h$ 
as the vector with complex conjugate components. Then
\[ 
\WF_h ( \bar \psi ) = \{ ( q, - p ) \; : \; ( q , p ) \in \WF_h ( \psi ) \}\,.
\]
\end{lem}
%%%%%%%%%%%%%%%%%%%%%%%%%%%
\begin{proof}
The definition  \eqref{eq:aw} of Weyl's quantization gives, for any
function $f\in\CI(\tn)$,
$$ 
\Op_h ( f )\, \bar \psi = 
f^w ( q , hD)\, \bar \psi = \overline { \bar f^w ( q, - h D)\, \psi }  \,. 
$$
The lemma follows from the definition~\eqref{eq:defwf} of the wave front set.
\end{proof}
Now let  $\Lambda\subset \tn $ be a 
union of Lagrangian submanifolds of $ \tn $ with piecewise 
smooth boundaries. 
%%%%%%%%%%%%%%%%%%
\begin{defn}\label{d:Lagrangian}
A sequence of states
$ \psi = \{\psi_h \in \Hh_h^n\} $  is a 
{\em Lagrangian state associated to $ \Lambda $}, which we denote by
$\psi \in I ( \Lambda )$,
if for any $M\in\NN$ and any sequence of functions,
$$
 f_j \in \CI ( \tn ) \,,  \quad 1 \leq j \leq M \,, \quad f_j \rest_\Lambda = 0  \,,
$$
we have 
\begin{equation}\label{e:def-WKB}
\mn{\Op_{h} (f_M)  \circ \cdots \circ \Op_{h} ( f_1)\, \psi_h} = 
\Oo ( h^{M} )\, \mn{\psi_h} \,. 
\end{equation} 
\end{defn}
%%%%%%%%%%%%%%%%%%
From the definition \eqref{eq:defwf} of the microsupport, we obtain that, if 
the sequence $\psi$ is tempered, then
\begin{equation}
\label{eq:wfi}
\psi \in I ( \Lambda ) 
\; \Longrightarrow \; \WF_h ( \psi ) \subset \Lambda \,.
\end{equation} 
Indeed, suppose that $ \rho \not \in \Lambda $. Then there exists 
$ f \in \CI ( \tn ) $ such that $ f\rest_\Lambda = 0 $ and $ f \equiv
1 $ in a neighbourhood of $ \rho $. We can also 
find $ a \in \CI ( \tn ) $ such that $ f = 1 $ on a 
neighbourhood of the support of $ a $, and $ a ( \rho ) \neq 0 $. 
The symbol calculus (see \cite[Chapter~7]{DiSj}) shows that for
any $ M $, $ \Op_h ( a ) \Op_h ( f ) ^M = \Op_h ( a ) + \Oo_M ( h ^\infty ) $. 
On the other hand 
$ \mn{ \Op_h ( f )^M \psi_h } = \Oo ( h^{M} \mn{\psi_h}) $, and as $ M $ is arbitrary
and $\psi$ tempered,
$ \mn{ \Op_h ( a )\, \psi_h }= \Oo ( h^\infty ) $. In view of 
\eqref{eq:defwf}, this gives \eqref{eq:wfi}.

We stress that the opposite implication in \eqref{eq:wfi}
is not true in general. To see that consider $n=1$ and the Lagrangian $ \Lambda = 
\{ ( 0 , p ) \; : \; p \in \II \}\subset \TT^2 $.  Let $ \psi_h \in \Hh^1_h $ be 
the ``torus coherent state at the origin'':
\[ 
\psi_h(Q_j)=\Big(\frac{2}{N}\Big)^{1/4}\,\sum_{r\in\ZZ}\exp\{- \pi  N (Q_j-r)^2 \}
 \,,\quad j=0,\ldots,N-1\,.
\]
Then one can check that $\mn{\psi_h}\hto0 1$, that 
$ \WF_h ( \psi ) = \set{ ( 0, 0) } \subset \Lambda $.
On the other hand, 
\[ 
\mn{\Op_h\big( \sin(2\pi q) \big) \psi_h}\sim \pi\sqrt{2 h}\,,
\]
which shows that $ \psi_h \notin I(\Lambda) $.

In the physics literature, Lagrangian
states are usually called WKB states, and are introduced as 
Ans\"atze for eigenstates of integrable systems, using
Bohr-Sommerfeld quantization formulae \cite{keller}.
For instance, in the case $n=1$, if $\Lambda$ is generated by the 
function $S\in \CI(\II)$:
\begin{equation}
  \label{eq:lamb} \Lambda_S=\set{(q,-S'(q)),\,q\in \II }\,, 
\end{equation}
then for any function $a(q)\in \CI (\II) $, 
the state $\psi_h\in \Hh_h^1 $ defined as
\begin{equation}
  \label{eq:qj}
  \psi_h(Q_j)=\frac{a(Q_j)}{\sqrt{N}}\,\exp(-2i\pi N S(Q_j))\,,\quad j=0,\ldots,N-1\,,
\end{equation}
is in $ I( \Lambda_S ) $. 
In the next proposition, we generalize this construction to any dimension.
%%%%%%%%%%%%%%%%%%
\begin{prop}
\label{p:new}
Let $ \Lambda \subset \tn $ be an embedded Lagrangian
manifold. Then for any $ \rho_0 \in \Lambda $ 
there exist Lagrangian states $ \psi\in I(\Lambda) $, such that
$ \rho_0 \in  \WF_h (\psi) $. 
\end{prop}
%%%%%%%%%%%%%%%%%%
\begin{proof}
We take $ \rho_0=( q_0 , p_0 ) \in \Lambda $, and assume that 
there exists a neighbourhood $V$ of $\rho_0$, and a function $S\in \CI( \pi ( V) )$ 
(where $ \pi ( q , p ) = q $), 
such that $\Lambda\cap V =\{(q;-d_q S(q))\,,\,q\in \pi ( V) \}$. 
This is a particular case of Proposition~\ref{l:well}. The general case 
of a generating function $S(q'',p')$ can
be transformed to that of $S = S(q)$ using the symplectic rotation 
$(q',p')\mapsto (-p',q')$. 
On the quantum mechanical side,
this rotation is performed through a partial Fourier transform in the 
variable $q'$. 
Our construction below can be transposed to this general case through
this Fourier transform (which acts covariantly on the Weyl quantization).

We also assume that the neighbourhood
$V$ is contained in the interior of $\II^{2n}$, and we identify $ \pi ( V ) 
$ with
a subset of $\II^n$. 
We first construct a Lagrangian state in $L^2(\IR^n)$:
\bequ\label{eq:u_h}
u_h(q)= a ( q )\,e^{-\frac{i}{h} S (q)} \,,
\end{equation}
with a symbol $a\in \CI(\IR^n)$ compactly supported inside $ \pi ( V ) $, and such that $a(q_0)\neq 0$. 
This state admits the norm $\mn{u_h}_{L^2}=\mn{a}_{L^2}$. 
For any $f\in \CI (\tn )$, we apply the operator $f^w(q,hD)$ to that state.
Although we could do it directly using \eqref{eq:aw}, we prefer to 
reduce the problem to the case of $ S = 0 $ by conjugation with
the unitary multiplication operator 
\begin{equation}
\label{eq:tru}
 v ( q )  \longmapsto  [e^{\frac{i}{h} S^w (q)} v] ( q )=e^{\frac{i}{h} S (q)} v ( q ) \,, 
\end{equation}
where we can assume that
$S\in\CIb ( \RR^n)$. 
We then apply the
operator 
$$
G^w(q,hD) \defeq 
e^{\frac{i}{h} S^w (q)}\,f^w(q,hD)\,e^{-\frac{i}{h} S^w (q)}\,,
$$
to the function $a(q)$. 
The symbol calculus shows that 
$G(q,p)$ admits an $h$-expansion, with
principal symbol $g(q,p)=f(q,p+d_q S(q))$: if $f$ vanishes on $\Lambda$, then
$g$ vanishes on $\{(q,0)\,:\,q\in \pi ( V ) \}$. We get
$$
[f^w(q,hD)\,u_h](q)=e^{-\frac{i}{h} S^w (q)}\, G^w(q,hD)\,a(q)
=e^{-\frac{i}{h} S^w (q)}\, g^w(q,hD)\,a(q)
+\Oo(h)\,,
$$
The explicit integral
$$
[g^w(q,hD)\,a](q)=\frac{1}{( 2 \pi h )^n } 
\int \! \int g \Big( \frac{q+r}{2}, p\Big)\, a (r)\, 
e^{ \frac{i}{h} \la q - r, p \ra}\, dr\,dp\,
$$
can be evaluated through the stationary phase method.
The derivative of the phase vanishes at $r=q$, $p=0$, 
so the integral admits the following expansion \cite[\S 7.7]{Hor1} for 
$q\in \pi ( V) $:
\bequ\label{eq:Weyl-expansion}
[g^w(q,hD)\,a](q)=
L_0 (g\,a) (q) +h L_1 (g\,a)(q)+\Oo(h^2)\,.
\end{equation}
Here each function $L_j(g\,a)$ is obtained by applying a certain differential operator (in $(r,p)$)
on the function $g( (q+r)/{2}, p )\, a (r)$, taking the output
at the point $(r=q,\,p=0)$. The first term is simply $L_0(g\,a)(q)=g(q,0)\,a(q)$.
For $q$ outside $ \pi ( V) $,  the nonstationary
phase estimates show that 
\bequ\label{e:outside}
f^w(q,hD)\,u_h(q) = \Oo\left(\Big(\frac{h}{h+{\rm dist}(q, \pi ( V ))}\Big)^\infty\right) \,.
\end{equation}

If $f(\rho_0)\neq 0$, then $L_0 (g\,a)$ is 
nonzero in a neighbourhood $W$ of $q_0$, and we obtain
\bequ\label{eq:Weyl-integ}
\mn{f^w(q,hD)\,u_h}_{L^2(\IR^n)}=\mn{g^w(q,hD)\,a}_{L^2(\IR^n)}+\Oo(h)
\geq \mn{L_0 ( g\,a) }_{L^2(W)}+\Oo(h)\,.
\end{equation}
The left hand side is thus bounded from below by a positive constant.

On the opposite, 
if $f$ vanishes on $\Lambda$, then at each point $q\in \pi ( V) $ we get
$L_0 ( g\,a) (q)=0$, which implies that
$\mn{f^w(q,hD)\,u_h}_{L^2(\IR^n)}=\Oo(h)$.
The same procedure can be iterated to show that,
for any family of functions $f_i\in \CI ( \tn )$ vanishing on $\Lambda$, 
the function
\bequ\label{e:uhM}
u_h^{(M)}(q) \defeq h^{-M}\, [f_M^w \circ \cdots \circ  f_1^w\, u_h](q)\,, 
\end{equation}
is uniformly bounded and smooth on $\RR^n$, and very small outside $\pi(V)$,
as in \eqref{e:outside}. As a result,
\bequ\label{eq:lagr-L2}
\mn{f^w_M(q,hD)  \circ \cdots \circ f^w_1(q,hD) u_h}_{L^2(\IR^n)} = h^M\,\mn{u_h^{(M)}}= \Oo ( h^{M} )\,.
\end{equation} 

We can now carry over the estimates (\ref{eq:Weyl-integ},\ref{eq:lagr-L2}) 
onto the state $\psi_h=P_{\tn}\,u_h\in\Hh_h^n$,
where $P_{\tn}$ is the periodizing operator \eqref{eq:Ptn}. 
Since $a(q)$ was supported inside $ \pi ( V) \subset \II^n$, this state admits 
the following representation, which generalizes \eqref{eq:lamb}:
\bequ\label{eq:psi_h}
\psi_h ( Q_j ) = \frac{u_h ( Q_{j})}{N^{n/2}}=
\frac{a ( Q_{j})}{N^{n/2}}\,\, \exp ( - 2 i \pi N S ( Q_{j})) \,, \qquad
j\in(\ZZ/N\ZZ)^n\,.
\end{equation}
The norm of this state is therefore the sum 
$$
\mn{\psi_h}^2=N^{-n}\sum_{j\in (\ZZ/N\ZZ)^n}|a ( Q_{j})|^2=\int dq\,|a(q)|^2 +\Oo(h^\infty)\,,
$$
where we used the smoothness of $a(q)$.
Similarly, the projection on $\Hh_h^n$ of the function $u_h^{(M)}$ 
defined in \eqref{e:uhM} satisfies
\bequ\label{eq:sum}
P_{\tn}\,u^{(M)}_h (Q_j) 
= \frac{u^{(M)}_h ( Q_{j})}{N^{n/2}}+\Oo(h^\infty),\quad j\in(\ZZ/N\ZZ)^n\,.
\end{equation}
From the smoothness of $u^{(M)}_h$, we obtain
the ``projected version'' of \eqref{eq:lagr-L2}:
$$
\mn{\Op_h ( f_M ) \circ \cdots \circ 
\Op_h ( f_1 )\psi_h}=h^M \mn{P_{\tn}\,u^{(M)}_h}
=h^M \mn{u^{(M)}_h}_{L^2(\RR^n)}+\Oo(h^\infty) = \Oo(h^M ) \,.
$$
On the other hand, if $f(\rho_0)\neq 0$, one easily deduces from \eqref{eq:Weyl-integ} that 
$$
\mn{\Op_h(f)\psi_h}=\mn{f^w(q,hD)\,u_h}_{L^2(\IR^n)}+\Oo(h^\infty)\geq C+\Oo(h),\quad C>0\,.
$$
These estimates show that 
the family $ \psi \in I ( \Lambda ) $, and that $ \rho_0 \in \WF_h ( \psi )$.
\end{proof}

\begin{rem}The definition of $ I ( \Lambda ) $
mimicks the H\"ormander-Melrose definition of Lagrangian distributions
\cite[Definition 25.1.1]{Hor} (see \cite{Al} for an adaptation to the
standard semiclassical setting). The requirement that $ \Lambda $ is
Lagrangian reflects the uncertainty principle, in the following
sense. A Lagrangian submanifold is the lowest dimensional submanifold
for which the conclusion of Proposition 
\ref{p:new} holds, that is, for any $ \rho \in \Lambda $,
there exists a state $ \psi$ satisfying $ \psi \in I ( \Lambda ) $ and 
$\rho\in \WF_h ( \psi ) $.

Indeed, let $ \Lambda $ be
an embedded submanifold of $\tn$. Let us assume that $\psi\in I(\Lambda)$, 
so \eqref{e:def-WKB} must hold for 
any family of functions $f_j\rest_\Lambda =0$.
From the identity
$$
\frac{i}{h} [\Op_h( f_i ) , \Op_h( f_j ) ] = \Op_h ( \{ f_i , f_j \}) + \Oo( h)  \,.
$$
we see that $\mn{\Op_h ( \{ f_i , f_j \})\,\psi_h}=\Oo(h)$. As in the
proof of \eqref{eq:wfi}, we can show that 
if $\{ f_i , f_j \}(\rho)\neq 0$ for some $\rho\in\Lambda$, then 
$\rho \not\in \WF_h(\psi)$. Hence,
if we want the conclusion of Proposition \ref{p:new} to hold for $\Lambda$,
then this submanifold must satisfy
$$ 
\forall f_i,\,f_j\in C^\infty(\tn),\qquad f_i \rest_\Lambda \,,\; f_j \rest_{\Lambda } = 0 
\ \Longrightarrow 
\ \{ f_i , f_j \} \rest_\Lambda = 0 \,.
$$
This property means that $\Lambda$ is {\em co-isotropic}, and 
must be of dimension $\geq n$. 
Lagrangian manifolds are co-isotropic manifolds of minimal dimension.
\end{rem}

%%%%%%%%%%%%%%%%%%%%%%%%%%%%%%%%%%%%%%%%%%%%%%%%%%%%%%%%%%%%%%%

\subsubsection{Singular Lagrangian states}\label{s:sing-Lagr}
We now give an example where $ \Lambda $ is a union of Lagrangians
with piecewise smooth boundaries (in \S\ref{s:sympl.rela.} we called such 
$\Lambda$ a singular Lagrangian). 
Let $ \Lambda_S $ be given by 
\eqref{eq:lamb} and $ \psi_h $ by \eqref{eq:qj}. 
Let us truncate $ \psi_h $ to some proper subinterval $[Q,Q']\subset \II$, 
that is, replace
the symbol $a(q)$ by the discontinuous function
$\tilde a(q)=a(q)\bbbone_{[Q,Q']}(q)$. That gives a state 
$ \tilde \psi_h \in\Hh_h^1$. One could expect 
$\tilde \psi_h$ to be a Lagrangian state in $I(\Lambda_S)$ (as is $\psi_h$), 
or rather in $I(\tilde\Lambda_S)$, where 
$$
\tilde\Lambda_S\defeq \Lambda_S \cap ([Q , Q'] \times \II)\,.
$$
This is not
the case: one needs to include in the Lagrangian the {\em singularity set}
$$ 
\Lambda_{\rm sing} = \{ ( Q, p) \; : \; p \in \II \} 
\cup \{ ( Q' , p ) \; : \; p \in \II \}\,, 
$$
which is the ``periodized'' conormal bundle of 
the boundary $\partial \tilde\Lambda_S$. We will indeed prove that
$\tilde \psi_h \in  I(\tilde\Lambda_S\cup\Lambda_{\rm{sing}})$, which 
can be considered as a semiclassical, discrete analogue of singular Lagrangian 
distributions of Guillemin-Uhlmann \cite{GuUh} and Melrose-Uhlmann 
\cite{MeUh}. We have the following

%%%%%%%%%%%%%%%%%%%%%%%
\begin{lem}\label{l:singular-lagr}
Let us truncate the state \eqref{eq:psi_h} to
a hypercube $\cube\subset\II^n$, $ \cube = \prod_{\ell=1}^n [ \alpha_\ell, \beta_\ell] $:
\begin{equation}
  \label{eq:qj2}
  \tilde \psi_h(Q_j)=\frac{a(Q_j)\,\bbbone_\cube(Q_j)}{{N^{n/2}}}\,\exp(-2i\pi NS(Q_j))\,,
\quad j \in (\ZZ/ N\ZZ)^n \,. 
\end{equation}
Then $ \tilde \psi_h $  is associated with the singular Lagrangian 
$\tilde \Lambda_S\cup\Lambda_{\rm sing}$, where 
$\tilde \Lambda_S=\{(q,-d_q S(q))\,,\ q\in \cube\}$ and
\begin{multline}
\Lambda_{\rm sing} = \bigcup_{\ell=1}^n% \beta_\ell - \alpha_\ell \not \equiv 1 } 
\Big(\big\{ ( q, p ) \; : \; q_\ell = \alpha_\ell, \ p_\ell \in \II, \ 
q_m\in [\alpha_m, \beta_m],\ 
 p_m =-d_{q_m} S ( q ), \ m \neq \ell   \big\} \\
 \cup \, \big\{ ( q, p ) \; : \; q_\ell = \beta_\ell, \ p_\ell \in \II, \ 
q_m\in [\alpha_m, \beta_m],\ 
 p_m =-d_{q_m}S ( q ), \ m \neq \ell \big\} \Big)\,.
\end{multline}
\end{lem}
%%%%%%%%%%%%%%%%%%%%%%%
\medskip
\noindent
{\bf Remark.} It would be tempting to generalize the lemma by replacing the hypercube $\cube$ by
an arbitrary set $\cS$ with smooth boundaries. However, if 
$ n=2 $, $ S \equiv 0$, $ a \equiv 1 $, and  $ \partial \cS $ 
does not contain a segment with rational slopes then 
$$
WF_h ( \tilde\psi_h ) = (\cS\times \{0\}) \cup (\partial \cS \times \II^2) \,.
$$
The second component being $3$-dimensional, this 
set is certainly not contained in a finite union of Lagrangians.

\begin{proof}
As in the proof of Proposition ~\ref{p:new}, we can, by conjugation with the
operator \eqref{eq:tru}, reduce the proof to the case $ S = 0 $. 
We first consider states defined on $ \RR^n $, localized on the hypercube $\cube\subset \RR^n$:
\begin{equation}\label{e:flat}
u_h(q)=  \bbbone_\cube(q)\, a(q) \,, \quad a \in  \CI ( \RR^n )\,.
\end{equation}
We use the following 
%%%%%%%%%%%%%%
\begin{lem}
\label{l:wrong}
Let $\tilde \Lambda_0=\cube\times\{0\}$ and 
$\Lambda_{\rm sing}$ be as in Lemma~\ref{l:singular-lagr}. 
The ideal $\cJ$ of periodic functions vanishing on the singular Lagrangian 
$\tilde \Lambda_0 \cup \Lambda_{\rm sing}$ is (infinitely) generated by 
\begin{align*}
& g_j ( p , q )  \defeq  \sin\big(\pi(q_j-\alpha_j)\big)\,
\sin\big(\pi(q_j-\beta_j)\big)\,\sin (\pi p_j)\,, \\
& g_{ij} ( q, p )   \defeq \sin (\pi p_i) \sin ( \pi p_j )
\ \ i \neq j \,, \ 1 \leq i, j \leq n \,, \\
& \phi_j ( q, p ) = \phi ( q_j , p_1 , \cdots, p_{j - 1}, p_{j+1} , \cdots , 
p_n ) \,, \ \ \text{ \ where $ \phi ( q_j , \bullet ) \equiv 0 $, $ 
\alpha_j \leq q_j \leq \beta_j $,} \\
& \psi ( q  )  \,,  \ \ \text{\ where $ \psi  \in\CI(\II^n)$  
vanishes on $  \cube $.}
\end{align*}
\end{lem}
%%%%%%%%%%%%%%
\begin{proof}
We only give the proof for the following model ($n=2$), which contains all the basic
ingredients of the general case.
Let us study the ideal of functions vanishing on
\begin{equation}
\label{eq:cap} 
\big( \{ q_1 = p_2 = 0 \} \cup \{ q_2 = p_1 = 0 \} 
\cup \{ p_1 = p_2 = 0 \} \big) \cap \{ q_1\geq 0 \,, q_2 \geq 0 \} \,.
\end{equation}
The functions vanishing on the first factor in the intersection 
are generated by $ q_1p_1$ , $ q_2 p_2$, and $ p_1 p_2 $. Writing 
an arbitrary function $F(q,p)$ as 
\begin{multline*}
F(q_1 , q_2 , p_1 , p_2 )=F_0 ( p_1,p_2) + q_1  F_{1} ( q_1, q_2 , p_2 ) 
+ q_2  F_{2} ( q_1, q_2 , p_1 ) +\\
+ q_1 p_1 F_{11}( q_1 , q_2 , p_1 , p_2 ) + q_2 p_2 F_{22}( q_1 , q_2 , p_1 , p_2 )\,,
\end{multline*}  
we need to find conditions for $ q_1 F_{1} ( q_1, q_2 , p_2) $ and 
$ q_2 F_{2} ( q_1 , q_2 , p_1 ) $ to vanish on \eqref{eq:cap}. We treat the
first function by expanding it as
$$ 
F_{1} ( q_1,q_2 , p_2 ) = 
F_{10} ( q_1, q_2 ) + p_2 F_{12} ( q_1, p_2 ) + q_2 p_2 F_{122} ( q_1, q_2 , p_2 ) \,.
$$
This forces $ F_{10} ( q_1, q_2 ) $ to vanish identically in $ \{q_1, q_2 \geq 0\}$ 
and $ F_{12} ( q_1 , p_2 ) $ to vanish identically in $ \{q_1 \geq 0\} $. 

The function $F_2( q_1, q_2 , p_1 )$ is treated identically.
Hence the functions vanishing on \eqref{eq:cap} are generated by 
$ q_1 p_1 $, $ q_2 p_2 $, $ p_1 p_2 $, and all the smooth fuctions $\psi( q_1,q_2)$, 
$ \phi_1 ( q_1 , p_2 ) $, $ \phi_2 ( q_2 , p_1 ) $ vanishing on $\{q_1, q_2 \geq 0\}$.
The transposition to the torus setting gives the lemma for that case. The general
case can be proven similarly.
\end{proof}

This lemma
means that any $F\in \cJ$ can be decomposed as
$$ 
F=\sum_{j \neq i} f_{ij}\, g_{ij} + \sum_{ j}( f_{jj}\, g_j +  f_j\, \phi_j  + 
\psi )\,,
$$
where the functions $f_\bullet $ are smooth and
either periodic or antiperiodic in each variable, so that
$f_\bullet g_\bullet $
are periodic in all variables.

The action of each term $(f \,g )^w(q,hD)$ on the state \eqref{e:flat} can be written 
$$
(f g )^w\,u_h= \big( (f\,a)^w\circ g^w +h\,L(f,a,g) \big) \bbbone_\cube\,,
$$
where $L(f,a,g)$ is a pseudodifferential operator of norm $\Oo(1)$.
Therefore, we are reduced to study the action of the generators $g^w(q,hD)$,
$ g = g_{ij}\,, \; g_j \,, \; \phi_j, \psi $, on the characteristic
function $\bbbone_\cube(q)$. 

We first note 
that $  \psi\,\bbbone_\cube = \phi_j^w \bbbone_\cube \equiv 0 $, so
there is nothing to prove in this case.

For each $j\in\set{1,\ldots,n}$, 
the generator $g_j$
contains a factor $\sin(\pi p_j)$. Up to an error $\Oo(h)$, we first 
quantize this factor and
apply it to $\bbbone_\cube$:
$$
\sin ( \pi h D_j ) \bbbone_\cube (q)=
\frac{1}{2 i}\big(\bbbone_\cube(q_j+\pi h,q')-\bbbone_\cube(q_j-\pi h,q')\big)\defeq b_j(q)\,.
$$ 
The function $b_j(q)$ is supported in the strips 
$S_j=\set{|q_j-\alpha_j|\leq\pi h}\cup\set{|q_j-\beta_j|\leq\pi h}$,
where it takes values $\pm 1$. We now apply the remaining factors of $g_j$:
this amounts to multiplying $b_j(q)$ by the product 
$\sin\big(\pi(q_j-\alpha_j)\big)
\sin\big(\pi(q_j-\beta_j)\big)$, and gives
a function $\Oo(h)$. Taking the error into account,
we obtain $\mn{g_j^w\,\bbbone_\cube}_{L^2(\RR^{n})}=\Oo(h)$.

In the case of $g_{ij}$, $ i \neq j $,  we apply $\sin(\pi h D_{i})$ to $b_j(q)$: the 
resulting function
takes values $\pm 1$ on its support $S_i\cap S_{j}$, so that
$\mn{g_{ij}^w\,\bbbone_\cube}_{L^2(\RR^{n})}=\Oo(h)$.

We have now proved that 
$\mn{F^w\,u_h}_{L^2(\RR^{n})}=\Oo(h)$ for any $F\in\cJ$. The procedure
can be iterated to any finite product of 
functions $F_i\in\cJ$, yielding 
the estimate \eqref{eq:lagr-L2}. 

The proof is completed by 
the periodization argument as in the proof of Proposition~\ref{p:new}. The only slight 
difference lies in the fact
that the analogues of the functions $u^{(M)}_h(q)$ of \eqref{e:uhM} may now have discontinuities 
near $\partial D$,
so that
$\mn{P_{\tn}\,u^{(M)}_h}-\mn{u^{(M)}_h}_{L^2(\RR^n)}=\Oo(h)$ instead of $\Oo(h^\infty)$.
\end{proof}

%%%%%%%%%%%%%%%%%%%%%%%%%%%%%%%%%%%%%%%%%%%%%%%%%%%%%%%%%%%%%%%%%%%%%%%%%%%%%%%
%%%%%%%%%%%%%%%%%%%%%%%%%%%%%%%%%%%%%%%%%%%%%%%%%%%%%%%%%%%%%%%%%%%%%%%%%%%%%%%

\subsection{Quantum relations}
\label{s:qu-relat.}

Suppose that $ \Lambda \subset 
\tn\times\tn $ is a Lagrangian submanifold.
The basic example is given by the twisted graph $\Gamma'_\kappa$ of a symplectic 
diffeomorphism $\kappa$ on $\tn$ (see \secref{s:sympl.maps}):
$$ 
\Gamma'_\kappa=\set{( q' , q; p', -p ) \; :\;( q', p' ) = \kappa ( q ,p ),\;(q,p)\in\tn}\,. 
$$
As we noticed in that section, the choice of change of sign depends 
on the choice of the 
splitting of variables $ ( q, p ) $, which is itself related with the choice of
a {\em polarization} in the
quantization $ a \mapsto \Op_h ( a ) $ \cite[\S 25.2]{Hor}. 
This somewhat cumbersome convention is explained as follows.

Any state $v\in \Hh_h^n $ is naturally identified to a linear form
$f_v \in ( \Hh_h^n)^* $ through  $ f_v ( w ) = \la v , w \ra $. 
In our notations\footnote{This is the physicists' convention.}, 
this scalar product is antilinear in the {\em first} component. To make
the identification linear, we choose instead 
\begin{equation}\label{e:identif.}
  v\in\Hh_h^n\Longrightarrow  f_v ( \bullet ) = \langle \bar v, \bullet \rangle \,,
\end{equation} 
where states $v$ are written as vectors in the basis \eqref{eq:bas}.

Let $\cL (\Hh_h^n) \simeq \Hh_h^n \otimes (\Hh_h^{n} )^* $ 
be the space of linear operators on $ \Hh_h^n $. 
The linear identification \eqref{e:identif.} of 
$ \Hh_h^n $ with $ (\Hh_h^n)^* $ gives the
identification,
\begin{equation}\label{e:ident.}
  \cL (\Hh_h^n ) \simeq \Hh_h^{2n}\,,\quad\text{through}\quad 
  ( u \otimes v ) ( w ) = u \,\langle \bar v, w \rangle\,,\quad u,v,w\in \Hh_h^n\,.
\end{equation}
We observe that the norm on $\Hh_h^{2n}$ is the same as the Hilbert-Schmidt norm
on $ {\mathcal L} ( \Hh_h^n ) $:
\begin{equation}\label{e:HS}
  \mn{ T }_{ \Hh_h^{2n} }= 
  (\tr_{ \Hh_h^n  } ( T^* T ) )^{\frac12}\,,
\end{equation}
It is related to the operator norm on ${\mathcal L}(\Hh_h^n)$ as 
follows:
\begin{equation}\label{e:HS-op}
  \,\mn{T}_{\cL (\Hh_h^n )}\leq \mn{ T }_{ \Hh^{2n}_h }\leq N^{ n/2 }\mn{T}_{\cL (\Hh_h^n )}\,.
\end{equation}
In particular, unitary operators have Hilbert-Schmidt norm $ N^{n/2}=(2\pi h)^{-n/2} $.

The identification \eqref{e:ident.} dictates the way an operator 
of the type $A_1 \otimes A_2$ (with $A_i\in\cL(\Hh_h^n)$) acts on 
$u\otimes v\in \Hh_h^{2n}\simeq\cL(\Hh_h^n)$. Indeed, if we 
take any $w\in\Hh_h^n$, we have
\begin{align*} 
  [(A_1\otimes A_2) ( u \otimes v )] ( w ) 
  &=[A_1 u\otimes A_2 v] (w)\\
  &=A_1 u\;\la \overline{A_2 v}, w\ra\\
  &=A_1 u\;\la \bar v, A_2' w  \ra\\
  &= [(A_1 u \otimes v )\circ A_2'](w)\,.
\end{align*} 
Here $A_2'$ is the transposed of the operator $A_2$, written as a matrix in the
basis \eqref{eq:bas}. In the case $A_1=\Op_h(a)$, $A_2=\Op_h(b)$ for some real
functions $a,b\in \CI (\tn)$, one checks that $A_2'=\Op_h(\tilde b)$,
with the same twisted function as in the proof of Lemma~\ref{l:wfo}:
$ \tilde b ( q , p ) = b ( q , - p ) $. 
By linearity, for any $C_h\in \Hh_h^{2n}\simeq \cL(\Hh_h^n)$, we have
\begin{equation}\label{e:action}
  \Op_h ( a \otimes b )\, C_h = \Op_h ( a) \circ C_h \circ \Op_h ( \tilde b )\,.
\end{equation}
The sign change in the tilting $\Gamma\leadsto\Gamma'$ parallels 
the transformation $a(\rho')\,b(\rho)\leadsto a(\rho')\,\tilde b(\rho)$.

\medskip

We are now in position to quantize a symplectic map, more generally a symplectic
relation $\Gamma$ as defined in \secref{s:sympl.rela.}.

%%%%%%%%%%%%%%%%%%%%%%%%%%%%%
\begin{defn}\label{d:q.relation}
  A semiclassical sequence $U=\set{U_h \in \Hh_h^{2n} }_{h\to 0}$ satisfying 
\begin{equation}
\label{eq:temp} 
 \mn{U_h}_{ \Hh_h^{2n} } =\Oo( h^{K} ) \,, \quad \text{for some fixed $ K\in\RR $, }
\end{equation}
  is a quantum relation associated with the symplectic relation $\Gamma$
  if $U$ is a Lagrangian state in $I(\Gamma')$, in
  the sense of Definition~\ref{d:Lagrangian}.

  Explicitly, for any $M\in\NN$ and any sequence of functions
  $$
  g_j \in \CI ( \tn\times\tn) \,, \quad g_j \rest_{\Gamma'} = 0  \,, \quad 1 \leq j \leq M \,,
  $$
  we must have
  \begin{equation}
\label{eq:Uhit}
    \mn{\Op_{h} (g_M)  \circ \cdots \circ \Op_{h} ( g_1)\, U_h}_{
      \Hh^{2n}_h } = \Oo ( h^{M} )\, \mn{U_h}_{ \Hh_h^{2n} } \,.
  \end{equation}
\end{defn}
%%%%%%%%%%%%%%%%%%%%%%%%%%%%%
The assumption that $ U_h $ is tempered in the sense of \eqref{eq:temp}
(which also implies temperedness in the operator norm)
is necessary to assure that composing $ U_h $ with residual ($\Oo(h^\infty)$) terms
produces residual terms. That is a 
standard assumption in $ \CI $ semiclassical calculi --- see \cite{Al,SjZw02}, and 
will be used in the proof of Prop.\ref{p:cumb}. 
The quantum \emph{weighted} relations defined in \S~\ref{s:qwr} will naturally
be tempered, having norms $\mn{U_h}_{ \Hh_h^{2n} }=\Oo(h^{-n/2})$. 

If a function $g\in\CI(\TT^{4n})$ vanishes on $\Gamma'$, then the function 
$\tilde g$ defined as $\tilde g(q',p';q,p)=g(q',p';q,-p)$ vanishes on $\Gamma$.
The condition $g_j\rest_{\Gamma'}=0$ can thus be written
$\tilde g_j\rest_{\Gamma}=0$.

We also note that \eqref{eq:Uhit} entails a version of {\em Egorov's theorem}.
If $f_L,\ f_R\in\CI(\tn)$ satisfy
$$ 
( \rho', \rho ) \in \Gamma \ \Longrightarrow \ 
f_L ( \rho' ) = f_R ( \rho )  \,,
$$
then we have
\begin{equation}\label{e:egorov}
\mn{\Op_h ( f_L ) U_h - U_h \Op_h (  f_R )}_{\Hh_h^{2n}} = \Oo ( h )\,\mn{ U_h }_{\Hh_h^{2n} }\,.
\end{equation}
Indeed, the function $ f \defeq f_L \otimes 1 - 1 \otimes f_R $ vanishes on $ \Gamma $, so that
$\tilde f$ vanishes on $\Gamma'$. We then simply apply the 
definition \eqref{eq:Uhit} with $g_1=\tilde f$ and use \eqref{e:action}. 
When $ \Gamma $ is a graph of a symplectic transformation, $ f_R $ is the pullback 
of $ f_L $, and we get a statement similar with the standard
Egorov's theorem. 
%%%%%%%%%%%%%%%%%%%%%%%%%%%%%
\begin{rem}\label{r:singular}
  Following \secref{s:lagrangian}, 
  in the case when $\Gamma'$ is a Lagrangian with boundaries projecting on a hypercube, it is useful
  to include in the definition sequences $U=\set{U_h}$ in the (larger) space 
  $I(\Gamma'\cup\Lambda_{\rm sing})$;
  the quantum baker's relation we define in next section will belong to such an
  enlarged space.
\end{rem}
%%%%%%%%%%%%%%%%%%%%%%%%%%%%%

Through the identification \eqref{e:ident.},  $U_h$ is
an operator on $\Hh_h^n$. 
We now show that this operator
``classically transports'' the microsupport of a sequence $w=\set{w_h\in  \Hh^n_h}$.
%%%%%%%%%%%%%%%%%%%%%%
\begin{prop}
\label{p:cumb}
Take $ U=\set{U_h \in \Hh_h^{2n}\simeq \cL (\Hh_h^n )}$ a
quantum relation $U \in I (\Gamma' )$. 
Then for any sequence $ w=\set{w_h\in \Hh^n_h}$, 
$ \mn{w_h} \asymp 1 $, the microsupport of the image
sequence $U(w)=\set{U_h(w_h)}$ satisfies:
$$  
\WF_h ( U ( w ) ) \subset \Gamma\big(\WF_h ( w)\big)=
\set{ \rho' \in \tn \;:\; \exists \;\rho \in \WF_h ( w) \,, 
\quad ( \rho' , \rho ) \in \Gamma  }\,.
$$
\end{prop}
%%%%%%%%%%%%%%%%%%%%%%
\begin{proof}
Assume that $\rho'_0\not\in \Gamma(WF_h ( w))$, which means that
$\Gamma^{-1}(\rho'_0)\cap \WF_h ( w)=\emptyset$. Then there exists a function
$f\in\CI(\tn)$ with $f\equiv 1$ near $\rho'_0$ but with $\supp(f)$ 
sufficiently small so that 
$\Gamma^{-1}(\supp(f))\Subset \complement \WF_h(w)$. Consequently, there
exists a function $g\in\CI(\tn)$ with $g\equiv 1$ near $\WF_h(w)$ but $g\equiv 0$ on
$\Gamma^{-1}(\supp(f))$. The function $f\otimes \tilde g\in\CI(\TT^{4n})$ then
automatically vanishes on $\Gamma'$. 

Our aim is to show that $\rho'_0\not\in \WF_h(U(w))$. For this, we
introduce one further function $a\in\CI(\tn)$ 
such that $a(\rho'_0)>0$ and
$f \equiv 1$ on $\supp(a)$. As in the proof of \eqref{eq:wfi} we see that
for any $M\in\NN$, $ \Op_h ( a ) \Op_h ( f ) ^M = \Op_h ( a ) + \Oo ( h ^\infty ) $. 
Hence 
\begin{align*}
    \mn{\Op_h(a)U_h w_h}&=\mn{\Op_h ( a ) \Op_h ( f )^M\,U_h w_h}+\Oo ( h ^\infty )\\
    &\leq \mn{\Op_h ( a ) \Op_h ( f )^M\,U_h\,\Op_h ( g )^M\,w_h}\\
    &\qquad+\mn{\Op_h ( a ) \Op_h ( f )^M\,U_h\,(1-\Op_h ( g )^M)\,w_h}+\Oo ( h ^\infty )
\end{align*}
To bound the second term on the right hand side, we notice that
the function $(1-g^M)$ vanishes near $\WF_h(w)$, so from Lemma~\ref{l:wfs} we get
$\mn{(1-\Op_h ( g )^M)\,w_h}=\Oo ( h ^\infty )$; from the temperedness of $U_h$, 
the second term is thus residual. 

The first term on the right hand side is estimated using the identity
$$
\Op_h ( f )^M\,U_h\,\Op_h ( g )^M=\Op_h(f\otimes\tilde g)^M\,U_h\,.
$$
Because $f\otimes \tilde g$ vanishes on $\Gamma'$, the Hilbert-Schmidt norm of
that operator is $\Oo(h^{M+K})$, where $K$ comes 
from the temperedness of $U_h$, \eqref{eq:temp}.
Using \eqref{e:HS-op}, we thus get
$
\mn{\Op_h ( a ) U_h\,w_h}=\Oo(h^{M+K})
$
for an arbitrary $M\in \NN$, which shows that $\rho'_0\not\in \WF_h(U(w))$.
\end{proof}

%%%%%%%%%%%%%%%%%%%%%%%%%%%%%%%%%%%%%%%%
%%%%%%%%%%%%%%%%%%%%%%%%%%%%%%%%%%%%%%%%
\subsection{Quantized weighted relations}
\label{s:qwr}
%%%%%%%%%%%%%%%%%%%%%%%%%%%%%%%%%%%%%%%%
In \secref{s:wr} we equipped a symplectic relation $\Gamma$ with a measure, or weight $\mu$.
In order to associate to
the weighted relation $(\Gamma,\mu)$ a sequence of operators $U_h\in\Hh_h^{2n}$, we
need to elaborate on Definition~\ref{d:q.relation}, thereby defining a subfamily 
$I(\Gamma',\mu)\subsetneq I(\Gamma')$.

In the standard microlocal
context \cite[Section 25.1]{Hor}, a Lagrangian state $\psi\in I(\Lambda)$
has a well defined amplitude, or \emph{symbol}, which 
is a section of the Maslov
half density bundle over the Lagrangian submanifold --- see 
\cite[Theorem 25.1.9]{Hor}. The local 
aspects of this procedure have recently been adapted to 
the semiclassical case \cite{Al}, and a similar approach can be 
used in the case of $ \TT^{4n} $. 

Although one could characterize the operators quantizing $(\Gamma,\mu)$
in terms of their symbols (grossly speaking, the absolute square of the symbol
should equal the weight $\mu$), we won't do it here, in order to avoid technical 
issues involved in the description of the symbol map. Instead, in the definition below
we use bilinear expressions in $U_h$, which allows us to avoid introducing symbols.

%%%%%%%%%%%%%%%%%%%%%%%%%%%%%%%%%%
\begin{defn}\label{d:q.wr}
Let $ (\Gamma, \mu ) $ be a weighted piecewise smooth 
relation as defined in \S \ref{s:wr} and let
 $ U \in I (\Gamma' \cup \Lambda_{\rm sing} )$, in the sense of
Definition~\ref{d:q.relation} and Remark~\ref{r:singular}.
For any  $ \chi_\alpha \in \CI (\tn;[ 0,1]) $, 
$ \alpha = L,\,R$, we define 
$$ 
U_{ \chi_L \chi_R } \defeq \Op_h ( \chi_L ) \; U_h \; \Op_h ( \chi_R ) \,.
$$
We say that $ U $ {\em quantizes the weighted relation} $ ( \Gamma , \mu ) $ 
if for all $ \chi_L $, $\chi_R$  with sufficiently small supports 
satisfying  $ \supp ( \chi_L \otimes \chi_R ) \cap 
\Lambda_{\rm sing}' = \emptyset $,
\begin{equation}\label{eq:ucc}
  \begin{split} 
U_{\chi_L \chi_R } U_{\chi_L \chi_R }^* & = \Op_h ( g_L^{ \chi_L \chi_R }) + 
\Oo ( h )  \\
U_{\chi_L \chi_R }^* U_{\chi_L \chi_R } & = \Op_h ( g_R^{ \chi_L \chi_R }) + 
\Oo ( h )  \,, 
\end{split}
\end{equation}
where $ g_\alpha^{\chi_L \chi_R } $ are the functions given in \eqref{eq:meas}, and
the remainder is $\Oo(h)$ in the operator norm on $\cL(\Hh_h^n)$.
We then write
$$  
U=\set{U_h} \in  I (\Gamma' \cup \Lambda_{\rm sing} , \mu ) \,.
$$
\end{defn}
%%%%%%%%%%%%%%%%%%%%%%%%%%%%%%%%%%

The conditions on the smallness of supports of $ \chi_\alpha $ guarantee
that the operators appearing on the left in \eqref{eq:ucc} are 
of the form $ \Op_h ( f ) $, $ f \in \CI ( \tn )$. 
That follows from the fact that 
$\Gamma$ is locally a graph --- see \S \ref{s:wr}.

If $\Gamma$ is the graph of a symplectic diffeomorphism $\kappa$
and $ \mu = \pi_L^* (\omega^n/n!) = \pi_R^* (\omega^n/n!) $, then 
$ U_h $ is unitary to leading order:
$$ 
U_h^* U_h = I + C_h \,, \quad  U_h U_h^* = I + D_h \,, 
\qquad 
\mn{C_h}_{\cL (\Hh_h^n )}=\Oo(h),\ \mn{D_h}_{\cL (\Hh_h^n )}=\Oo(h)\,.
$$
For $ h $ small, $ ( I + C_h)^{-\frac12} $,  $ ( I + D_h)^{-\frac12} $
exist, therefore a possibility to make the quantization strictly unitary
is to replace $U_h$ by 
$ U_h ( I + C_h)^{-\frac12} $ or 
$ ( I + D_h )^{-\frac12} U_h $.
%(as in the methods developed in \cite{Prosen96,Zel97}
%to quantize symplectic diffeomorphisms into unitary operators).

The condition \eqref{eq:ucc} can be interpreted as follows.
Suppose that $\psi\in \Hh_h^{n}$, $\mn{\psi}=1$,
is microlocalized at a single ``regular'' point $\rho_0$:
$$ 
\WF_h ( \psi) = \{ \rho_0 \} \subset \tn
\setminus \pi_R(\Lambda'_{\rm sing}) \,, 
$$
and $\Gamma(\rho_0)=\cup_{j=1}^J\rho'_j$, $\rho'_j=\kappa_j(\rho_0)$.
Then, 
\begin{gather*} 
 U_h \psi = \sum_{j=1}^J \psi_j +\Oo(h^\infty)\,, \\
\mn{\psi_j}^2  = P_j(\rho_0)+\Oo(h) \,,\qquad
\WF_h ( \psi_j ) \subset \set{\rho'_j}\,.
\end{gather*}
From Lemma~\ref{l:wfs}, if  
$P_j(\rho_0)\neq 0$ then $\WF_h ( \psi_j ) = \set{\rho'_j}$.
A similar statement holds for $ U_h^*$.

Indeed, if for each $ j = 0 ,\cdots, J $ we take $\chi_j\in\CI(\tn;[0,1])$ supported in 
a small neighbourhood of $\rho_0$, resp. $\rho'_j$, and equal to $1$ near 
that point, \eqref{e:probas} shows that $g_R^{ \chi_j \chi_0 }(\rho_0)=P_j(\rho_0)$
for $j=1,\ldots,J$. On the other
hand, Proposition~\ref{p:cumb} gives
\begin{gather}
\label{eq:uhg}
\begin{gathered}
 U_h \psi = U_h \Op_h (\chi_0) \psi + \Oo ( h^\infty ) = 
\sum_{ j=1}^J U_{\chi_j\chi_0}\, \psi + 
\Oo ( h^\infty ) \,, \\ 
\WF_h ( U_{\chi_j\chi_0}  \psi ) \subset \set{\rho'_j} \,.
\end{gathered} \end{gather}
If we take $ \psi_j  \defeq  U_{\chi_j \chi_0}\psi$ then 
$$ 
 \mn{\psi_j}^2 = \langle 
U_{\chi_j \chi_0 }^* U_{\chi_j \chi_0 } \psi , \psi \rangle 
= \langle \Op_h ( g_R^{ \chi_j \chi_0} ) \psi , \psi \rangle  
+ \Oo ( h )= P_j(\rho_0)+\Oo(h)\,.
$$

\medskip

\noindent {\em Example.} We now consider a special case of quantum relations 
$ U_h $, of the form
\begin{equation}
\label{eq:uqq}
 \langle Q_j | U_h | Q_k \rangle = N^{-n/2} a ( Q_j , Q_k )\; \exp \big(  2 \pi i
 N S ( Q_j , Q_k ) \big) \,,
\end{equation}
where $a,\ S\in\CI(\tn\times \tn)$ and the generating function $S ( q' , q) $ satisfies 
the nondegeneracy condition 
$\det (\partial^2_{q' \, q} S) \neq 0 $ near the support of $ a ( q' , q )$.
Using Definition~\ref{d:q.relation} we see that $ U_h $ is 
associated to the graph $ \Gamma_S $ of the  symplectic transformation 
$ ( q ,- \partial_q S ) \mapsto ( q' , \partial_{q'}S) $. To be more precise,
\begin{equation}
\label{eq:uhs}
 U_h \in I ( \Gamma_S', \mu_S ) \,, \qquad \text{for}\quad \mu_S 
\defeq | a ( q', q) |^2\; dq'\, dq \,,
\end{equation}
where we used the coordinates $ ( q', q ) $ on $ \Gamma_S $.
Projecting this measure on the left and right tori, we get:  
\begin{equation}
\begin{split}
 \pi_{L *}\, \mu_S & = \Big(\sum_{q':\,p = -\partial_q S ( q', q )}
| a (q',q) |^2\;|\det (\partial^2_{q'\,q} S)|^{-1}\Big)\; dq\, dp \,, \\
 \pi_{R *}\, \mu_S & =  \Big(\sum_{q\,:\,p' = \partial_{q'} S ( q', q )}
| a (q',q) |^2\;|\det (\partial^2_{q'\,q} S)|^{-1}\Big)\;  dq'\, dp' \,.
\end{split} 
\end{equation}
The above sums are always finite. This example will be used to analyze the quantum
baker's relations studied in the next sections.

%%%%%%%%%%%%%%%%%%%%%%%%%%%%%%%%%%%%%%%%%%%%%%%%%%%%%%%%%%%%%%%%%%%%%%%%%%%%%%%
%%%%%%%%%%%%%%%%%%%%%%%%%%%%%%%%%%%%%%%%%%%%%%%%%%%%%%%%%%%%%%%%%%%%%%%%%%%%%%%

\subsection{Quantized baker's relation}
\label{s:qbr}

%\subsubsection{Definition}

We  explicitly construct quantum relations $B_h\in\cL(\Hh_h^1)$ 
associated with the ``open baker's maps''
described in \secref{e:open-baker-clas}.
For simplicity, we will assume that the coefficients $D_j$ and $\ell_j$ are integers.
Besides, we will only consider the subsequence of Planck's constants
of the form $h=(2\pi N)^{-1}$ such that 
${N}/{D_1}=M_1\in\NN$ and ${N}/{D_2}=M_2\in\NN$ (that is,
$N$ is a multiple of ${\rm lcm}(D_1,D_2)$). 

Restricting ourselves to this subsequence, we define the quantization
of the baker's relation~\eqref{e:classical-baker} as the following operators
(written as $N\times N$ matrices in the bases \eqref{eq:bas}): 
\begin{equation}\label{e:quantum-baker}
  B_h\defeq \F_{N}^* \circ
  \begin{pmatrix}
    0 &   0 & 0 & 0  & 0 \\
    0 & \F_{M_1} & 0 & 0 & 0  \\
    0 &  0 & 0 & \F_{M_2} & 0 \\
    0 & 0 & 0 & 0 & 0  \end{pmatrix} =B_{1,h}+ B_{2,h}.
\end{equation}
The numbers of columns in successive blocks are respectively given by 
$$
\ell_1 M_1  \,, \   M_1 \,, 
\ \ell_2  M_2 - (\ell_1 + 1 )M_1 \,, \ 
M_2 \,, \ ( D_2 - \ell_2 - 1 ) M_2 \,,
$$
and $ \F_{M} $ is the discrete Fourier transform
given in \eqref{e:DFT}. These matrices obviously generalize
the unitary matrices associated with the closed baker's map \cite{BaVo}.

We now check that the matrices \eqref{e:quantum-baker} satisfy the 
Definition~\ref{d:q.wr} if we select the appropriate Lagrangian surface on $\TT^4$,
namely by adjoining a singularity set $\Lambda_{\rm sing}$ to
the twisted graph $B'$ (see Remark~\ref{r:singular}), and equip $B$ with the weight
$\mu$ described in \eqref{eq:bmes}.
By linearity, we can separately consider the two blocks $B_{j,h}$. 
Let us study the left block $B_{1,h}$. 
Since the classical relation $B_1$ is generated by the function $S_1(q,p')$ of
\eqref{e:generating}, it
is natural to express the operator $B_{1,h}$ in the mixed representation $(p',q)$, that is
by a matrix from the basis $\set{|Q_j\ra}$ to the basis $\set{|P_k\ra}$. 
Since the change of basis matrix, $ ( |P_k \ra  \la Q_j | )_{j,k=0,\ldots,N-1} $,
equals  $ \F_N $, the operator $A_{1,h}$ defined as the matrix
$$ 
(\la Q_k|A_{1,h}|Q_j\ra)_{j,k=0,\ldots,N-1} \defeq 
 (\la P_k|B_{1,h}|Q_j\ra)_{j,k=0,\ldots,N-1} = \F_N \circ B_{1,h} 
$$ 
is given by the Fourier block $\F_{M_1}$ at the
same position as in \eqref{e:quantum-baker}, and zeros everywhere else.

The following lemma reduces finding the (weighted) Lagrangian relation associated to 
$B_{1,h}$ to finding the (weighted) Lagrangian associated to $A_{1,h}$. We denote by
$F$ the following transformation of $\tn$: $F(q,p)=(p,-q)$. It means, we rotate
by $-\pi/2$ around the origin in each plane $(q_i,p_i)$. 
We denote by $F_L$ the transformation of $\TT^{4n}$
acting through $F$ on the left coordinates $(q',p')$ and leaving the right coordinates
unchanged.
%%%%%%%%%%%%%%%%%%%
\begin{lem}
\label{l:find}
Suppose that $ U_h \in {\mathcal L} ( \Hh_h^n ) \simeq \Hh_h^{2n}$ and that
$ V_h \defeq\F_N \circ U_h$.
Then, for any (possibly singular) Lagrangian $\cC'\in \TT^{4n}$, 
$$ 
U_h \in I ( \cC')  
\; \Longleftrightarrow V_h \in I ( \cD ')\,, 
$$
where 
$$  
\cD' =F_L(\cC')%= \{ ( p' , q ; -q' , p) \, : \, ( q' , q ; p' , p ) \in  \cC' \} \,.
\,,\quad\text{equivalently}\quad
\cD =F_L(\cC)= \{ (p',-q';q,p) \, : \, (q',p';q,p)\in\cC \}\,.
$$
Furthermore, 
$$
U_h \in I ( \cC' ,\mu)  
\; \Longleftrightarrow V_h \in I ( \cD ',\nu )\,,\quad\text{with}\quad \nu=F_{L *}\,\mu\,.
$$
\end{lem}
%%%%%%%%%%%%%%%%%%%
\begin{proof}
The transformation $\cC\to\cD$ results from a general composition formula 
which can be proved
by mimicking the semiclassical proof in \cite{Al}. Here it follows
from the the covariance properties of
Weyl quantization with respect to the Fourier transform: for any $a\in\CI(\tn)$,
\begin{equation}\label{eq:Fourier-cov}
\F_h^{-1} \Op_h (a ) \circ \F_h = \Op_h ( a\circ F)\,.
\end{equation}
As a result, for any $ f \in \CI (\TT^{4n} )$,
$$
\Op_h ( f ) (\F_h \circ U_h )  = \F_h \circ \Op_h (f\circ F_L ) ( U_h ) \,. 
$$
This identity proves the first assertion. 

Using \eqref{eq:Fourier-cov}, we notice that for any
$\chi_L,\ \chi_R\in\CI(\tn;[0,1])$, the cutoff propagator
$V_{\chi_L\,\chi_R}$ satisfies
\begin{align*}
V_{\chi_L\,\chi_R}^*\,V_{\chi_L\,\chi_R}&=
U_{\chi_L\circ F\,\chi_R}^*\,U_{\chi_L\circ F\,\chi_R}=
\Op_h( g_R^{ \chi_L\circ F\, \chi_R }) + 
\Oo ( h )\,,\\ 
V_{\chi_L\,\chi_R}\,V_{\chi_L\,\chi_R}^*&=\F_h \,U_{\chi_L\circ F\,\chi_R}\,
U_{\chi_L\circ F\,\chi_R}^*\,\F_h^*=\Op_h( g_L^{ \chi_L\circ F\, \chi_R }\circ F^{-1})
+\Oo(h)\,. 
\end{align*}
Using the pushforward of functions $F_{L *}\,f=f\circ F_L^{-1}$ and the fact that
$\pi_R\circ F_L=\pi_R$, we get
\begin{align*}
g_R^{ \chi_L\circ F\, \chi_R} &= 
\pi_{R*} (\pi_L^*(F_{L *}^{-1}\chi_L) \;\pi_R^*\chi_R \; \mu  ) = 
%\pi_{R *}\,F_{L *}^{-1}\, ( \pi_L^*\chi_L\;\pi_R^* \chi_R \; F_{L *}\mu)=
\pi_{R *} ( \pi_L^*\chi_L\;\pi_R^* \chi_R \; F_{L *}\mu)\\
g_L^{ \chi_L\circ F\, \chi_R }\circ F^{-1}&= 
\pi_{L *}F_{L *} ( \pi_L^*(F_{L *}^{-1}\chi_L)\;\pi_R^*\chi_R \; \mu )=
\pi_{L *}(\pi_L^*\chi_L\;\pi_R^*\chi_R \;F_{L *}\mu )\,.
\end{align*}
This proves that $V_h$ is associated with the weight $\nu=F_{L *}\,\mu$ on $\cD'$.
\end{proof}
Let us now describe the weighted Lagrangian associated with the operator $A_{1,h}$.
The kernel of that operator vanishes outside the square $\cube=I_1\times I_1$, where
$I_1=\left[{\ell_1}/{D_1}, {\ell_1+1}/{D_1}\right]$,
and on $\cube$ it takes the values
\bequ\label{eq:A_1}
\la Q_k|A_{1,h}|Q_j\ra=\la P_k|B_{1,h}|Q_j\ra 
=\sqrt{\frac{D_1}{N}}\;\bbbone_\cube(Q_k,Q_j)\,
\exp\big( -2i\pi N\,S_1(Q_k,Q_j)\big)\,.
\end{equation}
The operator $A_{1,h}$
has the same form as in \eqref{eq:uqq}, with the (obviously nondegenerate) generating function $S=-S_1$
and symbol $a(q',q)=\sqrt{D_1}\, \bbbone_{\cube}(q',q)$. If we  
forget (for a moment) the discontinuities of the symbol, we find that
$A_{1,h}$ is associated with the graph 
$$
\Gamma_{S_1}=\set{\big(q',-(D_1q - \ell_1);\,q,(D_1 q'-\ell_1)\big)\,:\,q,q'\in I_1}\,,
$$
equipped with the weight
$$
\mu_{S_1}= D_1\, \bbbone_\cube(q',q)\;dq'\,dq\,.
$$
From Lemma~\ref{l:find}, the operator $B_{1,h}=\F_N^*\circ A_{1,h}$ 
is associated with the graph 
$$
F_L^{-1}(\Gamma_{S_1})=
\set{\big((D_1q - \ell_1),q';\,q,(D_1 q'-\ell_1)\big)\,:\,q,q'\in I_1}=B_{1}
$$ 
and the weight
$$
\mu_1\defi F^{-1}_{L *}\mu_{S_1}=D_1\, \bbbone_\cube(p',q)\;dp'\,dq\,,
$$
which can be expressed as
$$
\pi_{R *}\, \mu_1 =\bbbone_{I_1}(q)\;dq\,\,dp\,,\quad 
\pi_{L *}\, \mu_1 =\bbbone_{I_1}(p')\;dq'\,dp'\,.
$$
It represents the half part of the weight \eqref{eq:bmes}.

Let us now take the discontinuities of $a(q',q)$ into account. Since they occur
at the boundary of the square $\cube$, they 
have the same consequences as in
Lemma~\ref{l:singular-lagr}. Namely, we must add to the Lagrangian $\Gamma_{S_1}'$ 
a ``singular'' Lagrangian, which is the union of $4$ pieces,
each piece sitting above a side of $\cube$. This Lagrangian 
should then be rotated through $F_L^{-1}$ as well.

For instance, the side
$\{q'\in I_1,\,q={\ell_1}/{D_1}\}$ leads (after rotation) 
to the singular Lagrangian
%\[ \begin{split}
%\cD_{\rm sing,1} & =\set{ \big(q',\,q=\frac{\ell_1}{D_1}\,;
%\,p'=-\partial_{q'}S_1(q',\ell_1/D_1),\,p\big)\ :\ q'\in I_1,\, p\in\II} \\ & =
%\set{\big(q',\,\frac{\ell_1}{D_1}\,;
%\,0,\,p\big)\ :\ q'\in I_1,\,p\in\II}\,.\end{split} \]
$$
\Lambda_{{\rm sing},1}
=\set{\big(q'=0,\,q=\frac{\ell_1}{D_1};\,p'\,,p\big)\ :\ p'\in I_1\,,p\in\II}\,,
$$
which contains the corresponding side of $\partial B'_1$. 
Similar Lagrangians $\Lambda_{{\rm sing},i}$, $i=2,3,4$, 
contain the other sides of $\partial B_1'$.  

The same analysis applies to $ B_{2,h} $ and hence we have proved the
%%%%%%%%%%%%%%%%%%
\begin{prop}
\label{p:qbr}
The sequence of matrices $ \set{B_h} $ given in \eqref{e:quantum-baker} quantizes
the classical baker's relation $ B = B_1 \cup B_2 $ of  
\eqref{e:classical-baker}, in the sense of Definitions~\ref{d:q.relation},
\ref{d:q.wr},
and Remark~\ref{r:singular}:
$$ 
B_h \in I\left( B' \cup \bigcup_{j=1}^8 \Lambda_{\rm{sing},j} , \mu \right) \,,
$$
where the weight $\mu$ is given by \eqref{eq:bmes}.
\end{prop}
%%%%%%%%%%%%%%%%%%
This quantization of the baker's relation is very close to 
the ``quantum horseshoe'' defined by Saraceno-Vallejos in \cite{SaVa}. 
The operator $B_h$ is contracting, and
its eigenstates can be seen as ``metastable states'', ``decaying states'' or
``resonances''. This contraction mirrors the decay of a classical probability density
evolved through the open map $B$ (due to the ``escape'' of particles to infinity).
This classical decay can be analyzed in terms of a ``conditionally invariant measure''
on $\t2$ \cite{chernov1}, which decays according to the classical decay rate 
$\gamma_{\rm cl}=-\log(D_1^{-1}+D_2^{-1})$.

%%%%%%%%%%%%%%%%%%%%%%%%%%%%%%%%%%%%%%%%%%%%%%%%%%%%%%%%%%%%%%%%%%%%%%%%%%%%%%%
\subsection{Numerical check of the Weyl law for the baker's relation}
\label{nr}
%%%%%%%%%%%%%%%%%%%%%%%%%%%%%%%%%%%%%%%%%%%%%%%%%%%%%%%%%%%%%%%%%%%%%%%%%%%%%%%

We have numerically computed the spectra of the quantum baker relations for various 
symmetric and nonsymmetric baker's relations. Results for the symmetric
``3-baker'' ($D=3,\ell=0$) were presented in \cite{nQM9} (see also Table~\ref{table2}), 
some for the ``5-baker'' ($D=5,\ell=1$) were given in \cite{nzJPA}, while
a nonsymmetric map $(D_1=32,D_2=3/2)$ was studied in \cite{NonRub06}.
In the symmetric cases, the trapped set is a pure Cantor set of dimension
$2d=2 \frac{\log 2}{\log D}$, so that for any $1>r>0$, the number of 
resonances in the annulus $\set{|\lambda|>r}$ is expected to scale as
$$
\# \{\lambda\in \Spec(B_{h})\,:\, |\lambda|\geq r\}\sim C(r)\,N^{\frac{\log 2}{\log D}}
$$
in the limit $N\to\infty$. Our numerics for both maps shows that this scaling is 
roughly satisfied along any sequence $N\to\infty$; much better fits
are obtained for $N$ taken along geometric
sequences of the type $N=N_o\,D^k$, with $N_o$ fixed and $k\to\infty$ 
(as in Table~\ref{table2}), which lead us to the following weaker conjecture for
the symmetric maps:
$$
\# \{\lambda\in \Spec(B_{h})\,:\, |\lambda|\geq r\}\sim C(N_o,r)\,N^{\frac{\log 2}{\log D}}\,,
\qquad (2\pi h)^{-1}=N=N_o\,D^k\,,\quad k\to\infty\,.
$$
Here, the ``profile function'' $C(N_o,r)$ may (slightly) depend 
on the ``root'' of the geometric
sequence. The special role played by geometric 
sequences is probably due to the strong relationship between the symmetric $D$-baker
and the $D$-nary decomposition.

On the opposite, for the nonsymmetric map
the fractal Weyl law seems accurate for an ``arbitrary'' sequence $N\to\infty$ \cite{NonRub06}, 
which was also the case for the nonlinear map studied by
\cite{schomerus}.

%%%%%%%%%%%%%%%%%%%%%%%%%%%%%%%%%%%%%%%%%%%%%%%%%%%%%%%%%%%%%%%%%%%%%%%%%%%%%%%
%%%%%%%%%%%%%%%%%%%%%%%%%%%%%%%%%%%%%%%%%%%%%%%%%%%%%%%%%%%%%%%%%%%%%%%%%%%%%%%

%%%%%%%%%%%%%%%%%%%%%%%%%%%%%%%%%%%%%%%%%%%%%%%%%%%%%%%%%%%%%%%%%%%%%%%%%%%%%%%
%%%%%%%%%%%%%%%%%%%%%%%%%%%%%%%%%%%%%%%%%%%%%%%%%%%%%%%%%%%%%%%%%%%%%%%%%%%%%%%

\section{A toy model}
\label{s:tm}

Let us explicitly compute the matrix elements of the
two vertical blocks $B_{1,h}$, $B_{2,h}$ in \eqref{e:quantum-baker}, for the symmetric
3-baker. Both are matrices $N\times N/3$, which we index by
$0 \leq k \leq N -1$, $0 \leq l \leq N/3-1$: 
\begin{equation}  
\begin{split}\label{e:quantum3-b}
    ( B_{1,h} )_{ k\,l}&=\begin{cases}
    \sqrt{3}(1-e^{2 i \pi \frac{k-3l}{N}})^{-1} (1-\omega_3^k)/ N 
    & \text{if $k\neq 3l$},\\
    1/\sqrt{3}&\text{if $k= 3l$,}
    \end{cases} \,, \\
    ( B_{2,h} )_{ k\,l}&=\omega_3^{2k}\,( B_{1,h} )_{ k\,l}\,, 
 \qquad \text{where}\ \omega_3 = e^{2 i \pi/3} \,.
  \end{split}
\end{equation}

%On Fig.~\ref{fig:dens} (left) we represent the moduli 
%of these matrix elements.
The largest matrix elements are near
the ``tilted diagonals'' $k\approx 3l$, and decay as $1/{|k-3l|}$ away from them
(see the Figure~6 in \cite{nzJPA}).
%%%%%%%%%%%%%%%%%%%%%%%%%%%%%%
%\begin{figure}[htbp]
%    \centerline{\rotatebox{-90}{\includegraphics[height=14cm]{matrices-3-2.ps}}}
    %\includegraphics[height=6cm]{cont2.eps}
%    \caption{Matrices  $B_{27}$ (left) and 
%      its toy model $\tB_{27}$ (right) for the quantum 3-baker. 
%      The gray scale corresponds to the
%      moduli of the matrix elements, (white $=0$, black $=1$).}
%    \label{fig:dens}
%  \end{figure}
%%%%%%%%%%%%%%%%%%%%%%%%%%%%%%
Being unable to rigorously analyze the spectrum of $B_{h}$, 
we replace this matrix by the following 
  simplified model:
  \begin{gather}
    \label{eq:ttoy}
    \begin{gathered}
      \tBh =\tB_{N} = [\tB_{1,h}, 
	0,\tB_{2,h} ] \,, \\
      (\tB_{1,h})_{k\,l} =  
      \begin{cases}  
	1/\sqrt{3}& \text{if}\ l = \lfloor k/3\rfloor \\
	0 & \text{if}\  l \neq \lfloor k/3\rfloor \end{cases} \,,\qquad  
      (\tB_{2,h})_{k\,l}=\omega_3^{2k}\,(\tB_{1,h})_{k\,l}\,,
    \end{gathered}
  \end{gather}
where $\lfloor x\rfloor$ denotes the integer part of $x$. For $N=9$, this gives
  \begin{equation}
    \label{e:B9}
    \tB_{N=9} = \frac{1}{\sqrt 3} \left( \begin{array}{lllllllll}
      1 & 0 & 0 & 0 & 0 & 0 & 1 & 0 & 0 \\
      1 & 0 & 0 & 0 & 0 & 0 & \omega_3^2 & 0 & 0 \\
      1 & 0 & 0 & 0 & 0 & 0 & \omega_3 & 0 & 0 \\
      0 & 1 & 0 & 0 & 0 & 0 & 0 & 1 & 0 \\
      0 & 1 & 0 & 0 & 0 & 0 & 0 & \omega_3^2 & 0 \\
      0 & 1 & 0 & 0 & 0 & 0 & 0 & \omega_3 & 0 \\
      0 & 0 & 1 & 0 & 0 & 0 & 0 & 0 & 1 \\
      0 & 0 & 1 & 0 & 0 & 0 & 0 & 0 & \omega_3^2 \\
      0 & 0 & 1 & 0 & 0 & 0 & 0 & 0 & \omega_3 
  \end{array} \right), \ \omega_3 = e^{ 2 \pi i / 3 }\,.
\end{equation}
The model has been obtained ``by hand'', by replacing ``lower order'' terms 
in the matrix $ B_h $ by $ 0 $, keeping only nonzero elements on the
``tilted diagonals'', and replacing 
$({1-e^{2\pi i(\pm 1)/3}})/({N(1-e^{2 i \pi (\pm 1)/N})})$ 
by $1$. 

The new 
matrix $ \tBh $ retains some qualitative features
of $ B_h $ but there is no immediate connection between 
their spectra: the ``lower order'' terms are not
small enough for that, and $B_h$ cannot be considered as a ``small perturbation'' of
$\tBh$. 

%Still, a comparison of density plots of 
%  both matrices for $N=27$  
%  (Fig.~\ref{fig:dens}) gives a visual motivation for 
%  the toy model.

The simplicity of the matrices  $ \tBh $ will allow us to 
prove (in the case $N=3^k$, $k\in\NN$) the fractal Weyl law which we 
could numerically observe for $ B_{h} $ (see \secref{s:reson-walsh}). 
It is interesting to notice that the
simplified operator $ \tBh $ is in fact not associated with the
same classical relation as $B_h$: 
%%%%%%%%%%%%%%%%%%%%%%%%%%%%%%%%%%%
\begin{prop}
\label{p:class}
In the notations of \secref{s:lagrangian}, the 
quantum relation $\{\tBh\}$ is associated
with the weighted relation $ ( \tB , \tilde \mu ) $ given by 
\eqref{eq:tbc} and \eqref{eq:tbm}:
\begin{gather*}
\tB_{h} \in I (\tB' \cup \widetilde \Lambda_{\rm{sing}} , \tilde \mu ) \,,\qquad where\\
\widetilde \Lambda_{\rm{sing}}
= \bigcup_{j=0}^2 \widetilde\Lambda_{\rm{sing},j}\,,\qquad
\widetilde\Lambda_{\rm{sing},j}=\set{ \big(q'=0,q=j/3\,;\, p',\, p \big),\:\ p',\,p\in\II}\,. 
\end{gather*}
\end{prop}
%%%%%%%%%%%%%%%%%%%%%%%%%%%%%%%%%%%
\begin{proof}
In place of $ \tBh $ we will consider $ \widetilde A_h = \F_N \circ 
\tBh $, and apply Lemma \ref{l:find}. From the structure
of $\tBh$, the operator  
$\widetilde A_h$ can obviously be split into $\widetilde A_{1,h}+\widetilde A_{2,h}$. 
We will analyze the first
component in detail, the analysis for the second one being similar. 
The matrix $\la Q_k|\widetilde A_{1,h}|Q_j\ra$ is
nonzero in the vertical strip $I_1\times \II$, with $I_1=[0,1/3)$:
$$
\la Q_k|\widetilde A_{1,h}|Q_j\ra=\frac{\bbbone_{I_1}(Q_j)}{\sqrt{3N}}\,\big(\sum_{\ell=0}^2 
e^{-2i\pi Q_k\ell}\big)\;\exp( -6i\pi N Q_k Q_j)\,.
$$
Like $A_{1,h}$ (see \S \ref{s:qbr}), this operator is of the form \eqref{eq:uqq}, 
with generating function 
$S(q',q)=-S_1(q',q)=-3q'q$ and discontinuous symbol 
$$
a(q',q)=\bbbone_{I_1}(q)\, 
\frac{e^{-2 i \pi q'}}{\sqrt{3}}\frac{\sin(3\pi q')}{\sin(\pi q')}\,.
$$ 
Forgetting about discontinuities, $\widetilde A_{1,h}$ is therefore associated with 
the graph
$$
\widetilde\Gamma_{S_1}=\set{(q',\,p'=-3q;q\,,\,p=-3q'),\ :\ q'\in\II,\,q\in I_1}\,,
$$
and the weight
$$
\mu_{S_1}=|a(q',q)|^2\,dq'\,dq=\bbbone_{I_1}(q)\,\frac{\sin^2(3\pi q')}{3\sin^2(\pi q')}\,dq'\,dq\,.
$$
After applying the transformation of Lemma~\ref{l:find}, this leads to the 
graph
$$
F_L^{-1}(\widetilde\Gamma_{S_1})=\set{(q'=3q,\,q\,;p',\,p=3p')\ :\ q\in I_1,\,p'\in\II}=
\bigcup_{j=0}^2\tB_{1j}\,,
$$
and the weight 
$$
F_{L *}^{-1}\,\mu_{S_1}=\bbbone_{I_1}(q)\,\frac{\sin^2(3\pi p')}{3\sin^2(\pi p')}\,dp'\,dq\,.
$$
Through the
change of variable $(q,p')\mapsto (q,p)$, we see that this is the weight \eqref{eq:tbm}
on the component $\tB_1$.

The discontinuities of $a(q',q)$ only occur along the two segments
$\{(q'\in\II,\,q=0)\}$, $\{(q'\in\II,\,q=1/3)\}$: they generate the singular
Lagrangian
$$
\cD_{{\rm sing},j}=\set{\big(q'=0,\,q=\frac{j}{3};\,p'\in\II,\,p\in\II\big)}\,,\quad j=0,1\,,
$$
which transforms under $F_L^{-1}$ into the components 
$\widetilde\Lambda_{\rm{sing},0}$, $\widetilde\Lambda_{\rm{sing},1}$. 

Similarly, the second part of the matrix, $\tB_{2,h}$,
is associated to the twisted graph $\tB_2'$ with weight $\tilde\mu_{\rest\tB_2}$ and 
the two singular components $\widetilde\Lambda_{\rm{sing},2}$, 
$\widetilde\Lambda_{\rm{sing},0}$.
\end{proof}

As explained in \secref{s:wr}, the graph $\tB$ can be obtained by adjoining to
each point $(\rho';\rho)\in B$ the points $(\rho'+(0,1/3);\rho)$ 
and $(\rho'+(0,2/3);\rho)$. This ``aliasing'' is due to the diffraction created 
by the sharp cutoff in the matrix $\tB_h$, as opposed to the ``smooth'' decay of 
coefficients in $B_h$.
A similar aliasing was observed in \cite{TraSco02} for the graph associated with the
unitary matrices $A_{2^k}$ defined in \eqref{e:unitary-toy}: instead of quantizing the 
standard $2$-baker
\eqref{e:2-baker}, they are associated with a multivalued map obtained from it by aliasing.
This observation was obtained using the propagation of coherent states.

Both $B$ and $\tB$
share the same forward trapped set $\widetilde\Gamma_-=\Gamma_-=C\times\II$ 
(see \S\ref{e:open-baker-clas}), but the
backwards trapped set of $\tB$ is easily shown to be $ \widetilde\Gamma_+=\t2$, which drastically
differs from $\Gamma_+$. This asymmetry between $\widetilde\Gamma_-$ and $\widetilde\Gamma_+$
reflects the fact that, unlike $B$, the relation $\tB$ is not time reversal symmetric.

The fact that $\tB_h$ is not associated with the relation $B$ 
should not bother us too much though.
In the next section, we will give a more ``formal'' construction of the
matrix $\tB_h$, in the case where $N$ is a power of $3$
(this construction will also hold for any symmetric $D$-baker, for
$N$ a power of $D$). We will show that this matrix naturally appears through
a ``nonstandard'' (Walsh)
quantization of the open $3$-baker relation $B$.

%%%%%%%%%%%%%%%%%%%%%%%%%%%%%%%%%%%%%%%%%%%%%%%%%%%

\subsection{Walsh quantization of the baker's relation}

  The {\em Walsh model} of harmonic analysis has been originally devoted to
  fast signal processing \cite{lifermann}.
  It has been used recently in mathematics to obtain simpler (and provable)
  versions of statements of the
  usual harmonic analysis --- see \cite{MTT} for an application 
  in scattering theory and for pointers to the recent literature. 
  The major advantage of Walsh harmonic analysis is the possibility to 
  completely localize a wavepacket both in position and momentum: for our
  problem, this has the effect of avoiding diffraction problems 
  due to the discontinuities of the map, which spoil the usual semiclassics 
  \cite{SaVo}.
  Closer to our context, Meenakshisundaram and Lakshminarayan recently
  used the Hadamard Fourier transform (which is related with the
  Walsh transform we give below) to analyze the multifractal
  structure of some eigenstates of the
  (unitary) quantum 2-baker $B_h$ \cite{arul}.

%%%%%%%%%%%%%%%%%%%%%%%%%%%%%%%%%%%%%%%%%%%%%%%%%%%

\subsubsection{The quantum torus as a system of quantum $D$its}

We first fix the
coefficient $D\in\NN$ ($D\geq 2$) of the symmetric baker's relation~\eqref{e:symmetric}, 
and will consider in this section only the inverse Planck's constants of the form
$N=D^k$ for some $k\in\NN$. In this
case, integers $j\in\ZZ_{D^k}=\{0,\ldots,D^k-1\}$ are in one-to-one correspondence with
the words $\bep=\ep_1\ep_2\cdots \ep_k$ made of symbols (or ``$D$its'') 
$\ep_\ell\in \ZZ_{D}$:
\begin{equation}\label{e:Dit}
\ZZ_{D^k}\ni j=\sum_{\ell=1}^k \ep_\ell\,D^{k-\ell}\,.
\end{equation}
The natural order for $j\in\ZZ_{D^k}$
corresponds to the lexicographic order for the symbolic words $\{\bep\in(\ZZ_D)^k\}$.
This way, each position eigenstate $|Q_j\ra$ of the basis \eqref{eq:bas}
can be associated with the unique symbolic sequence $\ep_1\ep_2\cdots \ep_k$ which
gives its $D$nary expansion 
\bequ\label{e:Q_j}
Q_j=\frac{j}{N}=0\cdot\ep_1\ep_2\cdots \ep_k\,.
\end{equation} 
Let us denote the canonical basis of $\CC^D$ by
$\{e_0,e_1,\ldots, e_{D-1}\}$. Then,
each $|Q_j\ra$ can be written as
\begin{equation}\label{e:quDit}
|Q_j\ra= e_{\ep_1}\otimes e_{\ep_2}\otimes\cdots \otimes e_{\ep_k}\,.
\end{equation}
Following \cite{schack}, we denote each $|Q_j\ra$ by
$|\bep\ra=|\ep_1\ep_2\cdots\ep_k\ra$ to emphasize the above
tensor product decomposition. 
This way, the quantum space $\Hh_h^1$ is naturally identified with the {\em tensor product} 
of $k$ spaces $\CC^D$:
$$
\Hh^1_h= (\CC^D)_1\otimes(\CC^D)_2\otimes\cdots\otimes(\CC^D)_k\,.
$$
In the quantum computating framework,
each space $(\CC^D)_\ell$ is interpreted as a ``quantum $D$it'', 
or `` qu$D$it'', and the basis $\set{|\bep\ra}$ is called the 
{\em computational basis} \cite{miquel}.
Viewed
in our toral phase space, the qu$D$it $(\CC^D)_\ell$ is associated
with the {\em scale} $D^{-\ell}$ in the position variable, so $(\CC^D)_1$ is
called the ``most significant qu$D$it''.

%%%%%%%%%%%%%%%%%%%%%%%%%%%%%%%%%%%%%%%%%%%%%%%%%%%%%%%%%%%%
\subsubsection{Walsh Fourier transform}
The discrete Fourier transform of \eqref{e:DFT} (with $n=1$, $N=D^k$) 
is the Fourier transform
(in the sense of abstract harmonic analysis) on  the 
group $\ZZ_{D^k}$. 
More explicitly, 
each row of $\F_{D^k}$ corresponds to the  character 
$ j' \mapsto \exp\Big(- 2i\pi j j'/{D^k} \Big)$
of $\ZZ_{D^k} $. Using \eqref{e:Dit}, the matrix elements can be factorized:
\begin{equation}\label{e:fourier-decompo}
(\F_{D^k})_{jj'}=D^{-k/2}\,\exp\Big(-2i\pi \frac{j j'}{D^k}\Big)
=D^{-k/2}\,\prod_{\ell=1}^k \exp\Big(-2i\pi \frac{\ep_\ell(j j')}{D^{\ell}}\Big)\,,
\end{equation}
Notice that each $\ep_m(jj')$ can be easily expressed in terms of 
the symbols of $j$ and $j'$:
$$
\ep_{m}(j j')=\sum_{\ell+\ell'=k+m} \ep_\ell(j)\,\ep_{\ell'}(j')\,.
$$
The Walsh Fourier transform is the 
Fourier transform on the group $(\ZZ_{D})^k $. It can be defined by
keeping only the first factor 
on the right hand side of \eqref{e:fourier-decompo}: one obtains the matrix
\begin{equation}\label{e:walsh}
(W_k)_{j j'}=D^{-k/2}\,\exp\Big(-2i\pi \frac{\ep_{1}(j j')}{D}\Big)
=\prod_{\ell=1}^k D^{-1/2}\,\omega_D^{-\ep_\ell(j)\ep_{k+1-\ell}(j')}\,,\qquad 
\omega_D=e^{2i\pi/D}\,.
\end{equation}
Using the identification $\Hh_h^1\simeq (\CC^D)^{\otimes k}$, 
this definition can be recast as follows. 
%%%%%%%%%%%%%%%%%%%%%%%%%
\begin{lem}\label{l:pk}
The Walsh Fourier transform $W_{k}$ acts simply
on tensor product states:
$$ 
W_{k} ( v_1 \otimes \cdots \otimes v_k ) 
=  \F_D v_k \otimes \cdots \otimes \F_D v_1  \,, \qquad v_\ell\in\CC^D,\ \ell=1,\ldots,k\,.
$$
Here $ \F_D = W_1 $ is the discrete Fourier transform on $\CC^D$. As a result,
$W_k$ is a unitary tranformation on $\Hh_h^1$.
\end{lem}
%%%%%%%%%%%%%%%%%%%%%%%%%
The proof consists in a straightforward algebraic check.

As opposed to the discrete Fourier transform, 
the Walsh Fourier transform {\em does not entangle} the different
qu$D$its: a tensor product state is sent to another tensor product state.

%\renewcommand\thefootnote{$\flat$}%

%\medskip
%\noindent
\begin{exmple} To illustrate this simple lemma we take $D=2$, and consider 
the following $ 2^k \times 2^k $ matrix, $k\geq 1$:
\begin{equation}\label{e:unitary-toy}
A_{2^k}  =  [ A_{0,2^k} , A_{1,2^k} ] \,,  \quad 
( A_{j,2^k} )_{0 \leq n \leq 2^k - 1 ,\, 0 \leq m \leq 2^{k-1}-1}  =  
\begin{cases}
(-1)^{jn}/\sqrt{2}\,, & m= \lfloor n/2\rfloor \\
0\,,  & m \neq \lfloor n/2\rfloor \,. 
\end{cases} 
\end{equation}
For instance when $ k=2 $ we get
\[ 
A_{2^2} = \frac{1}{\sqrt{2}} \begin{pmatrix}
 1 & 0 & \ \ 1 & \ \ 0 \\
 1 & 0 &  -1 &  \ \ 0 \\
 0 & 1 & \ \ 0 & \ \ 1  \\
 0 & 1 & \ \ 0 & -1  
\end{pmatrix} \,. 
\] 
This sequence of matrices
has been obtained as the ``extreme'' possibility
among a family of different quantizations of the
(closed) 2-baker's map \cite{schack}\footnote{We thank M.~Saraceno for 
pointing out this interpretation to us.}, and its semiclassical properties
were further studied in \cite{TraSco02}.
In a different context, this (unitary) matrix belongs to the family of
transfer matrices associated with the de Bruijn
graph with $2^k$ vertices \cite{tanner}.

The transformation $ A_{2^k} $ acts as follows on tensor product states:
\[  
v_1 \otimes \cdots \otimes v_k  \longmapsto  v_2 \otimes \cdots 
\otimes v_k \otimes \F_2 v_1 \,.
\]
%From this and the identity $(F_2)^2=I$ we see that 
%$ (A_{2^k})^k = F_2^{\otimes k} $, and $(A_{2^k})^{2k} = Id_{2^k}$: the
%matrix $A_{2^k}$ is periodic.
This implies that this matrix can be
easily expressed in terms of the Walsh Fourier transform (for $D=2$):
\begin{equation}\label{e:A_k}
A_{2^k} = W_k \begin{pmatrix} W_{k-1} &  \ 0 \\
\ 0 & W_{ k-1 } \end{pmatrix} \,,
\end{equation}
where the $2\times 2$ block structure corresponds to the most significant (leftmost) qubit.
This expression exactly parallels the one defining
the Balazs-Voros (unitary) quantum baker \cite{BaVo}. Compared to this ``usual''
quantum baker, $A_{2^k}$ is thus obtained by replacing the discrete Fourier 
matrices $\F_{2^k}$, $\F_{2^{k-1}}$
by their Walsh analogues $W_k$, $W_{k-1}$.

The matrix $A_{2^k}$ is unitary; as we will see in the next 
section, our toy model $\tBh$ for the quantum open 3-baker 
(see Eq.~\eqref{eq:ttoy}) is its subunitary analogue.
\end{exmple}

%%%%%%%%%%%%%%%%%%%%%%%%%%%%%%%%%%%%%%%%%%%%%%%%%%%%%%%%%%%%%%%%%%%%%%%%%%%%%%%
%%%%%%%%%%%%%%%%%%%%%%%%%%%%%%%%%%%%%%%%%%%%%%%%%%%%%%%%%%%%%%%%%%%%%%%%%%%%%%%

\subsection{Resonances for the Walsh quantization of the open baker relation}
\label{s:reson-walsh}

In this section we set $D=3$, and concentrate on the symmetric 3-baker \eqref{eq:exb}. 
By analogy with the example in last section, we
modify the quantization (\ref{e:quantum-baker},\ref{e:quantum3-b}), 
in the case $ N = 3^k $, 
by replacing the discrete Fourier matrices by their Walsh analogues. The resulting
operator exactly coincides with 
the toy model \eqref{eq:ttoy} introduced in the beginning of this section:
%%%%%%%%%%%%%%%%%
\begin{lem}
\label{l:coinc} In the case $N=3^k$, the matrix $ \tBh $ defined in  
\eqref{eq:ttoy} can be rewritten in terms of the Walsh Fourier transforms as follows:
\[ 
\widetilde B_{h } = W_{{k} }^* 
\begin{pmatrix} W_{k-1} & 0 & 0 \\
0 & 0 & 0 \\
0 & 0 & W_{k-1} \end{pmatrix} \,. 
\]
\end{lem}
%%%%%%%%%%%%%%%%%
We omit the simple algebraic proof.
If we define the ``truncated'' inverse Fourier matrix
\begin{equation}
\widetilde \F_3^* \defeq \frac{1}{\sqrt3} 
\begin{pmatrix} 
1 & 0 & 1 \\
1 & 0 & \omega_3^2 \\
1 & 0 & \omega_3 
\end{pmatrix}\,,
\end{equation}
the toy model $\tBh$ acts as follows on tensor product states:
\begin{equation}\label{e:toy-tensor}
\tBh ( v_1 \otimes \cdots\otimes v_k ) =
 v_2 \otimes v_3\otimes \cdots  \otimes \widetilde \F_3^* v_1\,.
\end{equation}
This form is particularly nice to compute the spectrum of $\tBh$.
We start by computing the spectrum of its power $ (\tBh) ^k $,
which is enough to obtain the radial distribution of resonances (that is, 
the distribution of {\em resonance widths}). 
%%%%%%%%%%%%%%%%%%%%%%%%%%%
\begin{prop}\label{p:toy}
Let $ \lambda_\pm$, $ |\lambda_- | < |\lambda_+ | $, be the 
eigenvalues of the matrix 
\[ 
\Omega_3 =\frac{1}{\sqrt3}
 \begin{pmatrix} 1 & \ 1 \\1 & \omega_3 \end{pmatrix}\,. 
\]
The non-zero eigenvalues of $ (\tBh)^{k} $ 
(for $N=(2 \pi  h)^{-1} = 3^k $)  
are given by $ \lambda_+^{k-p} \lambda_-^p $, $ 0 \leq p \leq k $, 
each occurring with multiplicity $\binom{k}{p}$. From this we
get the radial distribution of the eigenvalues of $\tBh$ (counted with multiplicities):
\begin{gather}
\begin{gathered}
\label{eq:last}
\forall \, r\in [0,1],\qquad \frac1{2^k}\;
\#\set{\lambda\in\Spec ( \tBh ) \,:\,|\lambda|\geq r} 
\xrightarrow{k\to \infty}  C ( r ) \,,\\
C ( r ) = \begin{cases} 1\,, & r < |\det\Omega_3|^{\half}\\
0\,, & r > |\det\Omega_3|^{\half} \,.
\end{cases}
\end{gathered}
\end{gather}
Hence the nontrivial resonances accumulate near the circle of
radius $r_0(\tB)=|\det\Omega_3|^{\half}$.
\end{prop}
%%%%%%%%%%%%%%%%%%%%%%%%%%%
This proposition gives Theorem~\ref{t:1}, where $\tB$ is the baker's relation 
described in Proposition~\ref{p:class}, $\tBh$ the matrices \eqref{eq:ttoy},  
and
Planck's constants are taken along the sequence 
$\{h_k=(2\pi\times 3^k)^{-1},\ k\in\NN\}$.
\begin{proof}
From the expression \eqref{e:toy-tensor}, 
we see that 
\[ 
(\tBh)^k  
  ( v_1 \otimes \cdots\otimes v_k )
= \widetilde \F_3^* v_1 \otimes \cdots  \otimes \widetilde \F_3^* v_k \,.
\]
That means that $(\tBh)^k = ( \widetilde \F_3^* )^{\otimes k} $, so one eigenbasis
is obtained by taking the tensor products of eigenstates of $\widetilde \F_3^*$, and the
eigenvalues of $(\tBh)^k$ are the corresponding products of eigenvalues of $\widetilde \F_3^*$. 
The nonzero eigenvalues $ \lambda_+$,
$\lambda_-$ of $\widetilde \F_3^*$ are the eigenvalues of $\Omega_3$, so 
the first part of the proposition follows.
To prove the second part, notice that each eigenvalue
$\lambda_+^{k-p} \lambda_-^p$ of $\tBh^k$ corresponds to an
eigenvalue (possibly in the generalized sense) of modulus 
$|\lambda_+^{1-p/k} \lambda_-^{p/k}|$
of $\tBh$. Therefore, we are able to count eigenvalues of $\tBh$ (with multiplicities)
in a given annulus.

Let $ H( t ) $ denote the Heaviside function, $ H( t) = 0 $ for $
t < 0 $, and $ H ( t ) = 1 $ otherwise. Then, for any $0<r<1$,
\[ 
\begin{split}
\#\set{\lambda\in\Spec ( \tBh ) \,:\,|\lambda|\geq r} & = \sum_{ p=0}^k 
H( |\lambda_+|^{1-p/k} |\lambda_-|^{p/k}- r ) \binom{k}{p} \\
& = \sum_{p=0}^k H ( -p/k + 1/2 + \rho ) \binom{k}{p} \,, \quad
\rho = \frac{\log( | \lambda_- \lambda_+ |^{\frac12} / r )}{ \log ( 
|\lambda_+ |/|\lambda_-| )}  \,.
\end{split} 
\]
Using Stirling's formula, 
one easily gets in the limit $k\to\infty$:
\[ 
\frac{1}{2^k}  \sum_{p=0}^k H ( -p/k + 1/2 + \rho ) 
 \binom{k}{p}  \sim \sqrt{\frac{2k}{\pi}}\int_{-\infty}^\rho e^{-2k x^2}\,dx\to H ( \rho ) \,.
\]
This expression shows that the distribution of resonances is 
semiclassically dominated by the degrees
$|p-k/2|=\Oo(k^{1/2})$, and proves the second part of the proposition.
\end{proof}
The explicit eigenvalues are 
$\lambda_\pm = \frac{1+i\sqrt{3}}{4\sqrt{3}}\pm\sqrt{\frac{11-i3\sqrt{3}}{24}}$, with 
approximate values
$$
\lambda_+\approx 0.8390+i 0.0942,\quad |\lambda_+|\approx 0.8443,\quad  
\lambda_-\approx -0.5504+ i 4058,\quad |\lambda_-|\approx 0.6838\,.
$$
The geometric mean of their moduli is 
$r_0(\tBh)=|\lambda_-\lambda_+|^{1/2}=\sqrt{|\det\Omega_3|}=3^{-1/4}$. 
%On Figure~\ref{f:lat}
%we plot the circles with radii $|\lambda_+|$, $|\lambda_-|$ and $|\lambda_-\lambda_+|^{1/2}$,
%together with the spectrum of $\tBh$.

\medskip

We need to analyze the spectrum of $\tBh$ more precisely to show that
the distribution of resonances is asymptotically uniform with 
respect to the angular variable.
%%%%%%%%%%%%%%
\begin{prop}
\label{p:last}
Let $ h = ( 2 \pi 3^k )^{-1} $. As a set, the nontrivial spectrum of 
$ \tBh $ is given by 
\[   
\{ \lambda_+ \} \cup \{ \lambda_- \} \cup 
\bigcup_{ \omega^k =1 } \{ \omega \lambda_+^{1 - p/k} \lambda_-^{p/k} 
\; : \; 1 \leq p \leq k-1 \} \,.
\]
For each $p\neq 0,k$, the $k$ eigenvalues asymptotically have the same
degeneracy $\frac1k\binom{k}{p}$, which shows that their distribution is
uniform in the angular variable. Therefore, for any continuous function $f\in C(D(0,1))$
we have (counting multiplicities in the LHS):
$$
\frac1{2^k}\sum_{0\neq\lambda\in\Spec(\tBh )} f(\lambda)
\xrightarrow{k\to \infty}
\int_0^{2\pi} f(| \lambda_- \lambda_+|^{\frac12},\theta)\,\frac{d\theta}{2\pi}\,.
$$
\end{prop}
%%%%%%%%%%%%%%

\begin{proof}
To classify the nontrivial 
spectrum of $ \tBh$, we will use the eigenvectors $v_\pm$ of 
$\widetilde F_3^*$ associated
with the eigenvalues $\lambda_\pm$.   Call 
$$
\set{\bet=\eta_1\eta_2\cdots\eta_k\,:\,\eta_\ell\in\{\pm\}}\simeq (\ZZ_2)^k
$$ 
the set of binary sequences of length $k$. The number of symbols $\eta_\ell=-$ 
in the sequence $\bet$ is called the {\em degree} of $\bet$.
The {\em cyclic shift} $\tau$ acts on these sequences as
$\tau ( \eta_1\cdots \eta_k)=\eta_2  \cdots  \eta_k  \eta_1\,.$
The shift allows us to partition $(\ZZ_2)^k$ into periodic
orbits, each orbit $\Oo=\set{\bet,\tau(\bet),\ldots,\tau^{\ell_\Oo-1}(\bet)}$ being
of (primitive) period $\ell_{\Oo}=\ell_{\bet}$. Since $ \tau^k = id $, 
the primitive period must divide $k$.
We call $\deg(\Oo)$ the common degree of the elements of $\Oo$
and observe that 
\begin{equation}
\label{eq:observ}
k \,  | \, \ell_\Oo \deg(\Oo)\,.
\end{equation}

To each sequence $\bet$ we associate the state 
$|\bet\ra\defeq v_{\eta_1}\otimes v_{\eta_2}\otimes\cdots\otimes v_{\eta_k}$,
which is obviously an eigenstate of $ (\tBh)^k $, with eigenvalue
$\lambda_+^{k-\deg(\bet)} \lambda_-^{\deg(\bet) }$. These
$2^k$ states form an independent family, which span the 
nontrivial eigenspaces of $\tBh$.
This operator acts very simply on these states:
$$
\forall \bet\in(\ZZ_2)^k,\qquad\tBh |\bet\ra=\lambda_{\eta_1}\,|\tau(\bet)\ra\,.
$$
Hence, for any orbit $\Oo$, 
$\tBh$ leaves invariant the $\ell_\Oo$-dimensional subspace
$
V_{\Oo}\defeq {\rm{span}}\set{|\bet\ra\,,\,\bet\in\Oo}\,.
$
To compute the spectrum of $\tBh\rest_{V_{\Oo}}$ we first 
observe that it is contained in the set of $ k$-th  roots of 
$ \lambda_+^{k - \deg(\Oo)} \lambda_-^{\deg(\Oo)} $, which in view
of \eqref{eq:observ} is equal to 
$$
S_{\Oo} \defeq \set{\omega_{\ell_\Oo}^j\,\lambda_+^{1 - \deg(\Oo)/k} \lambda_-^{\deg(\Oo)/k}\,,\ 
j=0,\ldots,\ell_\Oo-1}\,. 
$$
We claim that $ \Spec(\tBh\rest_{V_{\Oo}}) = S_{\Oo } $ (clearly with
no degeneracies). In fact, let
$ \Omega_{\Oo } \; : \; V_{\Oo} \rightarrow
V_{\Oo} $ be defined by $ \Omega_\Oo |\tau^\ell (\bet )\ra = 
\omega_{\ell_\Oo}^{-\ell} |\tau^\ell ( \bet )\ra $, for a choice of $ \bet \in\Oo$. 
This operator is invertible on $V_{\Oo}$.
By a verification on basis elements, 
\[ 
\tBh\rest_{V_{\Oo}} \circ \, \Omega_\Oo = 
\omega_{\ell_\Oo} \; \Omega_\Oo \circ \tBh\rest_{V_{\Oo}} \,, 
\]
and hence if $ \lambda \in  \Spec(\tBh\rest_{V_{\Oo}})$ then 
$ \omega_{\ell_\Oo} ^j \lambda \in \Spec(\tBh\rest_{V_{\Oo}}) $ for any $j$.

Since 
$\Oo\neq\Oo'\Longrightarrow V_{\Oo}\cap V_{\Oo'}=\set{0}$, enumerating
the orbits decomposition of $(\ZZ_2)^k$ yields
the full nontrivial spectrum of $\tBh$. In spite of the
large degeneracies, this nontrivial spectrum does not
contain any Jordan block.

The degree $p=0$ corresponds to the unique orbit $\Oo=\set{\bet=++\cdots+}$, 
so the eigenvalue $\lambda_+$ is simple. Similarly, the degree $p=k$
leads to the simple eigenvalue $\lambda_-$.

For any degree $1\leq p\leq k-1$, call $g={\rm gcd}(k,p)$. 
The sequences of degree $p$ will take all
possible periods $\ell_{\bet}=k/\ell$, where $\ell\in\NN$,  $\ell|g$.
We show below that, in the semiclassical limit, the huge majority of 
the sequences of any degree $p\neq 0,k$ have primitive period $k$. 
%%%%%%%%%%%%%%%%%%%%%%%%
\begin{lem}\label{l:nonperiodic}
There exists $C>0$, $K>0$ s.t., for any $k\geq K$ and any degree $1\leq p\leq k-1$,
$$
\frac{\#\set{\bet\in(\ZZ_2)^k\,:\,\deg(\bet)=p\,, \ell_{\bet}<k}}
{\#\set{\bet\in(\ZZ_2)^k\,:\,\deg(\bet)=p}}
\leq C\,\frac{\log k}{k}\,.
$$
\end{lem}
%%%%%%%%%%%%%%%%%%%%%%%%
\begin{proof}
We still use $g={\rm gcd}(k,p)$.
 
If $g=1$, then all orbits of degree $p$ are of primitive period $k$. 

If $g>1$, there exists $\ell>1$, $\ell|g$. 
For any $P$ prime divisor of $\ell$,  
any sequence of primitive period $\ell_{\bet}=k/\ell$ is also of 
(nonnecessarily primitive) period $k/P$.
Any sequence of degree $p$ and (nonnecessarily primitive) period $k/P$ can be
seen as the $P$ repetitions of a sequence of $k/P$ bits, among which $p/P$ take
the value $(-)$. Therefore, the 
number of such sequences is exactly $\binom{k/P}{p/P}$.
As a consequence, we have
\bequ\label{e:sum}
\frac{\#\set{\bet\in(\ZZ_2)^k\,:\,\deg(\bet)=p\,, \ell_{\bet}<k}}
{\#\set{\bet\in(\ZZ_2)^k\,:\,\deg(\bet)=p}}\leq 
\sum_{P\; {\rm prime},\; P|g}\frac{\binom{k/P}{p/P}}{\binom{k}{p}}\,.
\end{equation}
We will now estimate each term in the above sum.
From the symmetry $\binom{k}{p}=\binom{k}{k-p}$, we can assume $p\leq k/2$. 
Expanding the coefficient $\binom{k}{p}$ into
$$
\binom{k}{p}=
\frac{k(k-1)\cdots (k-p+1)}
{p(p-1)\cdots 1}\,,
$$
we notice that both the numerator and the denominator contain 
exactly $p/P$ factors which are multiples
of $P$. Their ratio gives $\binom{k/P}{p/P}$, while the ratio of the
remaining factors is
\begin{align*}
\frac{\binom{k}{p}}{\binom{k/P}{p/P}}&=\frac{(k-1)\cdots(k-P+1)(k-P-1)\cdots (k-p+1)}
{(p-1)\cdots(p-P+1)(p-P-1)\cdots 1}\\
&\geq \frac{k-p+1}{1}\geq \frac{k}{2}+1\,.
\end{align*}
Here we used the fact that each factor $({k-m})/(p-m)>1$, $0\leq m\leq p-2$,
and only kept explicit
the last factor. The last inequality comes from $p\leq k/2$. 

We have obtained
a uniform upper bound for each term in the sum of \eqref{e:sum}.
By standard arguments, there exists $K,\ \widetilde C>0$ s.t. 
the number of prime factors of any $k\geq K$ is $\leq \widetilde C\log k$,
so the number of terms in the sum is $\leq \widetilde C\log k$.
As a result,  \eqref{e:sum} is bounded from above by 
${\widetilde C\log k}/({k+2})$,
 which proves the lemma.
\end{proof}
\medskip

This lemma shows that the orbits of period 
$\ell_\Oo<k$ have a negligible contribution
to the asymptotic density of resonances. We can therefore act as if, 
for any $1\leq p\leq k-1$, each orbit of degree $p$ had period $k$,
leading to the $k$ eigenvalues 
$\set{\omega_k^j\,\lambda_+^{1 - p/k} \lambda_-^{p/k}\,,\ j=0,\ldots,k-1}$. 
In the semiclassical limit,
these $k$ eigenvalues are uniformly distributed on the circle of radius 
$|\lambda_+^{1 - p/k} \lambda_-^{p/k}|$, 
and each of them has multiplicity $\frac1k\binom{k}{p}$. 
This shows that the asymptotic resonance distribution is 
circular-symmetric, with the
radial distribution described in \propref{p:toy}.
\end{proof}
%%%%%%%%%%%%%%%%%%%%%%%%%%%%%%%%%%%
\begin{rem}\label{r:noRMT}
Several features of the (nontrivial) spectrum of $\tBh$ are very different from 
what one expects for a random subunitary matrix
of size $2^k\times 2^k$: the (logarithms of the) resonances form a regular lattice,
most eigenvalues are highly degenerate, and the radial density is 
a delta function at $r_0(\tB)$. 
Actually, the only generic feature seems to be the global fractal scaling of the Weyl law,
and the uniform angular distribution. 
\end{rem}

\begin{rem}\label{r:D=4}
The radial density of resonances 
is governed by $r_0(\tB)=\sqrt{|\det\Omega_3|}$, which seems
to depend on the subtelties of the quantization.
As an example of this fact, in \S \ref{cWm} we
will consider the open baker's map with $D=4$, which we call $B$, 
obtained by keeping only the second and third
strips. It has Lyapounov exponent $\log 4$, and the Cantor set $C$ 
(see \S\ref{e:open-baker-clas}) has dimension 
$\nu={\log 2}/{\log 4}=1/2 $. The open map $B'$ obtained by removing the first and third
strips has the same characteristics. However, if we Walsh-quantize these two maps, 
the spectra of $\tBh$ and $\tBh'$ are very different.
These spectra are obtained from the eigenvalues of different $2\times 2$ blocks of the inverse Fourier 
matrix $\F_4^*$. The first map leads to the block
$$
\Omega_4=\half\begin{pmatrix}\ \ i&-1\\-1&\ \ 1\end{pmatrix} \,, 
$$ 
with two nonzero eigenvalues
$\lambda_\pm$ of different moduli, so the spectrum of $\tBh$ will satisfy the fractal
Weyl law, and be concentrated around the circle of radius
$$
r_0(\tB)= \sqrt{|\det\Omega_4|}=2^{-3/4}\,. 
$$
In an opposite way, the second map leads to
the singular block
$$
\Omega'_4=\half\begin{pmatrix}1&1\\1&1\end{pmatrix}\,. 
$$
The nontrivial spectrum
of $\tBh'$ then reduce to a simple eigenvalue $\lambda_+=1$. 
In that case, the Weyl law is singular, corresponding to
the profile function $C(r)\equiv 0$. This qualitative difference between both spectra cannot
be explained from the classical maps, but is due to the phases in the matrix elements of
the quantum maps.
\end{rem}
%%%%%%%%%%%%%%%%%%%%

%%%%%%%%%%%%%%%%%%%%%%%%%%%%%%%%%%%%%%%%%%%%%%%%%%%%%%%%%%%%%%%%%%%%%%%%%%%%%%%
%%%%%%%%%%%%%%%%%%%%%%%%%%%%%%%%%%%%%%%%%%%%%%%%%%%%%%%%%%%%%%%%%%%%%%%%%%%%%%%

\section{Conductance in the Walsh model}
\label{cWm}

%%%%%%%%%%%%%%%%%%%%%%%%%%%%%%%%%%%%%%%%%%%%%%%%%%%%%%%%%%%%%%%%%%%%%%%%%%%%%%%

\subsection{Quantum transport}
\label{qtr}
In this section, we consider open baker's relations for which
the ``opening'' consists in two disjoint intervals, which are supposed to represent
two ``leads'' connecting a quantum dot to the outside world. 
We will prove Theorem~\ref{t:2} in this setting: \eqref{eq:t2.1} in 
\S \ref{cond} and \eqref{eq:t2.2} in \S \ref{noise}.

The baker's relations defined in \S\ref{e:open-baker-clas} can all be seen as
truncations of invertible maps on $\t2$. More precisely there exists an invertible
baker's  map, $\kappa:\t2\to\t2$, such that the graph $B=B_1\cup B_2$ 
of the open baker's map is 
$$
B=\Gamma_\kappa\cap \set{(q,p)\,:\,q\in I_1\cup I_2=I,\ p\in\II}\,.
$$
For admissible values of $N$, one can quantize the closed map $\kappa$ 
into a unitary transformation
$U_h=U_{\kappa,h}$ on $\Hh_h^1$ by straightforwardly generalizing the method
of \cite{BaVo}.  Multiplying this unitary operator by the quantum
projector $\Pi_{I}=\sum_{Q_j\in I}|Q_j\ra \la Q_j|$, we 
obtain the quantum open baker's map \eqref{e:quantum-baker}
$$
B_h=U_{\kappa,h}\,\Pi_{I}\,.
$$
To obtain an agreement with the notations of \S\ref{s:ben}, we can 
interpret the
set $I=I_1\cup I_2$ as the ``wall'' of the quantum dot, while the complementary interval
$L=\II\setminus I$ represents the ``openings'' of the dot, perfectly connected
with the ``leads''.
In the previous sections, we studied the resonances, that is, the eigenvalues of 
$B_h=U_{\kappa,h}\Pi_{I}$ (or of $\tBh$ when choosing the Walsh quantization). 
Now, we want to study the ``transport'' through the dot, using the formalism
presented
in \S\ref{s:ben}. We assume that the opening $L$ splits into 
two disjoint ``leads'' $L=L_1\cup L_2$, and we study the transmission matrix 
from lead $L_1$ to lead $L_2$ (for simplicity, both leads will have the same width). 
This matrix is
obtained by decomposing the scattering matrix \eqref{e:scattering}
$$
\tilde{S}(\vartheta)
=\Pi_L\sum_{n\geq 0} \big(e^{i\vartheta}\, U_h\,\Pi_I\big)^n\,e^{i\vartheta} U_h\,\Pi_L
$$
into $4$ blocks, according to the decomposition $\Pi_L=\Pi_{L_1}\oplus\Pi_{L_2}$.
The transmission matrix from $L_1$ to $L_2$ is defined as the block
\begin{equation}
\label{eq:def.t}
t(\vartheta)=\sum_{n\geq 1} e^{in\vartheta}\,\Pi_{L_2}\,U_h\,(\Pi_I\, U_h)^{n-1}\,\Pi_{L_1}
\defeq \sum_{n\geq 1} e^{in\vartheta}\,t_n\,.
\end{equation}
Because $\Pi_{L_1}$ and $\Pi_{L_2}$ have the same rank $M= N|L_1|$, $t(\vartheta)$ is
a square matrix of size $M$. 
According to Landauer's theory of coherent transport, 
each eigenvalue $T_i(\vartheta)$ of the matrix 
$t^*(\vartheta) t(\vartheta)$ corresponds to a ``transmission channel''. The 
dimensionless conductance of the system is then given by the sum over these
transmission eigenvalues:
\begin{equation}\label{eq:cond}
g(\vartheta)=\tr\big(t^*(\vartheta) t(\vartheta)\big)\,.
\end{equation}
A transmission channel is ``classical'' if the eigenvalue $T_i$ is very close to
unity (perfect transmission) or close to zero (perfect reflection). The
intermediate values correspond to the ``nonclassical channels'', the importance of which
is reflected in the noise power
\bequ\label{e:power}
P(\vartheta)=\tr\Big(t^* t(\vartheta)\big(Id-t^* t(\vartheta)\big)\Big)\,,
\quad\text{or the Fano factor}\quad F(\vartheta)=\frac{P(\vartheta)}{g(\vartheta)}\,.
\end{equation}
In general it may be necessary to
perform the ``ensemble averaging'' (averaging 
over $\vartheta$) to obtain significant results \cite{beenakker}. However, 
for the model
we will study below, both conductance and noise power will depend very little on $\vartheta$,
so this averaging will not be necessary. To alleviate notations 
we will supress the 
dependence in $\vartheta$ in the transmission matrix $t$.

%%%%%%%%%%%%%%%%%%%%%%%%%%%%%%%
\subsubsection*{Our model}
In the remainder of this section, we will compute the quantities characterizing 
the transport through the ``dot'' when 
$U_h$ is a Walsh-quantized baker's map similar to the
operator \eqref{e:unitary-toy}, but with $D=4$ instead of $D=2$. 
The sequence of values of $ h$ consequently is given by 
$$ 
2 \pi h_k  =  4^{-k } \,, \ \ k = 1 , 2 , \cdots \,.
$$
We will choose the two
leads $L_1=[0,1/4]$ and $L_2=[3/4,1]$: this way, the projectors 
$\Pi_{L_i}$ and $\Pi_I=Id-\Pi_{L_1}-\Pi_{L_2}$ can 
be represented as tensor product operators:
\begin{align*}
\Pi_{L_1}&=\pi_{0}\otimes Id_4\otimes\cdots\otimes Id_4\,,\\
\Pi_{L_2}&=\pi_{3}\otimes Id_4\otimes\cdots\otimes Id_4\,,\\
\Pi_I&=\pi_{I}\otimes Id_4\otimes\cdots\otimes Id_4\,,
\end{align*}
where $\pi_i=|e_i\ra\,\la e_i|$ is a rank-1 orthogonal projector acting on $\CC^4$,
and we note $\pi_I=\pi_1\oplus \pi_2$.

The ``inside'' propagator 
for this model, namely $\tBh=U_h \Pi_I$, is the first one among the two 
quantum maps
mentioned in Remark~\ref{r:D=4}: its nontrivial spectrum
satisfies the fractal Weyl law with
exponent $\nu=\half$, and is
concentrated near the radius $r_0(\tB)= 2^{-3/4}$.

The number of {\em scattering channels} in each lead is the 
rank of $\Pi_{L_1} $ (equal to that of $\Pi_{L_2}$). It is given by 
$ \frac14 $ of the total dimension, and we denote it by 
\begin{equation}
\label{eq:Mh}  M (h) = \frac{1}{4} ( 2 \pi h )^{-1}=4^{k-1} \,, \qquad  h\in\{h_k\} \,.
\end{equation} 
The number of channels is ``macroscopic'', and each channel
is ``fully coupled'' to the leads. We are therefore in a 
very nonperturbative r\'egime, 
where resonances have no memory at all of the 
eigenvalues of the closed (unitary) system.

%%%%%%%%%%%%%%%%%%%%%%%%%%%%%%%%%%%%%%%%
%%%%%%%%%%%%%%%%%%%%%%%%%%%%%%%%%%%%%%%%
\subsection{Conductance}
\label{cond}

We will crucially use the fact that all operators under consideration 
act nicely on the tensor
product structure $\Hh_h^1=(\CC^4)^{\otimes k}$, that is, 
they do not entangle the qu$D$its. It is then
suitable to compute the trace of $t^* t$ in a basis adapted to this
tensor product, and we naturally choose the computational (or position) basis.
The conductance is then given by
$$
\tr(t^* t)=\sum_{Q_j\in L_1} \la Q_j|t^* t|Q_j\ra 
=\sum_{j=0}^{4^{k-1}-1}\mn{t|Q_j\ra}^2\,.
$$
Let us consider an arbitrary $j=\ep_1\ep_2\cdots\ep_k$ with $\ep_1=0$, so that
$ 0 \leq j \leq 4^{k-1} - 1 $. 
Using \eqref{eq:def.t} we write
\bequ\label{eq:def.tQ_j}
t|Q_j\ra=
\sum_{n\geq 1} e^{in\vartheta}\,\Pi_{L_2}\,U_h\,(\Pi_I\,U_h)^{n-1}\,\Pi_{L_1}|Q_j\ra
=\sum_{n\geq 1} e^{in\vartheta}\,t_n|Q_j\ra\,,
\end{equation}
so that 
$$
\mn{t|Q_j\ra}^2=\sum_{m,n\geq 0} e^{i(n-m)\vartheta}\,\la Q_j|t_m^*\,t_n|Q_j\ra\,.
$$
From now on, we replace the notation $|Q_j\ra$, $j\in[0,4^{k-1}-1]$, by the symbolic
notation $|\bep\ra$, where the sequence $\bep=0\,\ep_2\cdots \ep_k$ corresponds to $j$.

%%%%%%%%%%%%%%%%%%%%%%%%%%%%%%%%%%%%%%%%%%%%%%%%%%%
\subsubsection{Classical transmission channels}
To understand the action of $t_n$ on $|\bep\ra$ 
we notice that $ \Pi_I\,U_h$ acts on tensor products as
$$
\Pi_I\,U_h(v_1\otimes\cdots\otimes v_k)=\pi_I v_2\otimes\cdots\otimes v_k
\otimes\F_4^* v_1\,.
$$
If $n<k$, we obtain 
\bequ\label{e:tn}
t_n|\bep\ra=
\pi_3 e_{\ep_{n+1}}\otimes e_{\ep_{n+2}}\otimes \cdots e_{k}\otimes 
\F_4^*e_{0}\otimes\F_4^*\pi_I e_{\ep_2}\otimes\cdots \otimes\F_4^*\pi_I e_{\ep_{n}}\,,
\end{equation}
frow which we draw the
%%%%%%%%%%%%%%%%%%%%%%%%%
\begin{lem}\label{l:classical-state}
Consider a sequence $\bep=0\,\ep_2\ep_3\cdots\ep_k$, and
assume that there exists an index $2\leq\ell\leq k$
such that $\ep_\ell\in\{0,3\}$. Let $\ell_0$ be the smallest such index. 
Then 
\[ 
\mn{t|\bep\ra}= \begin{cases} 0 & \text{ if }\  \epsilon_{\ell_0} = 0\,,\\
1 &  \text{ if }\ \epsilon_{\ell_0} = 3 \,.
\end{cases}
\]
This shows that $|\bep\ra$ is a classical transmission channel.
\end{lem}
%%%%%%%%%%%%%%%%%%%%%%%%%
\begin{proof}
For any 
$1\leq n\leq\ell_0-2$,  $\ep_{n+1}\in\{1,2\}$ by assumption. 
Hence the first qu$D$it on the right hand side 
 of \eqref{e:tn} vanishes and $t_n|\bep\ra=0$. 
Furthermore, the state $(\Pi_I U_h)^{\ell_0-1}|\bep\ra$ admits as first qu$D$it
$\pi_I\,e_{\ep_{\ell_0}}=0$, so that 
$t_{n}|\bep\ra=0$ for any $n\geq \ell_0$. The only 
remaining term in \eqref{eq:def.tQ_j} is $t_{\ell_0-1}|\bep\ra$:
\begin{itemize}
\item if $\ep_{\ell_0}=0$, the first qu$D$it of that term is 
$\pi_3 e_{\ep_{\ell_0}}=0$, so $t_{\ell_0-1}|\bep\ra=t|\bep\ra=0$.
\item if $\ep_\ell=3$, 
$t_{\ell_0-1}|\bep\ra=e_{\ep_{3}}\otimes e_{\ep_{\ell_0+1}}\otimes \cdots\otimes 
\F_4^*e_{0}\otimes\F_4^*e_{\ep_2}\otimes\cdots \F_4^*e_{\ep_{\ell_0-1}}$ . 
Since $\F_4^*$ is unitary, $\mn{t_{\ell_0-1}|\bep\ra}=\mn{t|\bep\ra}=1$.
\end{itemize}
\end{proof}
The number of classical channels discussed in Lemma \ref{l:classical-state} 
is easy to compute: it is obtained by removing from 
the set
$[0,4^{k-1}-1]\equiv \set{\ep_2\cdots\ep_k\in(\ZZ_4)^{k-1}}$ 
the sequences such that $\ep_\ell\in\{1,2\}$
for all $2\leq\ell\leq k$ (these will be called ``nonclassical sequences''). 
The number of the latter is  $2^{k-1}$, so
the number of classical channels is $4^{k-1}-2^{k-1}$.  Among them, half are
fully reflected, $ t | \bep \ra = 0 $, and half are fully transmitted, 
$ \mn{t | \bep \ra} = 1 $. Hence the conductance through these classical channels is
$$
\tr_{\rm cl}(t^* t)=\frac{4^{k-1}-2^{k-1}}{2}\,.
$$
%%%%%%%%%%%
\begin{rem}
Such classical channels are mentioned in the analysis of \cite{beenakker} for the transmission
through an open kicked rotator. 
They sit in the phase space regions above the lead $L_1$ which are
either sent back to $L_1$, or sent to $L_2$ through the classical dynamics, in 
a time smaller than the Ehrenfest time $T_{\rm Ehr}={\log N}/({\log 4})=k$. 
For our baker's map $B$,
these regions are vertical strips of widths $4^{-\ell}$, $\ell=2,\ldots,k$ which
exit to $L_1$ or $L_2$ at time $\ell$.
The particularity of the Walsh quantization 
is the {\em exact} full transmission (or reflection) through these channels.
\end{rem}
%%%%%%%%%%%

%%%%%%%%%%%%%%%%%%%%%%%%%%%%%%%%%%%%%%%%%%%%%%%%%%%%%%%

\subsubsection{Nonclassical transmission channels}
The nonclassical channels are necessarily combinations of the 
position states $|\bep\ra$ with  $\ep_\ell\in\{1,2\}$
for all $2\leq\ell\leq k$ (``nonclassical'' sequences or states). 
The associated positions $4Q_j=0\cdot\ep_2\ep_3\cdots\ep_k$ 
lie close to
the Cantor set $C$, such 
that  $\Gamma_-=C\times\II$ is the set of points never escaping through $B$
or $\widetilde B$ (see Eq.~\eqref{e:Gamma+}). 

For such a state $|\bep\ra$, the term \eqref{e:tn} vanishes for all $n<k$, due to the first qu$D$it
$\pi_I e_{\ep_{n+1}}=0$. That state therefore accomplishes $k$ ``unitary bounces''
inside the cavity, before it starts to decay out of it.
We first consider the terms $t_{k+m}|\bep\ra$ for 
$0\leq m<k$: 
\begin{equation}
\label{e:tn2}
\begin{split}
&  t_{k}|\bep\ra= (e_3/2)\otimes  \F_4^*e_{\ep_{2}}\otimes \F_4^*e_{\ep_{3}} \cdots\ \F_4^*e_{\ep_k} \,, \quad\mbox{while for $0<m<k$}\\
&  t_{k+m}|\bep\ra= \pi_3 \F_4^*e_{\ep_{m+1}}\otimes \F_4^*e_{\ep_{m+2}} \cdots\ \F_4^*e_{\ep_k}\otimes
\F_4^*(e_{3}/2)\otimes\F_4^*\pi_I\F_4^* e_{\ep_2}\cdots\F_4^*\pi_I\F_4^* e_{\ep_m}\,.
\end{split}
\end{equation}
An explicit computation shows that 
$$
 \mn{\pi_3 \F_4^*e_j}^2=\frac14\,,
\qquad\mn{\pi_I\F_4^*e_j}^2=\frac12\,, \ \ \  j =0, \cdots , 3\,,
$$
so that
\bequ\label{e:norms-tn}
\mn{t_{k+m}|\bep\ra}^2=\frac1{4\times2^m}\,, \ \ 0 \leq m \leq k-1 \,.
\end{equation}
For larger times $n=pk+m$, $p> 1$, $m\in[0,k-1]$,
the state $t_n|\bep\ra$
is obtained from \eqref{e:tn2} by inserting  the operator $(\pi_I\F_4^*)^{p-1}$
in front of each qu$D$it $e_{\ep_\ell}$.
Since $\pi_I\F_4^*$ has a spectral radius $|\lambda_+|<1$, 
the norms of these states will
decay exponentially fast as $n\to\infty$. This argument gives the following 
%%%%%%%%%%%%%
\begin{lem}\label{l:cutoff-Theta}
For any $0<\Theta<1$, there exists $C>0$ such that, for any $k\geq 1$ and any 
nonclassical state $|\bep\ra$, we have
$$
\sum_{m> \lfloor \Theta k \rfloor} \mn{t_{k+m}|\bep\ra}\leq C\,2^{-\Theta k/2}\,.
$$
\end{lem}
%%%%%%%%%%%%%
Neglecting errors of order $\Oo(2^{-\Theta k/2})$, we just need to
compute
$\mn{\sum_{m=0}^{\lfloor \Theta k\rfloor} t_{k+m}|\bep\ra}^2$.
From \eqref{e:norms-tn} we already know the diagonal terms:
\begin{equation}\label{e:sum1}
\sum_{m=0}^{\lfloor \Theta k\rfloor} \mn{t_{k+m}|\bep\ra}^2=\half + \Oo(2^{-\Theta k})\,.
\end{equation}
In the next lemma we will show that 
the contribution to the conductance of the nondiagonal terms 
is negligible in the semiclassical limit.
%%%%%%%%%%%%%%
\begin{lem}\label{l:nonperiodic1}
Let $0<\Theta\leq 1/5$. There exists $ C = C( \Theta )  >0$
such that for any $k\geq 1$,
$$
\frac{\#\set{{\rm nonclassical}\ \bep,\ \exists m,m'\in[0,\Theta k],\ \ 
\la \bep|t_{k+m}^* t_{k+m'}|\bep\ra\neq 0}}
{\#\set{{\rm nonclassical}\ \bep}}\leq C\, 2^{-k/2}\,.
$$
\end{lem}
%%%%%%%%%%%%%%
In other words, in the semiclassical limit,
a ``generic'' nonclassical state $|\bep\ra$ will satisfy
$$
\forall m,m'\in[0,\Theta k],\quad m\neq m'\Longrightarrow \la \bep|t_{k+m}^* t_{k+m'}|\bep\ra=0\,.
$$
\begin{proof}
Take an arbitrary nonclassical state $|\bep\ra$, and any $m,m'\in[0,\Theta k]$, $m>m'$.
From \eqref{e:tn2}, the first $(k-m)$  
qu$D$its of the states $t_{k+m}|\bep\ra$ and  $t_{k+m'}|\bep\ra$) are respectively
\begin{align*}
&\pi_3\F_4^*e_{\ep_{m+1}}\otimes \F_4^*e_{\ep_{m+2}}\otimes \cdots\otimes \F_4^*e_{\ep_k}\,,\\
&\pi_3\F_4^*e_{\ep_{m'+1}}\otimes \F_4^*e_{\ep_{m'+2}}\otimes \cdots\otimes \F_4^*e_{\ep_{k+m'-m}}\,.
\end{align*}
Due to the unitarity of $\F_4^*$ and the fact that the $e_i$ form an orthonormal
basis of $\CC^4$, 
the two states $t_{k+m}|\bep\ra$, $t_{k+m'}|\bep\ra$ will be orthogonal 
if the sequences $\ep_{m+2}\cdots \ep_k$ and $\ep_{m'+2}\cdots \ep_{k+m'-m}$ are not
equal.
Since we took $m<\Theta k$, these two sequences are 
subsequences of length $(k-m-1)\geq (1-\Theta)k$ 
of the sequence $\bep$, shifted from one another by $(m-m')$ steps.

If the two subsequences are equal, then all subsequences 
\begin{gather*}  \ep_{k-(p+1)\Delta+1}\cdots \ep_{k -p \Delta }\,, \ \
p = 0, \cdots, R \,, \\
\Delta\defeq (m-m') \,, \ \ 
R\defeq\Big[\frac{k-m-1}{\Delta}\Big]\,,
\end{gather*}
have to be equal to each other.
Hence $\bep$ contains a subsequence of length $(R+1)\Delta$
which is periodic with period $\Delta$. 

Let us count the number of such sequences $\bep$.
Once we have fixed the $\Delta$ bits $\ep_{k-\Delta+1}\cdots \ep_k$,
the remaining free bits are $\ep_2\cdots \ep_{k-(R+1)\Delta}$. The number $\#(m,m')$ of such
sequences is therefore $2^{\Delta}\times 2^{k-(R+1)\Delta-1}$. From the definition
of $R$, we get
$$
\#(m,m')<2^{2m-m'}\leq 2^{2\Theta k}\,.
$$
Taking into account all possible pairs $(m,m')$, we obtain the following bound
for the number of nongeneric nonclassical channels:
$$
\#\set{\text{nonclassical}\ \bep,\ \exists\,m,m',\ 0\leq m'<m\leq\Theta k,\ 
\la \bep|t_{k+m}^* t_{k+m'}|\bep\ra\neq 0}\leq (\Theta k)^2\,2^{2\Theta k}\,.
$$
Since $\#\set{\text{nonclassical}\ \bep}=2^{k-1}$ and  $2\Theta\leq 2/5$,
we have proven the lemma.
\end{proof}
From Lemma~\ref{l:cutoff-Theta} and Eq.~\eqref{e:sum1}, a generic 
nonclassical sequence $\bep$ will satisfy
$\mn{t|\bep\ra}^2=\half +\Oo(2^{-\Theta k/2})$.
For a nongeneric nonclassical sequence $\bep$, we use the simple
bound $\mn{t|\bep\ra}^2\leq 1$. 
As a result, we get the following estimate for the ``nonclassical conductance'':
\bequ\label{e:nonclas-conduct}
\tr_{\rm{noncl}}(t^* t)=\sum_{\substack{{\rm nonclassical}\\{\rm generic}}}\mn{t|\bep\ra}^2
+\sum_{\substack{{\rm nonclassical}\\{\rm nongeneric}}}\mn{t|\bep\ra}^2
=\frac{2^{k-1}}{2}\big(1+\Oo(2^{-\Theta k/2})\big)\,.
\end{equation}
Adding this to the ``classical conductance'', we get the full conductance
\bequ
g(\vartheta)=\tr(t^* t(\vartheta))=\frac{4^{k-1}}{2}+\Oo(2^{(1-\Theta/2) k})\,.
\end{equation}
The implied constant is independent of $\vartheta\in [0,2\pi)$ and $0<\Theta\leq 1/5$. 
The number of scattering channels in our model is given by 
$ M  ( h ) = 4^{k-1} $, see \eqref{eq:Mh}, so 
we have proven \eqref{eq:t2.1} in Theorem \ref{t:2}.

%%%%%%%%%%%%%%%%%%%%%%%%%%%%%%%%%%%%%%%%%%%%%%%%%%%%%%%%%%%%%%%%%%%%%%%%%%%%%%%

\subsection{Noise power}
\label{noise}

The conductance corresponds to the first moment of the distribution of transmission
eigenvalues. It can not distinguish between a purely classical transport ($T_i\in\set{0,1}$)
and a quantum one (some $T_i$ take intermediate values). 
To do so, we need to compute the second moment of these eigenvalues,
that is, the trace
$$
\tr((t^* t)^2)=\sum_{Q_j\in L_1}\mn{t^* t|Q_j\ra}^2\,,
$$
or equivalently the noise power \eqref{e:power}.
As in the previous section, we split the sum on the right hand side 
between the
classical and nonclassical states $|Q_j\ra=|\bep\ra$.

Lemma~\ref{l:classical-state} shows that half the classical states are
in the kernel of $t^* t$, half in the eigenspace of $t^* t$ associated
with the eigenvalue $1$ (full transmission).  
As a consequence, the trace over the classical states takes the value
$$
\tr_{\rm{cl}}((t^* t)^2)=\frac{4^{k-1}-2^{k-1}}{2}\,.
$$
Obviously, the classical states are noiseless.

The contribution of the nonclassical states is more delicate. 
According to Lemma~\ref{l:cutoff-Theta},
for any nonclassical state $|\bep\ra$ we have (for any $0<\Theta<1$)
$$
t|\bep\ra=\sum_{m=0}^{\lfloor\Theta k\rfloor}e^{in\vartheta}\,
t_{k+m}|\bep\ra +\Oo(2^{-\Theta k/2})\,.
$$
We now apply to each state
$t_{k+m}|\bep\ra$, $m\leq\Theta k$, the adjoint operator 
\bequ\label{e:exp-t*}
t^*=\sum_{n\geq 0} e^{-in\vartheta}\,t^*_n\,.
\end{equation}
According to \eqref{e:tn2}, the state $t_{k+m}|\bep\ra$ has the form
$$
t_{k+m}|\bep\ra=e_3\otimes \F_4^*\pi_I w_2\otimes\cdots\otimes \F_4^*\pi_I w_k\,,
$$
for some explicit set of qu$D$its $w_\ell\in\CC^4$, $2\leq\ell\leq k$.
From the expressions
\begin{align*}
t^*_n&=\Pi_{L_1}\,U^*_h\,(\Pi_I\,U_h^*)^{n-1}\,\Pi_{L_2}\,,\\
\Pi_I U^*_h(v_1\otimes\cdots\otimes v_k)&=\pi_I\F_4 v_k\otimes v_1\otimes\cdots\otimes v_{k-1}\,,
\end{align*}
we can easily write the action of 
the operators $t^*_n$
on $t_{k+m}|\bep\ra$:
$$
\mbox{if $n<k$, then }\quad t^*_n\,t_{k+m}|\bep\ra =
\pi_0\pi_I w_{k-n+1}\otimes\ldots =0\,.
$$
The first non-trivial case of  $ n = k $ is given by 
$$
t^*_k t_{k+m}|\bep\ra=\pi_0\F_4 e_3\otimes \pi_I w_2\otimes\cdots \pi_I w_k\,,
$$ 
while for any $1\leq m'\leq k-1$, we have 
\begin{multline*}
|\psi_{m',m}(\bep)\ra\defeq t^*_{k+m'} t_{k+m}|\bep\ra= \\
\pi_0\F_4\pi_I w_{k-m'+1}\otimes \pi_I\F_4\pi_I w_{k-m'+2}\otimes
\cdots \pi_I\F_4\pi_I w_k\otimes \pi_I\F_4 e_3\otimes\pi_I w_2\otimes\cdots \pi_I w_{k-m'}\,.
\end{multline*}
The above state is obtained by first evolving $|\bep\ra$ $k+m$ times through
the ``inside propagator'' $ U_h\Pi_I$, then projecting on the ``output lead'' $L_2$,
then evolving backwards (through $\Pi_I U^*_h$) $k+m'$ times, and finally projecting on
the ``input lead'' $L_1$.

As for the case of the operator $t$, we see that by increasing $m'$ we increase the number
of qu$D$its on which we apply the operator $\pi_I\F_4$. Therefore, for any index $m$, the
norm of $|\psi_{m',m}(\bep)\ra$ will decrease exponentially fast with $m'$. 
As in Lemma~\ref{l:cutoff-Theta}, we truncate the 
expansion \eqref{e:exp-t*} to the range $m'\leq \Theta k$, thereby omitting
a remainder $\Oo(2^{-\Theta k/2})$.

We now replace the qu$D$its $w_\ell$
by their explicit values, which depend on the index $m$. 
We introduce the following notations for operators on $\CC^4$:
$$ 
\cP_{\alpha\beta} \defeq \pi_\alpha\F_4\pi_\beta\F_4^* \,,\qquad\mbox{with}\quad
\alpha,\beta\in\{0,I,3\} \,.
$$
The explicit decomposition of  
$|\psi_{m',m}(\bep)\ra$ depends on  the sign of
$\Delta\defeq m-m'$, and on whether $m,m'=0$ or not (we will often omit to indicate the dependence
in $\bep$):
\begin{itemize}
\item for $m'=m$,
\bequ
\begin{split}
|\psi_{0,0}\ra&=\cP_{03}e_0\otimes e_{\ep_{2}}\otimes\cdots e_{\ep_k}\,,\\
|\psi_{m,m}\ra&=\cP_{0I} e_0\otimes
\cP_{II}e_{\ep_2}\otimes \cdots\cP_{II}e_{\ep_m}\otimes
\cP_{I3}e_{\ep_{m+1}}\otimes e_{\ep_{m+2}}\otimes\cdots e_{\ep_k}\,.
\label{eq:Delta=0}
\end{split}
\end{equation}
\item for $m=m'+\Delta$, $\Delta>0$, 
\bequ
\begin{split}
|\psi_{0,\Delta}\ra&=\cP_{03}e_{\ep_{\Delta+1}}\otimes e_{\ep_{\Delta+2}}\otimes\cdots
\cdots e_{\ep_k}
\otimes \pi_I\F_4^* e_0\otimes\pi_I\F_4^*e_{\ep_2}\otimes\cdots\pi_I \F_4^* e_{\ep_{\Delta}}\\
|\psi_{m',m'+\Delta}\ra&=\cP_{0I}e_{\ep_{\Delta+1}}\otimes
\cP_{II}e_{\ep_{\Delta+2}}\otimes
\cdots \cP_{II}e_{\ep_{\Delta+m'}}\otimes \\
&\qquad\otimes \cP_{I3}e_{\ep_{\Delta+m'+1}}\otimes e_{\ep_{\Delta+m'+2}}\otimes\cdots e_{\ep_k}
\otimes \pi_I\F_4^* e_0\otimes\pi_I\F_4^*e_{\ep_2}\otimes\cdots\pi_I \F_4^* e_{\ep_{\Delta}}
\label{eq:Delta>0}
\end{split}
\end{equation}
\item for $m=m'+\Delta$, $\Delta<0$, 
\bequ
\begin{split}
|\psi_{|\Delta|,0}\ra&=\pi_0\F_4 e_{\ep_{k-|\Delta|+1}}\otimes
\pi_I\F_4 e_{\ep_{k-|\Delta|+2}}\otimes\cdots
\pi_I\F_4 e_{\ep_{k}}\otimes\cP_{I3}e_{0}\otimes e_{\ep_{2}}\otimes
\cdots e_{\ep_{k-|\Delta|}}\,,\\
|\psi_{m+|\Delta|,m}\ra&=\pi_0\F_4 e_{\ep_{k-|\Delta|+1}}\otimes\pi_I\F_4 e_{\ep_{k-|\Delta|+2}}
\otimes\cdots
\pi_I\F_4 e_{\ep_{k}}\otimes \cP_{II} e_0\otimes\\
&\qquad\otimes\cP_{II} e_{\ep_2}\otimes\cdots
\otimes \cP_{II}e_{\ep_m}\otimes \cP_{I3}e_{\ep_{m+1}}\otimes e_{\ep_{m+2}}\otimes
\cdots e_{\ep_{k-|\Delta|}}\,.
\label{eq:Delta<0}
\end{split}
\end{equation}
\end{itemize}
We notice that each of these states contains subfactors
$e_{\ep_{m+2}}\otimes\cdots \otimes e_{\ep_{k}}$ if $m\geq m'$, and
$e_{\ep_{m+2}}\otimes\cdots \otimes e_{\ep_{k+m-m'}}$ if $m< m'$. 
Compared to its position in the 
tensor product expansion of $|\bep\ra$, this subfactor is shifted
by $m'-m=-\Delta$ steps. 
From this remark, and using similar methods as for Lemma~\ref{l:nonperiodic1},
we can easily prove the following
%%%%%%%%%%%%%%
\begin{lem}\label{l:nonperiodic2}
Let  $0<\Theta<1/6$
and for any pair of indices $ (m, m') $, denote 
$ \Delta = m-m'$. 
There exists $C = C( \Theta ) >0$ such that 
$$
\frac{\#\set{{\rm nonclass.}\ \bep\,:\, \exists \, 
m_1,m'_1,m_2,m'_2 \in[0,\Theta k],\ 
\Delta_1\neq \Delta_2,\ \la \psi_{m'_1,m_1}|\psi_{m'_2,m_2}\ra\neq 0}}
{\#\set{{\rm nonclass.}\ \bep}}\leq C\, 2^{-Ck }\,.
$$
\end{lem}
%%%%%%%%%%%%%%
In other words,
for a generic nonclassical sequence $\bep$,
any two states 
$|\psi_{m'_1,m_1}(\bep)\ra$,
$|\psi_{m'_2,m_2}(\bep)\ra$ 
with $m_i,\,m'_i\leq\Theta k$
will be orthogonal to each other if $\Delta_1\neq \Delta_2$, that is, if the 
shifts between their respective forward and backward evolution times are different.

From now on we assume that $\bep$ is a generic nonclassical sequence in the sense of
the above lemma. 
If we group the states $|\psi_{m+\Delta,m}(\bep)\ra$ into
$$
|\Psi_{\Delta}(\bep)\ra\defeq \sum_{\substack{0\leq m,m'\leq\Theta k\\m=m'+\Delta}}
|\psi_{m+\Delta,m}(\bep)\ra\,,
$$
then genericity implies
that $\la \Psi_{\Delta}(\bep)|\Psi_{\Delta'}(\bep)\ra = 0 $ if
$\Delta\neq\Delta'$. The square-norm of $t^*t|\bep\ra$
is then given by
\bequ\label{e:sum-t*t}
\mn{t^*t|\bep\ra}^2=\sum_{|\Delta|\leq\Theta k}\ \mn{\Psi_{\Delta}(\bep)}^2
+\Oo(2^{-\Theta k/2})\,.
\end{equation}
%The remainder comes from the truncation in $m,\ m'$. 
%and will be proven in the Remark~\ref{r:remainder}. 
As we will see, no
further simplification occurs in this expression, meaning that two different states 
$|\psi_{m',m}\ra$ with the same $\Delta$ will generally interfere with each other.
Our remaining task consists in computing each square norm on the right hand side 
of \eqref{e:sum-t*t}. 
We will
use the explicit tensor decompositions (\ref{eq:Delta=0}-\ref{eq:Delta<0}), and notice that 
the overlap
between two states $|\psi_{m',m}\ra$ is the product of the overlaps of their tensor factors.
We split the lengthy, yet straightforward computation according to the value of $\Delta$.

%%%%%%%%%%%%%%%%%%%%%%%%%%%%%%%%%
\subsubsection{Norm of $\Psi_{0}$}
We have 
\bequ\label{e:sum0}
\mn{\Psi_{0}}^2=\sum_{m\leq\Theta k}\mn{\psi_{m,m}}^2+2\sum_{0\leq m<n\leq \Theta k} 
\Re \la\psi_{m,m}|\psi_{n,n}\ra \,.
\end{equation}
The successive diagonal terms take the values
\begin{align*}
\mn{\psi_{0,0}}^2&=\mn{\cP_{03}e_0}^2=\frac1{16}\,,\quad\text{while for}\ m\geq 1\,,\\
\mn{\psi_{m,m}}^2&=\mn{\cP_{0I} e_0}^2
\big(\prod_{\ell=2}^{m}\mn{\cP_{II} e_{\ep_\ell}}^2\big)\,\mn{\cP_{I3}e_{\ep_{m+1}}}^2
=\frac14\,\big(\frac3{8}\big)^{m-1}\,\frac18\,.
\end{align*}
The sum over the diagonal terms is therefore 
$$
\sum_{m\leq\Theta k}\mn{\psi_{m,m}}^2
%=\frac1{16}+\frac1{32}\;\frac{1-\big(\frac{3}{8}\big)^{[\Theta k]}}{1-\frac{3}{8}}
=\frac{9}{80}
+\Oo\big((3/8)^{\Theta k}\big)\,.
$$
The nondiagonal terms read, for any $0<n\leq \Theta k$,
$$
\la\psi_{0,0}|\psi_{n,n}\ra=\la \cP_{03}e_0,\cP_{0I} e_0\ra\,
\big(\prod_{\ell=2}^{n}\la e_{\ep_\ell}, \cP_{II} e_{\ep_\ell}\ra\big)
\la e_{\ep_{n+1}}, \cP_{I3} e_{\ep_{n+1}}\ra
=\frac18\,\big(\frac1{2}\big)^{n-1}\,\frac1{4}\,,
$$
and for $0<m<n\leq \Theta k$, one similarly gets
$$
\la\psi_{m,m}|\psi_{n,n}\ra=
\frac14\,\big(\frac3{8}\big)^{m-1}\,\frac{1\pm i}{16}\,\big(\frac1{2}\big)^{n-m-1}\,\frac1{4}\,.
$$
The sign is $ + $ if
$\ep_{m+1}=1$, and $ - $ if $\ep_{m+1}=2$.
Adding up the real parts of these off-diagonal terms, we obtain
$$
2\sum_{0 \leq m<n\leq \Theta k} \Re \la\psi_{m,m}|\psi_{n,n}\ra
=\frac3{20}+\Oo(2^{-\Theta k})\,.
$$
We notice that this contribution is of the same order as the diagonal one.
Summing the diagonal and nondiagonal parts yields the norm
\bequ\label{e:Psi0}
\mn{\Psi_0}^2=\frac{21}{80}+\Oo(2^{-\Theta k})\,.
\end{equation}

%%%%%%%%%%%%%%%%%%%%%%%%%%%%%%%%%
\subsubsection{Norm of $\Psi_\Delta$ with $\Delta>0$}
\label{s:Delta>0}

From Eq.~\eqref{eq:Delta>0} we notice that all states 
$|\psi_{m,m+\Delta}\ra$, $0\leq m\leq \Theta k-\Delta$
share the same $\Delta$ last qu$D$its, which results in a common factor 
$$
\prod_{\ell=1}^{\Delta}\mn{\pi_I\F_4^*e_{\ep_\ell}}^2=\frac1{2^{\Delta}}\quad
\text{in the norm}\ \mn{\Psi_\Delta}^2\,.
$$
To avoid taking this factor into account at all steps, we rather consider
the states $|\psi'_{m,m+\Delta}\ra$ obtained by removing these last $\Delta$ qu$D$its
from $|\psi_{m,m+\Delta}\ra$.

We first compute the square-norm
$\mn{\psi'_{0,\Delta}}^2%=\mn{\cP_{03}e_{\ep_{\Delta+1}}}^2
=\frac1{16}$.

For all $m>0$, the first qu$D$it of  $|\psi'_{m,m+\Delta}\ra$ is 
$\cP_{0I}e_{\ep_{\Delta+1}}$.
From the explicit expression of $\cP_{0I}$, 
this qu$D$it vanishes if $\ep_{\Delta+1}=2$, so that
\bequ\label{e:1stcase}
\text{ if $\ep_{\Delta+1}=2$,}\qquad
\mn{\Psi_\Delta(\bep)}^2=\frac1{16}\;\frac1{2^{\Delta}} \quad
\text{for all}\  1\leq\Delta\leq\Theta k\,. 
\end{equation}
In the opposite case $\ep_{\Delta+1}=1$, the states $|\psi'_{m,m+\Delta}\ra$ 
are nontrivial: 
\begin{align}
\mbox{for any $0<m\leq\Theta k-\Delta$,}\qquad 
\mn{\psi'_{m,m+\Delta}}^2&=\frac18\,(\frac38)^{m-1}\,\frac18\,,\nonumber\\
\mbox{so that}\qquad\sum_{m=0}^{[\Theta k]-\Delta}\mn{\psi'_{m,m+\Delta}}^2
&=\frac7{80} +\Oo\big((3/8)^{\Theta k-\Delta})\big)\,.
\label{e:diag2}
\end{align}
We then compute the off-diagonal terms:
\begin{align*}
\la \psi'_{0,\Delta}|\psi'_{m,m+\Delta}\ra&=
\frac{-1-i}{16}\;\frac1{2^{m-1}}\;\frac14
\qquad\text{for $0<m\leq \Theta k-|\Delta|$,}\\
\la \psi'_{m,m+\Delta}|\psi'_{n,n+\Delta}\ra&=
\frac18\,(\frac38)^{m-1}\;\frac{1\pm i}{16}\,\frac1{2^{n-m-1}}\;\frac14
\qquad\text{for $1\leq m<n\leq \Theta k-|\Delta|$}
\end{align*}
(the sign $\pm$ in the last line depends on $\ep_{\Delta+m+1}$). 
Summing up the real parts yields:
$$
2\sum_{0\leq m<n\leq \Theta k-\Delta}\Re \la \psi'_{m,m+\Delta}|\psi'_{n,n+\Delta}\ra
=-\frac{1}{20}+\Oo(2^{-\Theta k+\Delta})\,.
$$
Adding this to the diagonal terms \eqref{e:diag2}, restoring the factor $2^{-\Delta}$, 
and using \eqref{e:1stcase}
results in the following norm (which explicitly depends on $\bep$):
\bequ\label{e:2dcase}
\mn{\Psi_\Delta(\bep)}^2=\frac1{2^\Delta}
\big(\frac3{80}\,\delta_{\ep_{\Delta+1}=1}+\frac1{16}\,\delta_{\ep_{\Delta+1}=2}\big)
+\Oo(2^{-\Theta k}) \quad \text{for any}\ \ 1\leq\Delta\leq\Theta k\,.
\end{equation}

%%%%%%%%%%%%%%%%%%%%%%%%%%%%%%%%%
\subsubsection{Norm of $\Psi_\Delta$ with $\Delta<0$}\label{s:Delta<0}
As in the previous case,
we notice from \eqref{eq:Delta<0} that all components of $\Psi_\Delta$ share the
same $|\Delta|$ first qu$D$its, which contribute a factor
\bequ\label{e:factor}
\mn{\pi_0\F_4 e_{\ep_{k-|\Delta|+1}}}^2
\Big(\prod_{\ell=2}^{|\Delta|}\mn{\pi_I\F_4 e_{\ep_{k-|\Delta|+\ell}}}^2\Big)
=\frac14\,\frac1{2^{|\Delta|-1}}\,.
\end{equation}
We call $\psi'_{m+|\Delta|,m}$ the states with these $|\Delta|$ qu$D$its removed.
They have the norms
$$
\mn{\psi'_{|\Delta|,0}}^2=
\frac18\,,\quad\text{and for $m\geq 1$,}\quad%\\
\mn{\psi'_{m+|\Delta|,m}}^2=
\frac18\,(\frac38)^{m-1}\,\frac18\,.
$$
Hence, the diagonal contribution reads
$$
\sum_{m=0}^{\lfloor \Theta k\rfloor-|\Delta|}\mn{\psi'_{m+|\Delta|,m}}^2=\frac3{20} 
+\Oo\big((3/8)^{\Theta k-|\Delta|})\,.
$$
The nondiagonal terms take the values
\begin{align*}
\la \psi'_{|\Delta|,0}|\psi'_{n+|\Delta|,n}\ra&=
\frac{-1+i}{16}\,\frac1{2^{n-1}}\,\frac14 
\qquad\text{for $0<n\leq \Theta k-|\Delta|$,}\\
\la \psi'_{m+|\Delta|,m}|\psi'_{n+|\Delta|,n}\ra &=
\frac18\,(\frac38)^{m-1}\,\frac{1\pm i}{16}\,\frac1{2^{n-m-1}}\,\frac14
\qquad\text{for $1\leq m<n\leq \Theta k-|\Delta|$}\,.
\end{align*}
These contributions sum up to
$$
2\sum_{0\leq m<n\leq \Theta k-|\Delta|}\Re \la \psi'_{m+|\Delta|,m}|\psi'_{n+|\Delta|,n}\ra
=-\frac{1}{20}+\Oo(2^{-\Theta k+|\Delta|})\,.
$$
Putting together this with the diagonal contributions and restoring the factor \eqref{e:factor},
yields  
\bequ\label{e:Delta<0}
\mn{\Psi_{\Delta}}^2
=\frac1{20}\,\frac1{2^{|\Delta|}}+\Oo(2^{-\Theta k})\,,\quad -\Theta k\leq \Delta\leq -1\,.
\end{equation}
%As opposed to the case of $\Delta> 0$, this norm does not depend on 
%$\bep$.

%\begin{rem}\label{r:remainder}
%In the course of the computations, we have obtained expressions for the
%norms of all states $|\psi_{m',m}\ra$, which allow us to estimate the remainder
%in \eqref{e:sum-t*t}. Indeed, the expressions obtained in the previous subsections
%are valid for any pair $m,m'< k/2$, and in this case the norms satisfy:
%$$
%\mn{|\psi_{m',m}(\bep)\ra}\leq C\,\left(\frac38\right)^{\min(m',m)/2}\,2^{-|m'-m|/2}\,,
%$$
%for a constant $C>0$ independent of $\bep$. Using this we easily proves 
%the estimate \eqref{e:sum-t*t} for a generic nonclassical sequence.
%\end{rem}

%%%%%%%%%%%%%%%%%%%%%%%%%%%%%%%%%%%%%%%%%%%%%%%%%%%%%%%%%%%%%%%%%%%%%%%%%%%%%%%
\subsubsection{Summing up}

We can now sum over all shifts $\Delta$, $|\Delta|\leq \Theta k$ for a
given generic nonclassical sequence $\bep$.
The sum over the shifts $\Delta\leq 0$ is simple, and independent on the sequence $\bep$:
$$
\sum_{-\Theta k\leq\Delta\leq 0}\mn{\Psi_{\Delta}(\bep)}^2
=\frac{25}{80}+\Oo(k\,2^{-\Theta k})\,.
$$
The sum over the shifts $\Delta>0$ is slightly more delicate, since the 
norm of $|\Psi_\Delta(\bep)\ra$
depends on $\bep$ --- see Eq.~\eqref{e:2dcase}. 
However, we
notice that the set of generic nonclassical sequences can be partitioned into ``mirror pairs''
$(\bep,\overline{\bep})$ such that 
$$
\forall \ell\in[2,k],\quad \overline{\ep}_{\ell}=1\Longleftrightarrow \ep_\ell=2\,.
$$
Summing the norms over a ``mirror pair'' is easy:
$$
\mbox{for}\  1\leq \Delta\leq\Theta k,\qquad
\mn{\Psi_{\Delta}(\bep)}^2+\mn{\Psi_{\Delta}(\overline{\bep})}^2=\frac1{2^\Delta}\,
\frac1{10}+\Oo(2^{-\Theta k})\,.
$$
This contribution is identical (up to the remainder) with
$\mn{\Psi_{-\Delta}(\bep)}^2+\mn{\Psi_{-\Delta}(\overline{\bep})}^2$, which shows a
sort of symmetry between positive and negative shifts.
Summing over all $|\Delta|\leq\Theta k$, we get, 
for any mirror pair $(\bep,\overline{\bep})$ of generic nonclassical sequences:
$$
\mn{t^* t|\bep\ra}^2+\mn{t^* t|\overline{\bep}\ra}^2=\frac{58}{80}+\Oo(2^{-\Theta k/2})\,.
$$
Using Lemma \ref{l:nonperiodic2}, we obtain the trace over the 
nonclassical states:
$$
\tr_{\rm{noncl}}((t^* t)^2)=2^{k-2}\,\Big(\frac{58}{80}+\Oo(2^{-Ck})+\Oo(2^{-\Theta k/2})\Big)\,.
$$
Substracting this expression from the ``nonclassical conductance'' \eqref{e:nonclas-conduct},
and calling $\widetilde C=\min(C,\half\Theta)$, we finally obtain the noise power:
\bequ
P=\tr(t^* t-(t^* t)^2)=\tr_{\rm{noncl}}(t^* t-(t^* t)^2)
=2^{k-1}\,\Big(\frac{11}{80}+\Oo(2^{-\widetilde C k})\Big)\,.
\end{equation}
This proves \eqref{eq:t2.2} in Theorem~\ref{t:2}.
As remarked in \S \ref{sotr}, the factor $11/80$
is close to the random-matrix prediction for 
this quantity, namely $1/8$ \cite{jalabert,beenakker}. 
This is in contrast with
our remark \ref{r:noRMT} that the semiclassical resonance spectrum of the 
propagator inside the dot, $\tBh=U_h\Pi_I$, is quite different from
that of a random subunitary matrix. Somehow, the matrix $t$, obtained by summing iterates
of $\tBh$, has acquired some ``genericity'', as far as the distribution of 
its singular values is concerned. It would be interesting (but quite cumbersome)
to compute the higher
moments of that distribution.

%%%%%%%%%%%%%%%%%%%%%%%%%%%%%%%%%%%%%%%%%%%%%%%%%%%%%%%%%%%%%%%%%%%%%%%%%%%%%%%
%%%%%%%%%%%%%%%%%%%%%%%%%%%%%%%%%%%%%%%%%%%%%%%%%%%%%%%%%%%%%%%%%%%%%%%%%%%%%%%

\end{document}